\PassOptionsToPackage{pdftex}{xcolor}

\documentclass[11pt, a4paper]{gdm_format}
\usepackage[authoryear, sort&compress, round]{natbib}

\usepackage{amsmath,amsfonts,bm}

\def\eqref#1{equation~\ref{#1}}

\def\1{\bm{1}}

\DeclareMathAlphabet{\mathsfit}{\encodingdefault}{\sfdefault}{m}{sl}
\SetMathAlphabet{\mathsfit}{bold}{\encodingdefault}{\sfdefault}{bx}{n}

\usepackage{hyperref}
\usepackage{url}
\usepackage{booktabs} 
\usepackage{array} 
\usepackage{multirow} 
\usepackage{amssymb}
\usepackage{graphicx} 
\usepackage{float}    
\usepackage{listings}
\usepackage[listings,theorems]{tcolorbox}
\usepackage{fontawesome5}
\usepackage{pgffor}

\usepackage{amsmath}
\usepackage{amssymb} 
\usepackage{xspace}
\usepackage{multirow}
\usepackage{booktabs}
\usepackage{color}
\usepackage{colortbl}
\usepackage{cleveref}
\usepackage{graphicx}
\usepackage[ruled, vlined, linesnumbered]{algorithm2e}
\usepackage{algpseudocode}
\usepackage{subcaption}
\usepackage{wrapfig}
\usepackage{sidecap}
\usepackage{soul}
\usepackage{enumitem}
\sidecaptionvpos{figure}{t}
\usepackage{multicol}
\usepackage{pifont}
\usepackage[normalem]{ulem}
\usepackage{titletoc}

\usepackage{longtable}
\usepackage{adjustbox}
\usepackage{marvosym}

\usepackage{xargs}       

\definecolor{kwblue}{RGB}{38,91,246}
\definecolor{taskblue}{RGB}{0,99,177}
\definecolor{refgreen}{RGB}{0,150,85}
\definecolor{subviolet}{RGB}{131,76,190}
\definecolor{templateblue}{RGB}{1, 128, 134}
\definecolor{refgreenDark}{RGB}{0, 90, 45}
\definecolor{analysisblue}{HTML}{1E90FF} 
\definecolor{algoyellow}{HTML}{A0522D}   

\lstdefinestyle{beforeoptimcode}{
  language=Python, numbers=left, frame=none,
  numberstyle   = \tiny,        
  numbersep     = 3pt,          
  xleftmargin   = 2pt,          
  framexleftmargin = 0pt,       
  basicstyle=\ttfamily\footnotesize, keywordstyle=\color{subviolet},
  breaklines=true, columns=fullflexible
}

\newcommand{\luz}[1]{{\color{magenta}{[Luz: #1]}}}

\def\ourmethod{EoK}

\newcommand{\commit}{\ensuremath{%
  \mathchoice{\includegraphics[height=1ex]{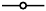}}
    {\includegraphics[height=1ex]{figures/assets/commits/commit.pdf}}
    {\includegraphics[height=0.75ex]{figures/assets/commits/commit.pdf}}
    {\includegraphics[height=0.5ex]{figures/assets/commits/commit.pdf}}
}}
\newcommand{\cmitmsg}{\ensuremath{%
  \mathchoice{\includegraphics[height=2ex]{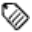}}
    {\includegraphics[height=2ex]{figures/assets/commits/commit_message.pdf}}
    {\includegraphics[height=1.5ex]{figures/assets/commits/commit_message.pdf}}
    {\includegraphics[height=1ex]{figures/assets/commits/commit_message.pdf}}
}}
\newcommand{\cmitcode}{\ensuremath{%
  \mathchoice{\includegraphics[height=2ex]{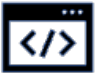}}
    {\includegraphics[height=2ex]{figures/assets/commits/commit_code.pdf}}
    {\includegraphics[height=1.5ex]{figures/assets/commits/commit_code.pdf}}
    {\includegraphics[height=1ex]{figures/assets/commits/commit_code.pdf}}
}}
\newcommand{\thought}{\ensuremath{%
  \mathchoice{\includegraphics[height=2ex]{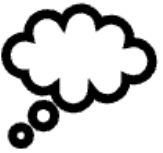}}
    {\includegraphics[height=2ex]{figures/assets/thoughts/thought.pdf}}
    {\includegraphics[height=1.5ex]{figures/assets/thoughts/thought.pdf}}
    {\includegraphics[height=1ex]{figures/assets/thoughts/thought.pdf}}
}}

\newcommand{\thmsg}{\ensuremath{%
  \mathchoice{\includegraphics[height=2ex]{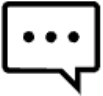}}
    {\includegraphics[height=2ex]{figures/assets/thoughts/thought_msg.pdf}}
    {\includegraphics[height=1.5ex]{figures/assets/thoughts/thought_msg.pdf}}
    {\includegraphics[height=1ex]{figures/assets/thoughts/thought_msg.pdf}}
}}

\newcommand{\thcode}{\ensuremath{%
  \mathchoice{\includegraphics[height=2ex]{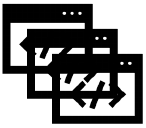}}
    {\includegraphics[height=2ex]{figures/assets/thoughts/thought_code.pdf}}
    {\includegraphics[height=1.5ex]{figures/assets/thoughts/thought_code.pdf}}
    {\includegraphics[height=1ex]{figures/assets/thoughts/thought_code.pdf}}
}}

\newcommand{\idea}{\ensuremath{%
  \mathchoice{\includegraphics[height=2ex]{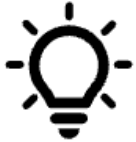}}
    {\includegraphics[height=2ex]{figures/assets/ideas/idea.pdf}}
    {\includegraphics[height=1.5ex]{figures/assets/ideas/idea.pdf}}
    {\includegraphics[height=1ex]{figures/assets/ideas/idea.pdf}}
}}

\newcommand{\ideamsg}{\ensuremath{%
  \mathchoice{\includegraphics[height=2ex]{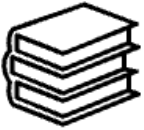}}
    {\includegraphics[height=2ex]{figures/assets/ideas/idea_msg.pdf}}
    {\includegraphics[height=1.5ex]{figures/assets/ideas/idea_msg.pdf}}
    {\includegraphics[height=1ex]{figures/assets/ideas/idea_msg.pdf}}
}}

\lstdefinestyle{mystyle}{
    commentstyle=\color{gray},
}

\definecolor{lighgreenbackground}{RGB}{200, 255, 200}
\definecolor{lighpurplebackground}{RGB}{241, 180, 241}

\definecolor{royalblue}{RGB}{65,105,225}
\definecolor{reference}{RGB}{128,128,128}
\definecolor{codegreen}{rgb}{0,0.6,0}
\definecolor{codegray}{rgb}{0.5,0.5,0.5}
\definecolor{codepurple}{rgb}{0.58,0,0.82}
\definecolor{backcolour}{rgb}{0.95,0.95,0.92}

\definecolor{framegreen}{HTML}{189399}
\definecolor{framegreencodebg}{RGB}{245,248,250}
\definecolor{strred}{HTML}{B80034}

\lstdefinestyle{mystyle}{
    backgroundcolor=\color{backcolour},   
    commentstyle=\color{codegreen},
    keywordstyle=\color{magenta},
    numberstyle=\tiny\color{codegray},
    stringstyle=\color{codepurple},
    basicstyle=\ttfamily\scriptsize,
    breakatwhitespace=false,         
    breaklines=true,                 
    captionpos=b,                    
    keepspaces=true,                 
    numbers=none,                    
    numbersep=5pt,                  
    showspaces=false,                
    showstringspaces=false,
    showtabs=false,                  
    tabsize=2,
}

\lstdefinestyle{pythonstyle}{
    language=Python,
    basicstyle=\ttfamily\scriptsize,
    keywordstyle=\color{blue},
    commentstyle=\color{green!50!black},
    stringstyle=\color{red},
    showstringspaces=false,
    numbers=none,
    numberstyle=\tiny\color{gray},
    frame=single,
    breaklines=true,
    tabsize=4,
}

\tcbset {
  base/.style={
    arc=0mm, 
    bottomtitle=-0.25mm,
    boxrule=0mm,
    colbacktitle=black!10!white, 
    coltitle=black, 
    fonttitle=\bfseries, 
    left=2.5mm,
    leftrule=1mm,
    right=3.5mm,
    title={#1},
    toptitle=0.25mm,
    breakable,
  }
}

\definecolor{brandblue}{rgb}{0.34, 0.7, 1}
\newtcolorbox{mybox}[1]{
  colframe=brandblue, 
  base={#1}
}

\definecolor{pink}{rgb}{1, 0.75, 0.8}
\newtcolorbox{safetybox}[1]{
  colframe=pink, 
  base={#1}
}

\title{Evolution of Kernels: Automated RISC-V Kernel Optimization with Large Language Models}

\correspondingauthor{Zhichao Lu (zhichao.lu@cityu.edu.hk)}

\author[ ]{Siyuan Chen}
\author[ ]{Zhichao Lu$^{\text{(\Letter)}}$}
\author[ ]{Qingfu Zhang}

\affil[ ]{City University of Hong Kong}

\begin{document}

\begin{abstract}

Automated kernel design is critical for overcoming software ecosystem barriers in emerging hardware platforms like RISC-V. 
While large language models (LLMs) have shown promises for automated kernel optimization – demonstrating success in CUDA domains with comprehensive technical documents and mature codebases – their effectiveness remains unproven for reference-scarce domains like RISC-V.
We present Evolution of Kernels (EoK), a novel LLM-based evolutionary program search framework that automates kernel design for domains with limited reference material.
EoK mitigates reference scarcity by mining and formalizing reusable optimization "ideas" (general design principles + actionable thoughts) from established kernel libraries’ development histories; 
it then guides parallel LLM explorations using these ideas, enriched via Retrieval-Augmented Generation (RAG) with RISC-V-specific context, prioritizing historically effective techniques. 
Empirically, EoK achieves a median 1.27$\times$ speedup—surpassing human experts on all 80 evaluated kernel design tasks and improving upon prior LLM-based automated kernel design methods by 20\%. 
These results underscore the viability of incorporating human experience into emerging domains and highlight the immense potential of LLM-based automated kernel optimization.

\end{abstract}

\maketitle


\begin{center}
  \href{https://github.com/Optima-CityU/Evolution_of_Kernels}{\faGithub  \xspace \texttt{https://github.com/Optima-CityU/Evolution\_of\_Kernels}}
\end{center}

\section{Introduction} \label{sec:intro}



RISC-V is democratizing innovation in CPU design through its open, free, and customizable instruction set architecture (ISA)~\citep{waterman2014risc}. 
This enables academic and industrial organizations to develop custom processors without incurring the high costs and licensing fees associated with proprietary ISAs like ARM~\citep{seal2001arm} or x86~\citep{guide2011intel}. 
However, real-world adoption of systems based on RISC-V faces significant barriers due to an immature software ecosystem, primarily stemming from the lack of highly-optimized kernels~\citep{cui2023riscsurvey}. 
Crucially, \emph{kernel}, the core software layer bridging hardware and software, provides essential functionalities such as process management, memory management, and system call handling~\citep{GASTER2012openCL}. 
Its current optimization gap remains a major impediment to broader deployment.

%
%
%
%

Developing efficient RISC-V kernels presents significant challenges, as the process remains largely manual and driven by human experts through trial-and-error~\citep{witteler2024muriscv, rivosinc2025veclibm}.
For proprietary ISAs like x86, where developments remain relatively stable, this approach has produced extensively optimized kernels that deliver significant computational advantages across diverse hardware and software stacks.
Widely adopted libraries such as BLAS~\citep{dongarra1990blas} and LAPACK~\citep{anderson1990lapack} exemplify this success. 
However, this manual process is not only time-consuming and tedious but also demands interdisciplinary expertise spanning hardware, software, and ISA. 
Furthermore, the rapid evolution of RISC-V extensions and hardware, accelerated by its open-source and free-licensing nature, exacerbates these difficulties~\citep{xiangshan2022develop}.  
Consequently, automating the design of efficient RISC-V kernels is critical for ensuring sustainable growth of the RISC-V ecosystem~\citep{igual2023automatic}. 
%

%
%
%
%

Early efforts to automate kernel design~\citep{ansel2014opentuner, schkufza2013stoke} focused on parameterization of the kernel design space, enabling systematic exploration through meta-heuristics~\citep{koza1993genetic, holand1975adaptation}. 
These approaches successfully identified reusable optimization patterns, reducing manual tuning efforts while achieving performance gains on kernels for stable ISAs.
Nevertheless, such methods remained fundamentally constrained by their reliance on human-crafted parameterization schemes, which were both labor-intensive to develop and inherently specific to individual kernels.
Recent advances have sought to overcome these limitations by leveraging large language models (LLMs), capitalizing on their inherent generative capabilities to produce or iteratively refine kernel implementations with minimal human guidance~\citep{lange2025ai, li2025cuda}.
This paradigm has shown particular promise in CUDA kernel optimization, where the availability of comprehensive technical documentation and mature codebases enables even off-the-shelf LLMs to surpass human experts' performance through few iterations of hill-climbing~\citep{nvidia2025AutomatingKernel} or prompt engineering~\citep{agrawal2025gepa}.
However, the effectiveness of this emerging paradigm in domains with scarce reference material, such as RISC-V, remains yet to be explored. 

To this end, we present \emph{Evolution of Kernels (\ourmethod{})}, a LLM-based evolutionary program search framework to automatically design kernels for domains with limited technical documentation or codebases. 
First, to mitigate the reliance on LLMs' intrinsic capabilities, \ourmethod{} systematically extracts and formalizes general design principles from past kernel optimization experience. 
This process involves mining the development history (i.e., git commits) of established kernel libraries to develop a structured pool of \emph{general ideas},  
where each idea encapsulates a general kernel design principle followed by a series of \emph{actionable thoughts},  
comprising succinct descriptions of actionable kernel optimization techniques, example codes, and historical effectiveness. 
With this idea pool, 
\ourmethod{} aims to provide an explicit and data-driven guidance that steers LLM responses, compensating for the absence of comprehensive RISC-V reference material presented in LLM training.
Second, to improve the efficiency of LLM-based iterative search, explorations along distinct kernel optimization directions (defined by ideas sampled from the pool) are carried out in parallel in each iteration of \ourmethod{}. 
Specifically, each idea-guided search samples actionable thoughts with probabilities weighted by their historical effectiveness to ensure prioritization of proven techniques while maintaining exploratory diversity;  
these thoughts are further enriched via retrieval-augmented generation (RAG)~\citep{lewis2020retrieval} with RISC-V specific context, including ISA manuals and hardware profiles, encouraging LLM responses tailored for RISC-V kernel optimization. 
The main contributions of this paper are as follows.

\begin{itemize}[topsep=0pt, itemsep=3pt, parsep=2pt, itemindent=0em]
    \item 
    We present a preliminary attempt, dubbed Evolution of Kernel (\ourmethod{}), to leverage past kernel optimization experience within a LLM-based evolutionary program search paradigm to automatically design kernels for domains with scarce reference material. 

    \item 
    \ourmethod{} systematically extracts and formalizes a set of general kernel design ideas with actionable kernel optimization thoughts from the development history of established kernel libraries; 
    then, \ourmethod{} further enriches these ideas via RAG with RISC-V specific context. 
    Furthermore, explorations along multiple directions defined by distinct ideas are carried out in parallel in each iteration of \ourmethod{} to expedite the overall search. 
    
    \item 
    Empirically, \ourmethod{} successfully identifies kernels with performance surpassing those designed by human experts on all 80 kernel design tasks evaluated in this work. 
    In particular, \ourmethod{} achieves a median speedup of $1.29\times$ on 66 neural network kernels, yielding a $20\%$ improvement over the median speedup of $1.08\times$ achieved by AI CUDA Engineer~\citep{lange2025ai}. 
    
\end{itemize}

\section{Related Work} \label{sec:related}

\subsection{LLM-based Evolutionary Program Search (EPS)}
The pursuit of automating the development of efficient program for computational tasks has historically focused on genetic programming~\citep{koza1993genetic}. 
Recent advances in large language models (LLMs) have enabled a new paradigm termed LLM-based evolutionary program search (EPS), which integrates LLMs within an evolutionary computation framework~\citep{zhang2024understanding}.
In this paradigm, a set of candidate solutions are represented as executable computer \emph{programs} (or codes) that undergo iterative refinement.
LLMs serve as the core search engine, i.e., creating new programs, introducing variations to existing programs, etc.
This approach achieved state-of-the-art results across diverse domains, including mathematical discoveries~\citep{romera2024mathematical}, combinatorial optimization~\citep{liu2024evolution, ye2024reevo}, Bayesian optimization~\citep{yao2024evolve, liu2024large}, machine learning~\citep{ma2024eureka, guo2024autoda}, among many others~\citep{liu2024systematic, novikov2025alphaevolve}.

Notably, kernels represent a specialized class of programs whose optimization could theoretically benefit from LLM-based EPS methodologies. 
While the conceptual compatibility suggests promising applications in automated kernel optimization, the practical effectiveness of this approach remains understudied and warrants systematic investigation.

\subsection{Automated Kernel Optimization}
Early automation efforts for kernel optimization have focused on auto-tuning kernel hyperparameters or configurations with representative methods like OpenTuner~\citep{ansel2014opentuner}, STOKE~\citep{schkufza2013stoke}, and Kernel Tuner~\citep{van2019kernel}.
With the rise of deep learning, substantial attention has turned to optimizing GPU kernels. 
AutoTVM ~\citep{chen2018learning} presents the initial attempt to optimize CUDA kernels, followed by subsequent methods such as Ansor~\citep{zheng2020ansor}, AMOS~\citep{shao2022tensor}, among others~\citep{zhao2024felix, wu2025mirage}. 
Concurrently, there is another line of work that focuses on streamline the performance evaluation of kernels into a standardized procedure such that fair and reproducible comparisons among different kernel optimization methods become fesible. 
Notable frameworks include KernelBench~\citep{ouyang2025kernelbench}, TritonBench~\citep{li2025tritonbench}, and MultiKernelBench~\citep{wen2025MultiKernelBench}.

The advent of LLMs has also redirected automated kernel optimization from meta-heuristics~\citep{koza1993genetic, holand1975adaptation} toward LLM-driven approaches.
This emerging paradigm leverages the inherent generative capabilities of LLMs within an iterative search process, progressively improving an initial kernel implementation towards ones with higher efficiency.
Empirically, this paradigm has consistently yielded more efficient kernels than those optimized by human experts, as demonstrated in AI CUDA Engineer~\citep{lange2025ai}, CUDA-LLM~\citep{chen2025cudallm}, and CUDA-L1~\citep{li2025cuda}, among others~\citep{novikov2025alphaevolve, baronio2025kevin}.

However, existing approaches focus predominantly on CUDA kernels, benefiting from extensive prior resources; their applicability to domains with scarce prior knowledge remains unexplored. 
In this work, we present the first LLM-based EPS approach for optimizing RISC-V kernels, addressing the challenges of limited prior knowledge and the constant emergence of new extensions and hardware variants of RISC-V.

\begin{figure*}[t]
    \centering
    \includegraphics[width=0.95\linewidth]{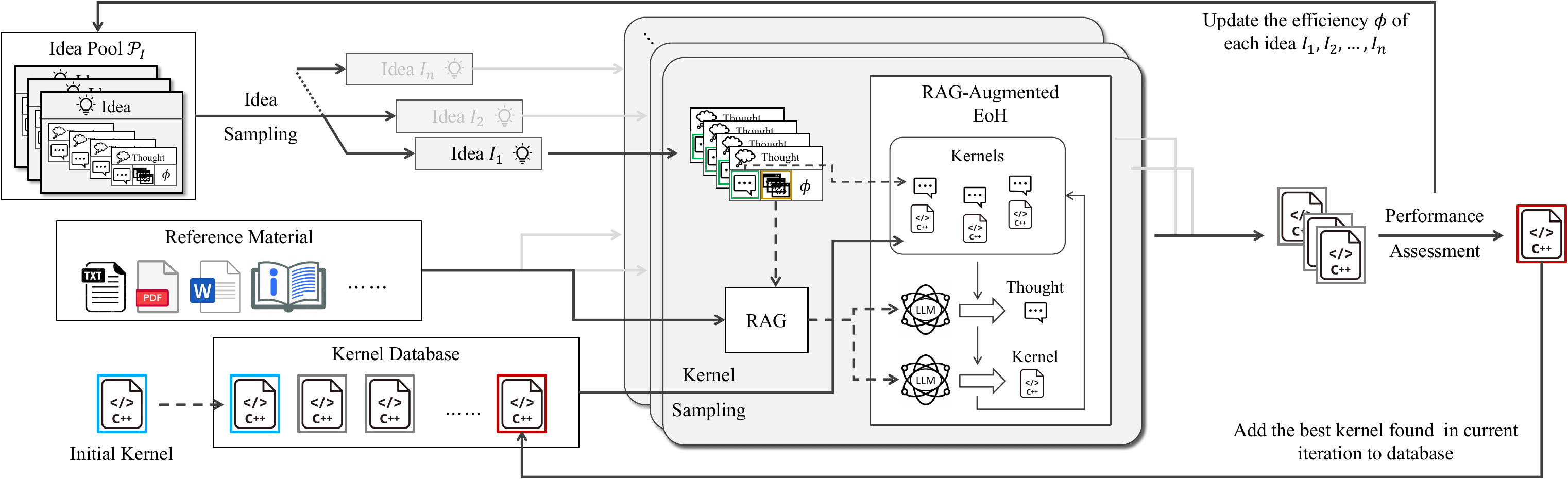}
    \caption{\textbf{Overview of \ourmethod{} Framework}. 
    \ourmethod{} follows the LLM-based evolutionary program search paradigm to automatically optimize kernels specified to RISC-V hardware and ISA extensions. 
    It starts with a pool of ideas learned from past kernel optimization history, a set of reference material provided by users, and an initial implementation of the target RISC-V kernel.
    In each iteration, explorations along distinct directions defined by ideas sampled from the pool are facilitated via multiple EoH~\citep{liu2024evolution} searches in parallel. 
    The implementation of the target kernel is gradually improved over iterations and stored in a database. 
    }
    \label{fig:overview}
\end{figure*}

\section{Our Approach: Evolution of Kernels (\ourmethod{})}

\subsection{Core Design Principle}
\ourmethod{} follows the LLM-based evolutionary program search paradigm to automatically optimize kernels specified to RISC-V hardware and ISA extensions. 
The core design principle behind \ourmethod{} is to \emph{leverage past kernel design experience},  
which is realized through the following two interconnected steps: 

\begin{enumerate}[topsep=0pt, itemsep=3pt, parsep=2pt, itemindent=1.5em]
    \item[\textbf{Step 1.}] Learning from past kernel design experience. 
    Essentially, we aim to transform raw reference materials (e.g., development history of well-established kernel libraries like OpenBLAS~\citep{zhang2012openblas}) into \emph{general ideas} with \emph{actionable thoughts}, which are triplets comprising a brief natural language description, example codes, and estimated effectiveness. 
    (Section \S\ref{subsec:idea-pool})
    
    \item[\textbf{Step 2.}] Applying the learned ideas to guide kernel optimization. 
    In general, we employ the Evolution of Heuristics (EoH)~\citep{liu2024evolution} framework to co-evolve higher-level guidance with lower-level code implementation of kernels simultaneously. 
    In particular, we sample the learned ideas from Step 1 to create an initial population of kernels instead of relying on those randomly generated by LLMs (as done in original EoH); 
    then, we use retrieval augmented generation (RAG)~\citep{lewis2020retrieval} to identify (potentially) related context as supplementary prompts to LLMs; 
    finally, we launch multiple EoH searches (each one with its own ideas) in parallel to expedite the overall procedure. (Section \S\ref{subsec:rag_eoh})
    
\end{enumerate}

\begin{algorithm}[t]
\SetAlgoLined
\SetKwInOut{Input}{Input}
\SetKwInOut{Output}{Output}
\SetKwFor{For}{for}{do}{end for}
\footnotesize
\Input{A reference implementation of the target kernel \texttt{ker}$_0$, reference material \texttt{Doc},\\
\# of ideas to explore in parallel $N$, 
\# of kernels to keep in each iteration $k$, 
maximum iterations $T$.}
$t$ $\leftarrow$ 0 \textcolor{gray}{// initialize an iteration counter}.\\
\texttt{DataBase}$_{\mathcal{K}}$ $\leftarrow$ \texttt{ker}$_0$ \textcolor{gray}{// initialize a kernel database with the provided initial kernel}.\\
$\mathcal{P}_{\mathcal{I}}$ $\leftarrow$ learn a pool of general ideas from past kernel optimization experience. \hspace{10em} $\triangleleft$ \underline{Sec.~\S\ref{subsec:idea-pool}}\\
\While{$t < T$}{
    $i$ $\leftarrow$ $0$, $\mathcal{K}$ $\leftarrow$ $\emptyset$ \textcolor{gray}{// initialize a counter and an empty set for storing candidate kernels}.\\
    \For{$i < N$}{
        \textcolor{gray}{// Generate a new candidate kernel based on idea-guided EoH with RAG}. \hspace{8em} $\triangleleft$ \underline{Sec.~\S\ref{subsec:rag_eoh}} \\
        \texttt{ker} $\leftarrow$ sample a reference kernel from \texttt{DataBase}$_{\mathcal{K}}$.\\
        \texttt{idea} $\leftarrow$ sample an idea from $\mathcal{P}_{\mathcal{I}}$.\\
        $\hat{\texttt{ker}}$ $\leftarrow$ EoH\_with\_RAG$(\texttt{ker}, \texttt{idea}, \texttt{Doc})$ \\
        $\mathcal{K}$ $\leftarrow$ $\mathcal{K}$ $\cup$ \{$\hat{\texttt{ker}}$\}, $i$ $\leftarrow$ $i + 1$\\
    }
    $\Delta_{\text{perf}}$ $\leftarrow$ performance assessment of kernels in $\mathcal{K}$.\\
    \texttt{DataBase}$_{\mathcal{K}}$ $\leftarrow$ \texttt{DataBase}$_{\mathcal{K}}$ $\cup$ \{top-$k$ kernels in $\mathcal{K}$ according to $\Delta_{\text{perf}}$\}.\\
    $\mathcal{P}_{\mathcal{I}}$ $\leftarrow$ update the efficiency of ideas utilized from $\mathcal{P}_{\mathcal{I}}$ according to $\Delta_{\text{perf}}$. \\
    $t$ $\leftarrow$ $t + 1$ \textcolor{gray}{// repeat above steps for $T$ iterations}.
}
\textbf{Return} Top-$1$ kernel in \texttt{DataBase}$_{\mathcal{K}}$.
\caption{Evolution of Kernels (\ourmethod{}) \label{algo:framework}}
\end{algorithm}

\subsection{\ourmethod{} Framework \label{subsec:framework}}
Figure~\ref{fig:overview} and Algorithm~\ref{algo:framework} provide a pictorial and algorithmic overview of \ourmethod{} with the main procedures summarized as follows.

\noindent\textbf{Initialization.} 
\ourmethod{} maintains a database of kernels (\texttt{DataBase}$_{\mathcal{K}}$) comprising one member (i.e., an initial implementation of the target kernel \texttt{ker}$_0$) to begin with.
\ourmethod{} initiates with a pool of general ideas ($\mathcal{P}_{\mathcal{I}}$) learned from past kernel optimization experience.  
A idea is defined by a general kernel design principle and a set of actionable thoughts, 
where each thought comprises a brief natural language description, example codes, and its efficiency estimated from past benchmark data.

\noindent\textbf{Generating Candidate Kernels.} 
In each iteration, $N$ new candidate kernels ($\mathcal{K}$) are generated via $N$ parallel idea-guided EoH searches~\citep{liu2024evolution}.
Specifically, for each EoH search, a general idea and a reference kernel (\texttt{ker}) are sampled from $\mathcal{P}_{\mathcal{I}}$ and \texttt{DataBase}$_{\mathcal{K}}$, respectively;  
then, a LLM is tasked to generate a set of new kernels by applying a sampled subset of actionable thoughts associated with the general idea to \texttt{ker}, 
and this set of (thought, kernel)-pairs serves as the initial population to start the EoH search;
subsequently, additional context information is retrieved from user-provided reference materials (\texttt{Doc}) to augment the generation of new kernels within EoH iterations~\citep{lewis2020retrieval}; 
finally, the best kernel found by each of the $N$ parallel EoH searches are merged together to create $\mathcal{K}$.  

\noindent\textbf{Global Management.}
At the end of each iteration, the performance ($\Delta_{\text{perf}}$) of all kernels in $\mathcal{K}$ (i.e., $N$ new candidate kernels generated in each iteration) are assessed on specified hardware;  
then, top-$k$ kernels from $\mathcal{K}$ (according to $\Delta_{\text{perf}}$) are archived into the \texttt{DataBase}$_{\mathcal{K}}$,  
and the efficiency of all ideas utilized from $\mathcal{P}_{\mathcal{I}}$ in each iteration are updated. 
At termination, the top-$1$ kernel in \texttt{DataBase}$_{\mathcal{K}}$ is regarded as the best kernel found by \ourmethod{}.

\begin{figure}[t]
     \centering
     \captionsetup[subfigure]{justification=centering}
     \begin{subfigure}[b]{0.33\textwidth}
         \centering
         \includegraphics[width=\textwidth]{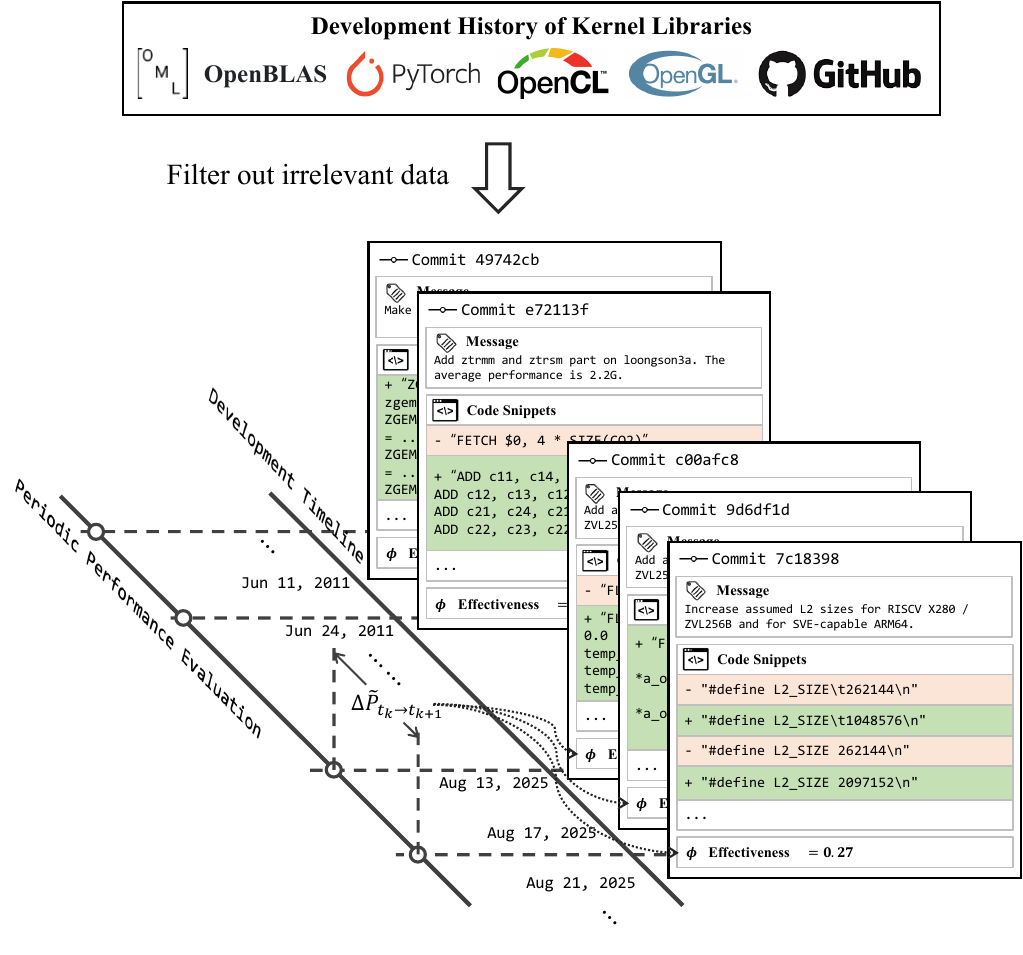}
         \caption{Preparation\label{fig:ip-prep}}
     \end{subfigure}
     \hfill
     \begin{subfigure}[b]{0.42\textwidth}
         \centering
         \includegraphics[width=\textwidth]{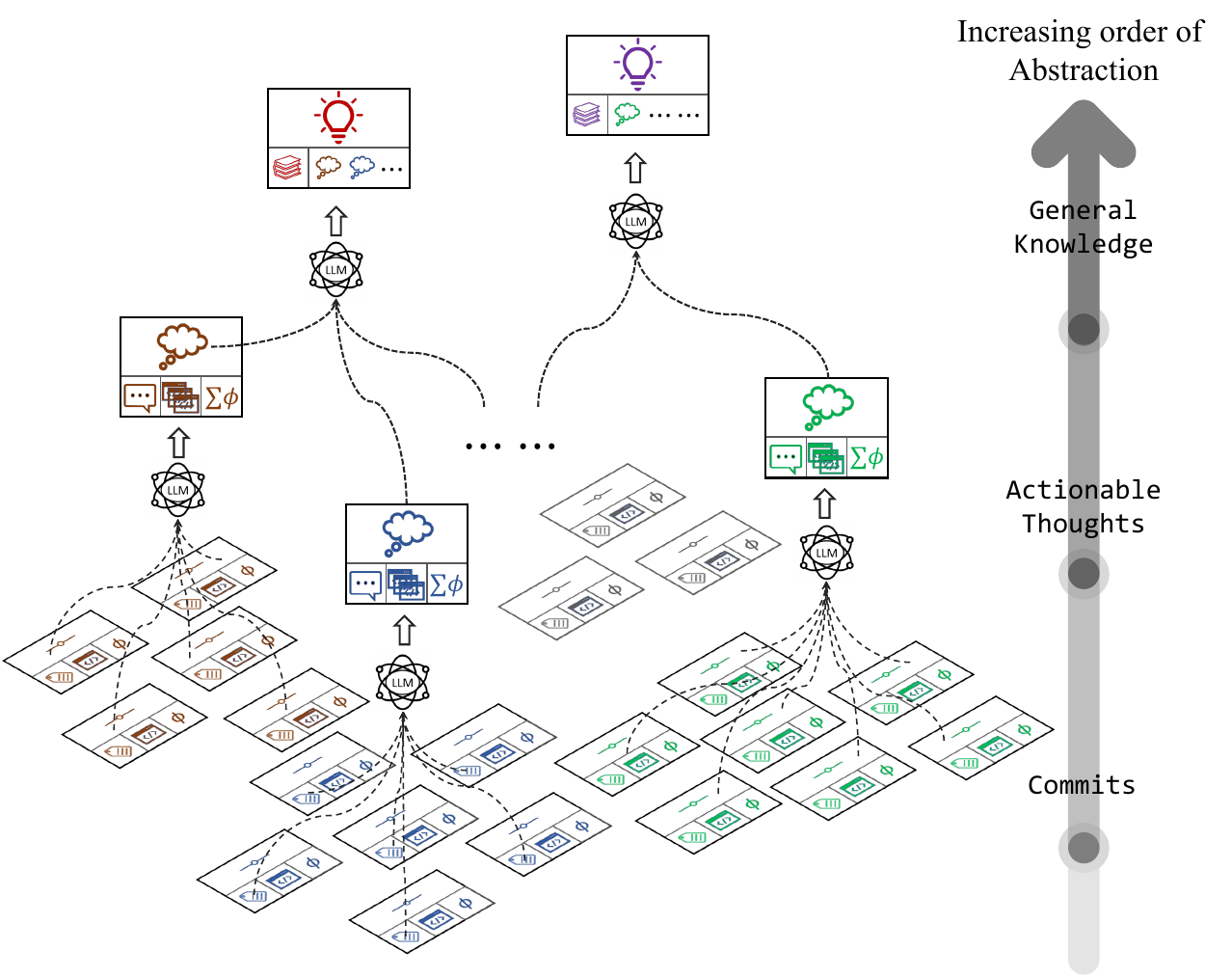}
         \caption{Two-level of Abstraction\label{fig:ip-abs}}
     \end{subfigure}
     \hfill
     \begin{subfigure}[b]{0.23\textwidth}
         \centering
         \includegraphics[width=\textwidth]{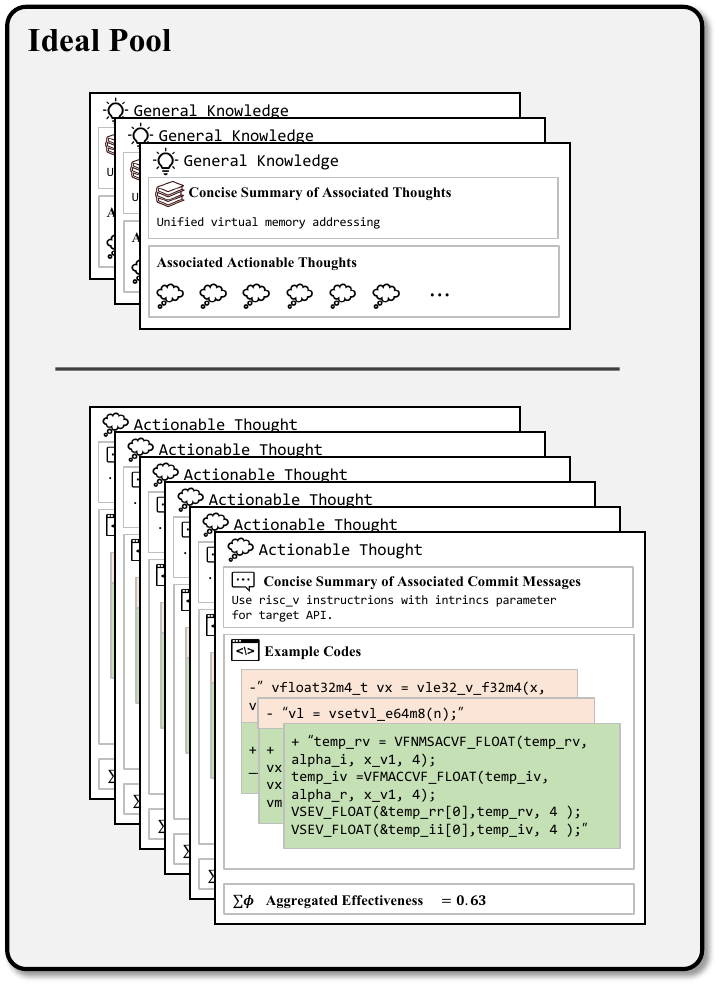}
         \caption{Idea Pool\label{fig:ip}}
     \end{subfigure}
    \caption{\textbf{Overview of Idea Pool Construction.}
    A structured pool of ideas (denoted as $\mathcal{P}_{I}$) is learned from past kernel optimization history to aid LLMs in generating new candidate kernels.
    Specifically, (a) raw commits from existing kernel libraries are filtered and processed; 
    (b) then, two levels of abstractions are applied to transform commits to actionable thoughts (denoted as $\thought$) from which general ideas (denoted as $\idea$) are subsequently distilled. 
    (c) Essentially, an idea comprises a general principle for kernel optimization (denoted as $\ideamsg$) and a set of actionable thoughts, i.e., $\idea \coloneq (\ideamsg, \{\thought, \thought, \ldots\})$;  
    and each thought is a triplet tuple of a concise description (denoted as $\thmsg$), example codes (denoted as $\thcode$), and an estimated effectiveness (denoted as $\phi$), i.e., $\thought \coloneq (\thmsg, \thcode, \phi)$. 
    }
    \label{fig:ip-overview}
\end{figure}

\subsection{Idea Pool: Learning from Past Kernel Optimization Experiences} \label{subsec:idea-pool}
The scarcity of knowledge and resources dedicated to RISC-V kernel design poses a significant challenge for existing automated kernel design methods that rely on LLMs.  
To address this issue, in this work, we aim to learn a structured pool of ideas to aid LLMs in generating new candidate kernels. 
Specifically, we 
%
%
trace back the development history of well-established kernel libraries to identify \emph{actionable thoughts},  
from which \emph{general kernel design ideas} are distilled. 
Figure~\ref{fig:ip-overview} provides a pictorial overview of the process. 
The main procedures are summarized as follows.

\noindent\textbf{Preparation.} 
The development history of a kernel library can be viewed as a sequence of commits (denoted as $\commit$), where each commit comprises \ding{172} code snippets showing the specific changes made to the kernel codes (denoted as $\cmitcode$) and \ding{173} a descriptive message summarizing the key changes (denoted as $\cmitmsg$).
Given a large volume of commits, the following two steps are performed to prepare commits for subsequent process.

\begin{enumerate}[topsep=0pt, itemsep=3pt, parsep=2pt, itemindent=1.5em]
    \item[\textbf{Step 1.}] 
    We filter out commits that are irrelevant to kernel optimization, such as code formatting and version updates. 
    We create vector representations of $\cmitmsg$ from filtered commits via an embedding-LLM. 
    
    \item[\textbf{Step 2.}] 
    We estimate the effectiveness of each commit (denoted as $\phi_{\text{cmit}}$) from the periodic performance evaluations (PPEs) of a kernel library done by the developers\footnote{An example is available from \url{https://tinyurl.com/7zxuc2ur}}.
    Specifically, given any two consecutive PPEs at $t_k$ and $t_{k+1}$,    
    we first identify all commits $\mathcal{C}_{\text{cmit}}$ that were merged between $[t_k, t_{k+1})$; 
    then, $\phi_{\text{cmit}}$ can be computed as $\phi_{\text{cmit}} \coloneq {\Delta \tilde{P}_{t_k \rightarrow t_{k+1}}} / {|\mathcal{C}_{\text{cmit}}|}$, 
    where $\Delta \tilde{P}_{t_k \rightarrow t_{k+1}}$ is the relative performance improvement (averaged over all kernels in a library) between two PPEs at $t_k$ and $t_{k+1}$\footnote{We assume that all commits equally contribute to the observed relative performance improvement between two PPEs for simplicity. More advanced scheme can be used to improve the reliability of the estimation of the effectiveness of each commit.}. 
\end{enumerate}

\noindent Thereafter, a processed commit is triplet tuple of $(\cmitmsg, \cmitcode, \phi_{\text{cmit}})$. 
A pictorial illustration of above-mentioned preparation steps is provided in Figure~\ref{fig:ip-prep}.

\noindent\textbf{Two-Level of Abstraction.}
While these commits provide valuable reference material, many of them exhibit highly similar or fragmented kernel optimization rationales. 
In addition, many commits leverage architecture-specific instructions, hence, their code snippets are not directly applicable to RISC-V. 
To this end, we apply two levels of abstraction to obtain \ding{182} \emph{actionable thoughts} that can be readily applied to optimize RISC-V kernels 
and \ding{183} \emph{general ideas} to provide higher-level guidance. 

\textbf{1. From Commits to Actionable Thoughts:} 
we group commits into clusters based on the similarities among their $\cmitmsg$;  
then we abstract the common kernel optimization technique behind a group of similar commits as actionable thoughts (denoted as $\thought$),   
where each thought comprises a concise description (denoted as $\thmsg$), example codes (denoted as $\thcode$), and an estimated effectiveness ($\phi$).
The main steps are outlined as follows.  

\begin{enumerate}[topsep=0pt, itemsep=3pt, parsep=2pt, itemindent=1.5em]
    \item[\textbf{Step 1.}] \emph{Initialization}. 
    We borrow a set of widely-used kernel optimization principles from the literature\footnote{For instance, ``pinned memory'' and ``unified virtual addressing''; full details are provided in Appendix~\ref{subappx:init-idea}.} as our initial set of thoughts, i.e., $\mathcal{P}_{\mathcal{T}} = \{\thought_1, \thought_2, \ldots\}$.
    Note that all thoughts now only have the $\thmsg$ part. 
    We create vector representations of $\thmsg$ from all thoughts in $\mathcal{P}_{\mathcal{T}}$ via an embedding-LLM. 

    \item[\textbf{Step 2.}] \emph{Associate commits to thoughts}.  
    For each commit, we compute the distances $\mathbf{d}_{\theta}$ of its $\cmitmsg$ to $\thmsg$ of all thoughts in $\mathcal{P}_{\mathcal{T}}$. 
    A commit is assigned to the thought with the shortest distance, provided this distance falls below a certain threshold ($\tau_{\theta} > \min(\mathbf{d}_{\theta})$)\footnote{Note that $\tau_{\theta}$ controls the generality/specificity of the thoughts in $\mathcal{P}_{\mathcal{T}}$. A smaller value of $\tau_{\theta}$ encourages more specific thoughts while a larger value of $\tau_{\theta}$ encourages more general thoughts.}.

    \item[\textbf{Step 3.}] \emph{Cluster remaining commits.} 
    Due to the inherent diversity in kernel optimization principles and techniques, the thoughts in $\mathcal{P}_{\mathcal{T}}$ may not cover all commits. 
    Therefore, commits not associated with any thought in $\mathcal{P}_{\mathcal{T}}$ are clustered using K-means based on the similarities of their $\cmitmsg$ in the embedding space.

    \item[\textbf{Step 4.}] \emph{Distill thoughts}. 
    Given commits from the same group, we employ a LLM to summarize the common technique from their $\cmitmsg$ as new thoughts (only the $\thmsg$ part). 
    We update $\mathcal{P}_{\mathcal{T}}$ by replacing the existing thoughts with these newly generated ones. 
    The prompt is provided in Appendix~\ref{subappx:prompt_idea}. 

    \item[\textbf{Step 5.}] \emph{Refine thought-commits pairing}.
    Given the updated thoughts in $\mathcal{P}_{\mathcal{T}}$, we re-assign all commits to the updated thoughts and re-distill thoughts based on the new associations. 
    We repeat these steps (i.e., Step 2 - 4) until \ding{172} every commit can be assigned to a thought in $\mathcal{P}_{\mathcal{T}}$ with distance shorter than $\tau_{\theta}$, 
    and \ding{173} the (thought, commits)-pairing becomes stable. 
    
\end{enumerate}

\noindent Given the finalized pairing of (thought, commits), we concatenate $\cmitcode$ from all commits associated with a thought to create the code examples $\thcode$ for the thought; 
we compute the effectiveness $\phi$ of a thought by aggregating the effectiveness $\phi_{\text{cmit}}$ of all commits that are associated with the thought, 
as $\phi = \sum{\phi_{\text{cmit}}}$. 
Thereafter, we have a set of actionable thoughts of $\mathcal{P}_{\mathcal{T}} = \{\thought_1, \thought_2, \ldots\} = \{(\thmsg_1, \thcode_1, \phi_1), (\thmsg_2, \thcode_2, \phi_2), \ldots\}$

\textbf{2. From Actionable Thoughts to General Ideas:} 
We further cluster thoughts in $\mathcal{P}_{\mathcal{T}}$ into groups for distilling more general ideas (denoted as $\idea$) that serve as valuable higher-level guidance for RISC-V kernel optimization. 
Specifically, we use K-means clustering with similarities among $\thmsg$ of all thoughts in $\mathcal{P}_{\mathcal{T}}$ as the distance metric; 
then, we employ a LLM to extract the general principle (denoted as $\ideamsg$) behind every cluster of commits from their $\thmsg$. 
Finally, we join $\ideamsg$ with its associated commits to define a general idea as a tuple of $(\ideamsg, \{\thought, \thought, \ldots\})$;  
then, the idea pool $\mathcal{P}_{\mathcal{I}}$ is defined by a set of general ideas structured as:  
\begin{equation} \label{eq:idea-pool}
\begin{split}
\mathcal{P}_{\mathcal{I}} & = \{\idea_1, \idea_2, \idea_3, \ldots\} \\
 & =\big\{(\ideamsg_1, \{\thought^{(1)}, \thought^{(1)}, \ldots\}), (\ideamsg_2, \{\thought^{(2)}, \thought^{(2)}, \ldots\}), \ldots\big\}\\
 & = \bigg\{\Big(\ideamsg_1, \big\{\big(\thmsg^{(1)}, \thcode^{(1)}, \phi^{(1)}\big), \ldots\big\}\Big), \Big(\ideamsg_2, \big\{\big(\thmsg^{(2)}, \thcode^{(2)}, \phi^{(2)}\big), \ldots\big\}\Big), \ldots\bigg\}
\end{split}
\end{equation}

\noindent This structured idea pool $\mathcal{P}_{\mathcal{I}}$ is a fundamental component in \ourmethod{}, providing the stepping-stone for the subsequent idea-guided evolutionary program search of RISC-V kernels.
Pictorial illustrations of the two-level of abstraction and $\mathcal{P}_{\mathcal{I}}$ are provided in Figures~\ref{fig:ip-abs} and \ref{fig:ip}, respectively.

\subsection{Parallel Idea-Guided EoH Searches with RAG \label{subsec:rag_eoh}} 
At core, \ourmethod{} is an evolutionary search process driven by a LLM. 
In each iteration, explorations along distinct directions are facilitated via multiple Evolution of Heuristics (EoH) searches in parallel. 

\noindent\textbf{Preliminaries.} 
EoH~\citep{liu2024evolution} itself is also an iterative search process. 
It maintains a population of $M$ individuals (i.e., heuristics in the original EoH paper and kernels in our context), 
where each individual comprises a natural language description and a corresponding code implementation.  
At each generation, five different prompt strategies are applied to every individual in the population to generate 5$M$ new individuals via a LLM; 
the better $M$ individuals survive to the next iteration, and this process repeats for $G$ generations. 

\noindent\textbf{EoH Initialization.} 
Conventionally, EoH starts with an initial population of individuals randomly generated by a LLM. 
In contrast, \ourmethod{} aims to guide EoH for optimizing kernels along a particular direction defined by an idea learned from past kernel optimization history (i.e., $\mathcal{P}_{\mathcal{I}}$ constructed in Section~\S\ref{subsec:idea-pool}). 
\ourmethod{} realizes this goal by providing a \emph{seeded} initial population to EoH. 
The main steps summarized as follows.

\begin{enumerate}[topsep=0pt, itemsep=3pt, parsep=2pt, itemindent=1.5em]
    \item[\textbf{Step 1.}] 
    We uniformly sample a general idea from $\mathcal{P}_{\mathcal{I}}$; 
    then, $M$ thoughts are sampled from all the actionable thoughts associated with the idea with probabilities in proportional to their efficiencies. 
    These probabilities are normalized via a softmax function over the efficiencies of all thoughts associated with the idea. 
    
    \item[\textbf{Step 2.}] 
    We sample a kernel from \texttt{DataBase}$_{\mathcal{K}}$ with a probability in proportional to its performance.
    Similarly, this probability is also normalized via a softmax function over the performance of all kernels in \texttt{DataBase}$_{\mathcal{K}}$.

    \item[\textbf{Step 3.}] 
    For every sampled thought, we employ a LLM to create a new kernel by asking it to improve the sampled kernel according to the thought. 
    In the end, a set of $M$ thought-kernel pairs is generated and provided to EoH as the initial population. 
    The prompt is provided in Appendix~\ref{subappx:prompt_init}. 
    
\end{enumerate}




\noindent\textbf{EoH with RAG.} 
During each iteration of EoH, a set of distinct prompt strategies are applied to a LLM for generating new individuals, i.e., kernels. 
On top of these prompt strategies, \ourmethod{} use retrieval augmented generation (RAG)~\citep{lewis2020retrieval} to provide additional context to augment the prompt inputs to the LLM. 
The main steps are summarized as follows.

\begin{enumerate}[topsep=0pt, itemsep=3pt, parsep=2pt, itemindent=1.5em]
    \item[\textbf{Step 1.}] 
    The example codes of sampled thoughts along with the reference material (\texttt{Doc}) provided by users serve as the external resources. 
    
    \item[\textbf{Step 2.}] 
    We use EoH's prompt strategies and kernels (both their descriptions and implementations) together as the query to RAG, from which relevant context is retrieved from the external resources. 

    \item[\textbf{Step 3.}] 
    We append the relevant context to EoH's prompt strategies for generating new candidate kernels. 
    The prompt is provided in Appendix~\ref{subappx:prompt_eoh}. 
    
\end{enumerate}

\noindent\textbf{Multiple EoH Searches in Parallel.} 
To expedite the explorations along multiple directions, $N$ EoH searches are performed in parallel. 
Each EoH search will sample a new reference kernel from \texttt{DataBase}$_{\mathcal{K}}$ and a new general idea from $\mathcal{P}_{I}$. 
In the end, we merge the best kernel identified from each EoH search together and select the top $k$ kernels to be stored in \texttt{DataBase}$_{\mathcal{K}}$. 
\section{Experiments}\label{sec:experiments}
In this section, we introduce the benchmark, evaluation metrics and baselines studied in this work, followed by the implementation details. 
Then, we present the empirical results on general-purpose and neural network kernels to evaluate the efficacy of \ourmethod{}. 
A case study on Mish kernel, ablative analysis, and generalization of \ourmethod{} are also studied.

\subsection{Experimental Setup} \label{subsec:setup}

\noindent\textbf{Benchmarks.} 
We consider a total of 80 distinct kernel design tasks, organized into two sets based on their functionalities and applications. 
The first set comprises 14 kernels for general-purpose operations such as sorting, searching, and data manipulation (e.g., insert, merge, move, expand);  
these kernels support the core functionalities in parallel computing environments.
The second set comprises 66 kernels from 38 operations commonly used in neural network computation; 
we group them into three types of operations for (i) feature extraction and transformation (e.g., convolution, pooling, inner product), (ii) activation, normalization, and regularization (e.g., ReLU, batch normalization, dropout), and (iii) tensor operations (e.g., clip, concatenation, flatten); 
these kernels provide the essential functions to run modern deep neural networks such as ResNets~\citep{he2016deep}, YOLOs~\citep{redmon2016you}, and vision Transformers~\citep{dosovitskiy2021an}. 
Readers are referred to Appendix~\ref{appx:benchmark} for details.

\noindent\textbf{Evaluation Metrics.}
The performance of a kernel is determined by its \emph{correctness} and \emph{runtime latency}. 
Specifically, we measure correctness both in compilation and functionality, i.e., a kernel is considered to be correct if \ding{172} it is successfully compiled without errors, 
and \ding{173} its output does not deviates more than a threshold $\epsilon$ from the output of the reference implementation written by human experts, i.e., \texttt{ker}$_0$ in Algorithm~\ref{algo:framework}. 
Thereby, in this work, we compare the number of correct kernels (\#\texttt{Success})\footnote{We exclude the reference kernels provided to each automated kernel design method from the correct kernel count.} identified by different automated kernel design methods. 
In terms of runtime latency, we measure the execution time of a kernel on a specified RISC-V hardware; 
then, we report its relative speedup ($\times$\texttt{SpeedUp}) that is normalized to the execution time of the reference implementation \texttt{ker}$_0$. 

\noindent\textbf{Baselines.} 
We consider the implementations from moderngpu~\citep{Baxter2022moderngpu} and NCNN~\citep{nihui2025Tencent} libraries as the reference implementations for the 14 general-purpose and the 66 neural network kernels, respectively. 
Then, we consider TVM~\citep{chen2018tvm} and Kernel Tuner~\citep{van2019kernel} as the two representative methods from the early automated kernel design approaches; 
and we consider AI CUDA Engineer~\citep{lange2025ai} as the representative method from the LLM-based automated kernel design approaches.
In addition, we also include EoH~\citep{liu2024evolution}, based on which our method is developed. 

\noindent\textbf{Implementation Details.}
In this work, we consider Spacemit K1 as the target RISC-V hardware; 
it features eight X60 cores cloked at 1.5GHz with 128KB L2 Cache per core and 16GB LPDDR4X memory;   
its architecture and block diagram are publicly accessible from \href{https://tinyurl.com/56u8n6pd}{here}.

We use OpenBLAS library~\citep{zhang2012openblas} as the primary data source for constructing the idea pool $\mathcal{P}_{\mathcal{I}}$; 
it contains a rich history of development records spanning 14 years of active maintenance with over 10,000 commits in the main branch alone. 
The efficiency ($\phi$) of ideas in $\mathcal{P}_{\mathcal{I}}$ is estimated from the past periodic performance evaluation of OpenBLAS as a part of the continuous integration workflow~\citep{BLASBenchmarks2025}. 
In addition, we provide the official RISC-V ISA manual~\citep{waterman2014risc}, the profiles of RISC-V extensions (i.e., RVV1.0 and RVA22), and the technical documents of Spacemit K1 as external reference materials for RAG. 

We use Qwen3-235b-a22b~\citep{yang2025qwen3} for generating kernels and Qwen3-Embedding-8B~\citep{zhang2025qwen3embed} for generating text embeddings. 
We compile RISC-V kernels using gcc 13.2 with \texttt{-O3 -march=rv64gcv} flags. 
The following three steps are performed to improve the validity of performance measurements of kernels:
\ding{172} each kernel execution is preceded by cache flushing and thermal calibration; 
\ding{173} runtime latency is averaged over 10 warm runs and 50 measurement runs.
%
%
All other hyperparameter settings are summarized in Table~\ref{tab:hyperparameter}. 


\begin{table}[htb]
\centering
\caption{Summary of hyperparameter settings.}
\label{tab:hyperparameter}
\begin{tabular}{@{\hspace{2mm}}lcl@{\hspace{2mm}}}
\toprule
\textbf{Notation} & \textbf{Value} & \textbf{Description} \\ \midrule
$N$ & $3$ & \# groups of ideas to explore in each iteration. \\
$k$ & $3$ & \# of kernels to keep in each iteration. \\
$T$ & $5$ & Maximum iterations of \ourmethod{}. \\
$M$ & $5$ & Population size of EoH. \\
$G$ & $5$ & Maximum generations of EoH. \\
$\tau_{\theta}$ & 0.05 & Max allowed distance for thought-commits association. \\
$\epsilon$ & 0.01 & \begin{tabular}[c]{@{}l@{}}Max allowed percentage differences in \\outputs for functional correctness.\end{tabular}\\ \bottomrule
\end{tabular}%
\end{table}

\begin{table}[ht]
\centering
\caption{Results on 14 general-purpose kernels. $\times$Speedup measures relative improvement over reference implementations; \#Success excludes the reference implementations from the count.}
\label{tab:results_general}
\begin{tabular}{@{\hspace{2mm}}l|ccccc|cc@{\hspace{2mm}}}
\toprule
\multirow{2}{*}{Method} & \multicolumn{5}{c|}{$\times$Speedup ($\uparrow$)} & \multicolumn{2}{c}{\#Success ($\uparrow$)} \\ \cmidrule(lr){2-6} \cmidrule(lr){7-8} 
 & Mean & Max & 75\% & 50\% & 25\% & (out of total) & ($>1.01\times$ out of total) \\ \midrule
EoH & 1.13 & 1.24 & 1.15 & 1.12 & 1.02 & 4/14 & 4/14 \\
AI CUDA Engineer & 1.06 & 1.14 & 1.07 & 1.03 & 1.01 & 4/14 & 3/14 \\
\textbf{\ourmethod{} (ours)} & \textbf{1.23} & \textbf{1.82} & \textbf{1.27} & \textbf{1.20} & \textbf{1.07} & \textbf{14/14} & \textbf{14/14} \\ \bottomrule
\end{tabular}%
\end{table}

\subsection{Main Results} \label{subsec:results}
\begin{wrapfigure}{r}{0.5\textwidth}
  \vspace{-2em}
  \begin{center}
    \includegraphics[width=0.48\textwidth]{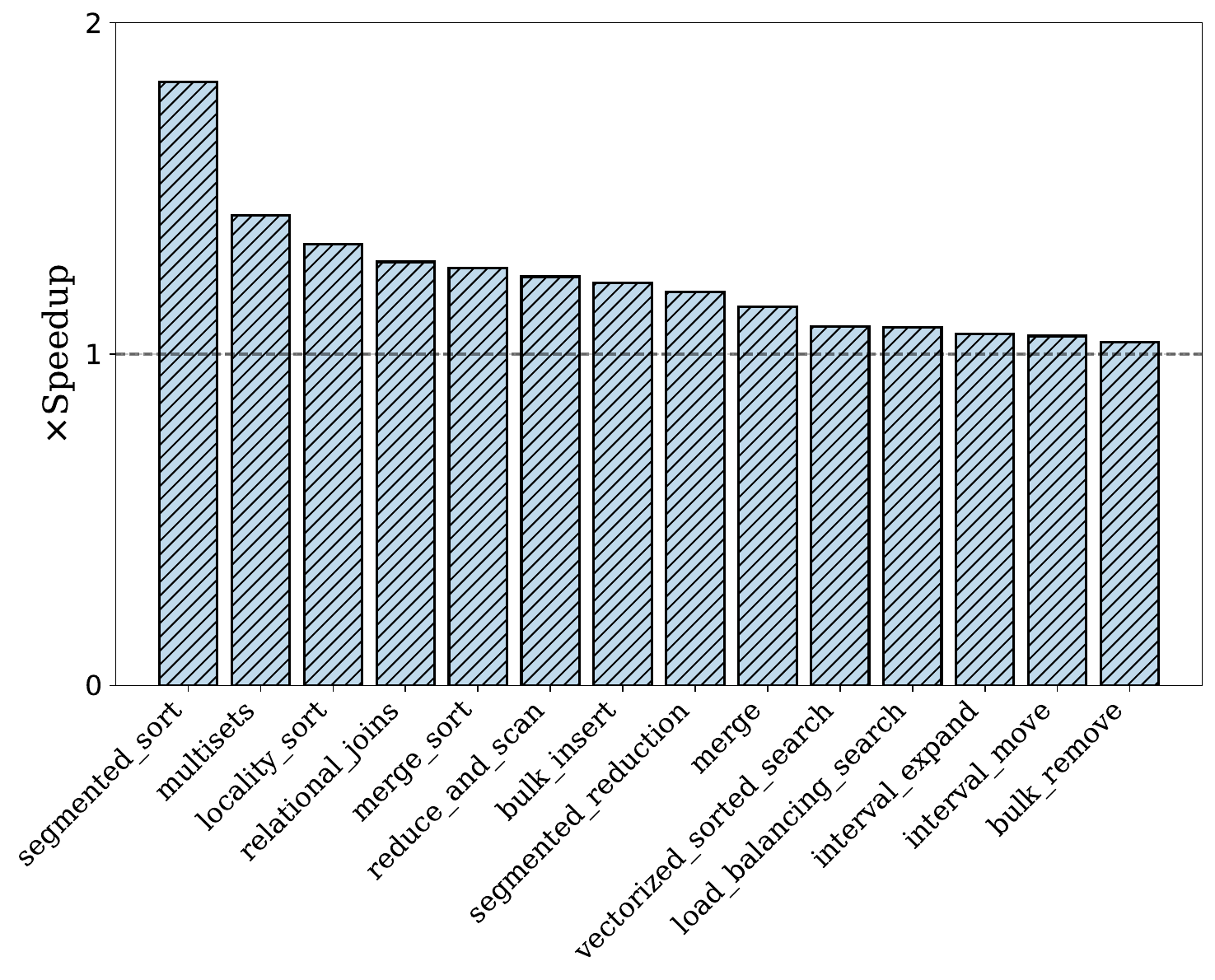}
  \end{center}
  \vspace{-2em}
  \caption{Relative speedup (over reference implementations) achieved by \ourmethod{} on 14 general-purpose kernels. We sort kernels based on descending speedup order.}
  \label{fig:general_kernel_barchart}
\end{wrapfigure}
In this section, we present experimental results comparing \ourmethod{} and baselines on general-purpose and neural network kernels. 

\noindent\textbf{General-purpose Kernels.}
Table~\ref{tab:results_general} compares the overall performance of \ourmethod{} with two other LLM-based automated kernel design methods in terms of relative speedup over reference implementations and \# of successful kernel designed. 
Notably, \ourmethod{} successfully identifies improved implementations on all 14 general-purpose kernel design tasks with a median speedup of 20\% over reference implementations.  
In particular, \ourmethod{} achieves 17\% higher median speedup and 10 more successful kernels than AI CUDA Engineer~\citep{lange2025ai}. 
Additionally, we provide the $\times$Speedup achieved by \ourmethod{} on each of the 14 kernels in Figure~\ref{fig:general_kernel_barchart}; %
and we present the $\times$Speedup over iterations achieved by \ourmethod{} against baselines in Figure~\ref{appxfig:gen _kernel_convergence_individual}. 

\begin{table}[ht]
\centering
\caption{Results on 66 neural network kernels. $\times$Speedup measures relative improvement over reference implementations; \#Success excludes the reference implementations from the count.}
\label{tab:results_ncnn}
\begin{tabular}{@{\hspace{2mm}}l|ccccc|cc@{\hspace{2mm}}}
\toprule
\multirow{2}{*}{Method} & \multicolumn{5}{c|}{$\times$Speedup ($\uparrow$)} & \multicolumn{2}{c}{\#Success ($\uparrow$)} \\ \cmidrule(lr){2-6} \cmidrule(lr){7-8} 
 & Mean & Max & 75\% & 50\% & 25\% & (out of total) & ($>1.01\times$ out of total) \\ \midrule
TVM & 1.24 & 2.20 & 1.28 & 1.21 & \textbf{1.14} & 64/66 & 60/66 \\ \midrule
Kernel Tuner & 1.11 & 1.30 & 1.19 & 1.12 & 1.01 & 64/66 & 50/60 \\
EoH & 1.20 & 1.54 & 1.26 & 1.18 & 1.09 & 31/66 & 31/66 \\
AI CUDA Engineer & 1.12 & 1.44 & 1.14 & 1.08 & 1.05 & 19/66 & 19/66 \\
\textbf{\ourmethod{} (ours)} & \textbf{1.47} & \textbf{7.87} & \textbf{1.56} & \textbf{1.29} & 1.08 & \textbf{66/66} & \textbf{66/66} \\ \bottomrule
\end{tabular}%
\end{table}

\begin{figure}[t]
    \begin{subfigure}[b]{0.95\textwidth}
         \centering
         \includegraphics[width=.95\textwidth]{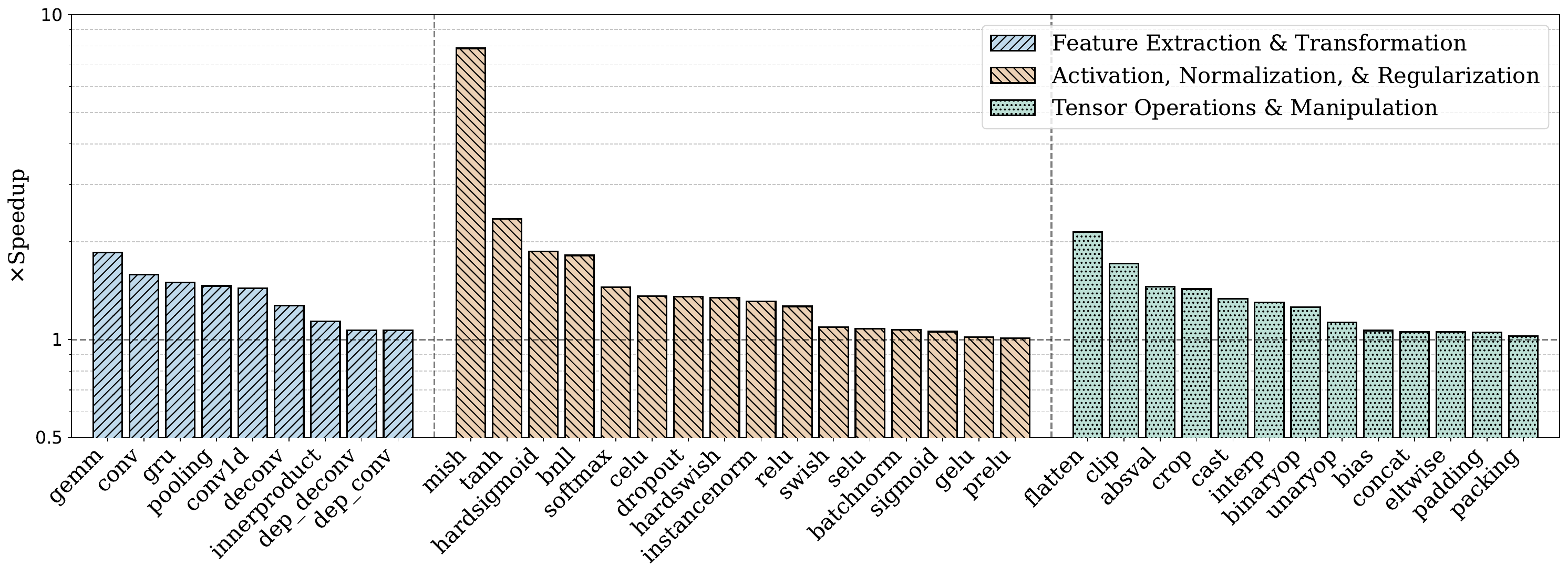}
         \caption{Full-precision Kernels. The median $\times$Speedup of each group (from left to right) is 1.44, 1.33, 1.26.}
         \label{fig:nn_kernel_barchart_fp}
     \end{subfigure}\\
     \begin{subfigure}[b]{0.95\textwidth}
         \centering
         \includegraphics[width=.72\textwidth]{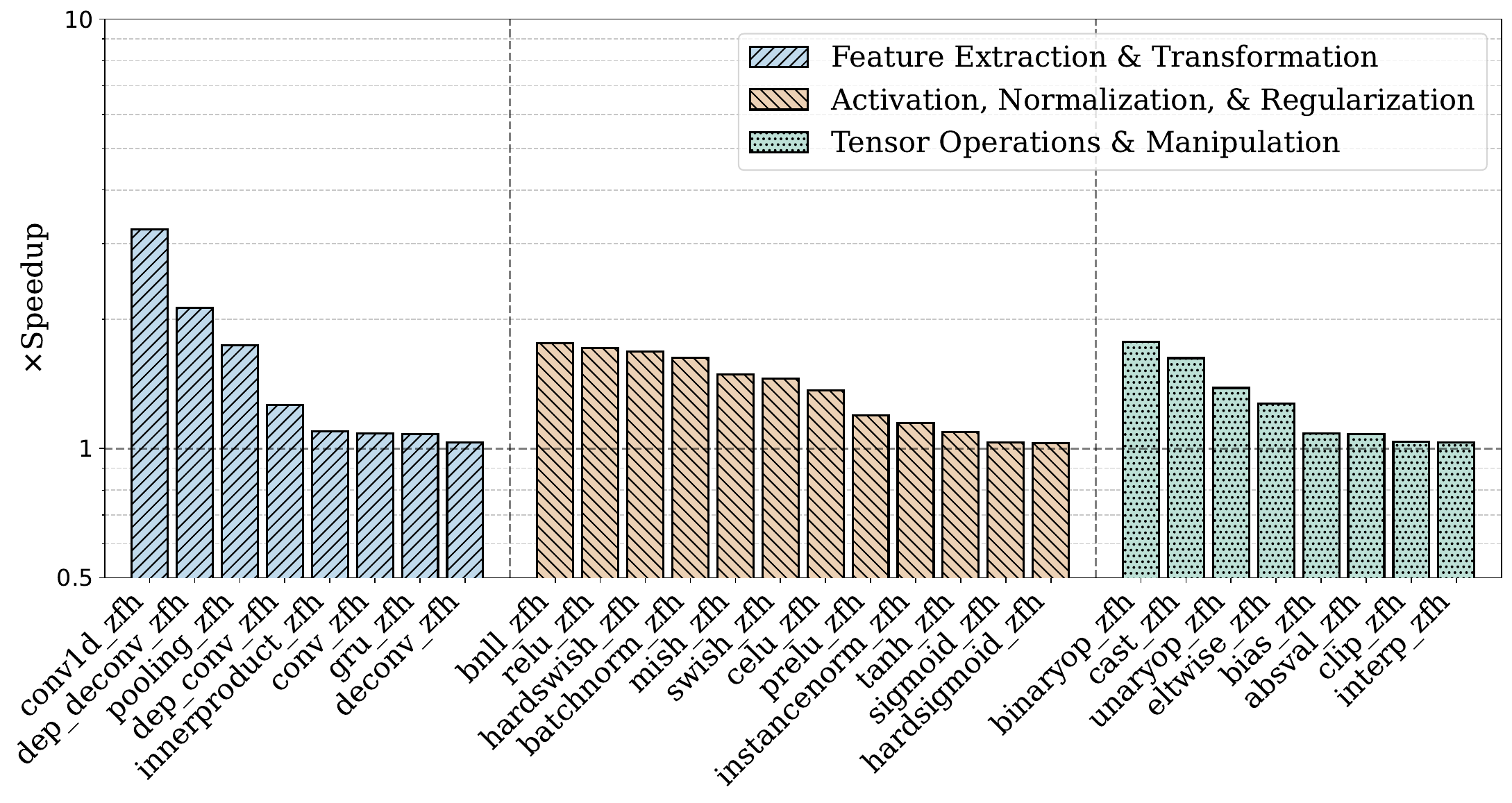}
         \caption{Half-precision Kernels with RISC-V Zfh Extension. The median $\times$Speedup of each group (from left to right) is 1.18, 1.41, 1.18.}
         \label{fig:nn_kernel_barchart_zfh}
     \end{subfigure}
     \caption{Relative speedup (over reference implementations) achieved by \ourmethod{} on 66 neural network kernels. Kernels are divided into two sets: (a) full-precision and (b) half precision kernels. For each set, kernels are further divided into three groups based on their functionalities and sorted based on descending speedup order within each group.}
     \label{fig:nn_kernel_barchart}
\end{figure}

\noindent\textbf{Neural Network Kernels.}
Table~\ref{tab:results_ncnn} summarizes the performance of \ourmethod{} and baselines on 66 neural network kernel design tasks, 
covering diverse neural network operations, including convolution, activation, pooling, element-wise operations, and normalization.
In general, \ourmethod{} successfully identifies improved implementations on all 66 neural network kernel design tasks with a median speedup of 29\% over reference kernels implemented by human experts. 
Notably, \ourmethod{} achieves 20\%+ higher median speedup and 35+ more successful kernels than existing LLM-based automated kernel design methods. 
Furthermore, we provide the $\times$Speedup achieved by \ourmethod{} on each of the 66 kernels in Figure~\ref{fig:nn_kernel_barchart}; 
and we present the $\times$Speedup over iterations achieved by \ourmethod{} against baselines in Figure~\ref{appxfig:nn_kernel_convergence_individual} and \ref{appxfig:nn_zfh_kernel_convergence_individual}.

\subsection{A Case Study: Mish Activation Kernel} \label{subsec:mish}

We examined the code for the Mish kernel as the best-improved kernel from neural network kernels, which is defined as $f(x) = x \cdot \tanh(\ln(1 + e^x)$. By comparing the reference NCNN implementation with the EoK optimized output, we identified the following optimization techniques applied by EoK:

\begin{enumerate}
    \item \textbf{ISA Extension Optimization:} Guided by the idea "Vectorization Heuristics," EoK leveraged custom instructions to accelerate the exponential approximation in \(\ln(1 + e^x)\). This replaces software-based Taylor series with hardware-accelerated computation, reducing latency by minimizing instruction overhead.

    \item \textbf{Instruction-Level Parallelism:} 
    EoK applied loop unrolling and fused operations to reduce dependency stalls, improving IPC through explicit SIMD chaining.
    
    \item \textbf{Arithmetic Optimization:} A dedicated FP16 arithmetic path was introduced to leverage half-precision computation via RISC-V's Zfh/Zvfh extensions, aligning with the idea "Precision-Specific Optimization."
    \item \textbf{Combine Multiple Functions:} Derived from the idea "Instruction Fusion," EoK merged adjacent operations into single calls. This reduced function call overhead and improved cache locality.
    
    \item \textbf{Precision-Specific Vectorization:} Guided by the idea "Avoidance of Slow Instructions", EoK optimized vector register usage for FP16 data types, doubling SIMD lane utilization and improving cache efficiency compared to FP32.
\end{enumerate}

\begin{figure}[ht]
    \centering
    \begin{subfigure}[b]{0.9\textwidth}
         \centering
         \includegraphics[trim={0 0 18cm 0}, clip, width=\textwidth]{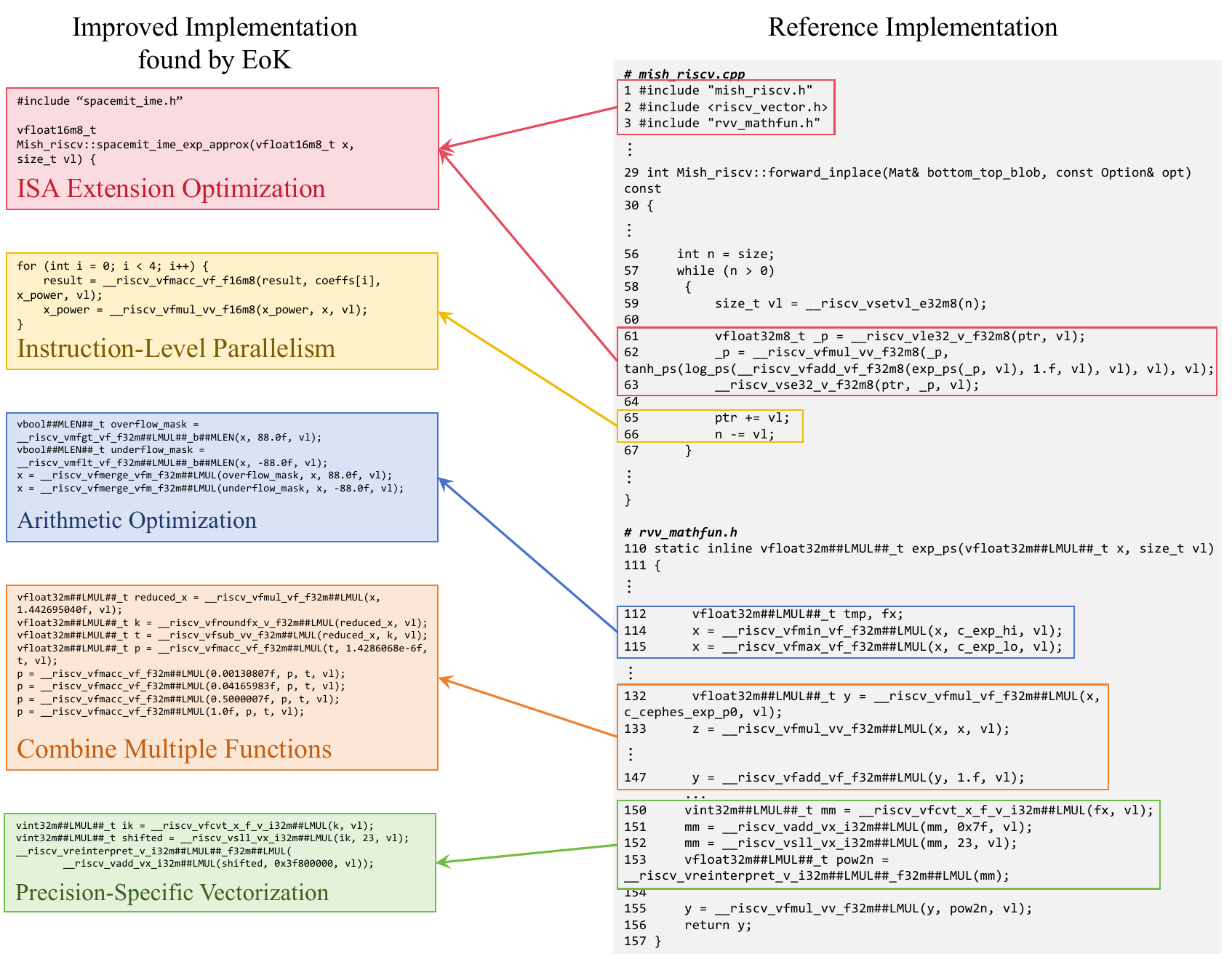}
     \end{subfigure}
     \caption{Visualization of the reference implementation (\emph{Right}) and the improved implementation (\emph{Left}) found by \ourmethod{} of the Mish kernel, a neural network activation function. }
     \label{fig:case_study}
\end{figure}

Figure~\ref{fig:case_study} represents some of the code snippets in detail, showing how ideas contribute to the final speedup.

\subsection{Ablation Study} \label{subsec:ablation}
In this section, we present ablative experiments on the Idea Pool and the RAG mechanism (for augmenting EoH search) introduced in \ourmethod{}. 
All experiments are conducted on the 66 neural network kernels;  
three independent runs are carried out for each kernel and the mean speedup over the reference implementation is recorded.  
Then, the speedup averaged over 66 kernels ($\times$Speedup) and the number of kernels with $>1.01\times$ speedup out of 66 total kernels (\#Success) are reported. 

\begin{table}[ht]
\centering
\caption{Comparison of variants of the Idea Pool on 66 neural network kernels. The final setting adopted in \ourmethod{} is highlighted in shades.}
\label{tab:results_abl_idea}
\begin{tabular}{@{\hspace{2mm}}l|c|c@{\hspace{2mm}}}
\toprule
\multirow{1}{*}{Variants of Idea Pool} & \multicolumn{1}{c|}{$\times$Speedup ($\uparrow$)} & \multicolumn{1}{c}{\#Success ($\uparrow$)} \\ \midrule
Commits + random sampling & 1.17 & 27/66 \\
Commits + clustering + random sampling & 1.26 & 41/66 \\
Commits + two-level abstraction + random sampling & 1.40 & 61/66 \\
\cellcolor{lightgray!60}Commits + two-level abstraction + weighted sampling & \cellcolor{lightgray!60}\textbf{1.47} & \cellcolor{lightgray!60}\textbf{66/66} \\ \bottomrule
\end{tabular}%
\end{table}

\noindent\textbf{Impact of Idea Pool.} 
As elaborated in Section~\S\ref{subsec:idea-pool}, a large volume of raw commits from existing kernel libraries are distilled into general kernel design ideas with actionable kernel optimization thoughts via a two-level abstraction with LLMs to form the Idea Pool; 
then ideas are sampled with probabilities proportional to their past effectiveness to guide the subsequent EoH searches. 
To understand the relative contribution of each step in the above process, three variants of the Idea Pool are studied, as follows. 
\begin{enumerate}[topsep=0pt, itemsep=3pt, parsep=2pt]
    \item \emph{Commits + random sampling}: Uniformly sample raw commits to to guide EoH searches.
    \item \emph{Commits + clustering + random sampling}: Group commits based on k-means clustering instead of the proposed two-level abstraction. 
    \item \emph{Commits + two-level abstraction + random sampling}: Uniform sample an idea from Idea Pool instead of weighted sampling by ideas' past effectiveness.
\end{enumerate}

Table~\ref{tab:results_abl_idea} summarizes the results comparing the above three variants along with the final setting of Idea Pool adopted in \ourmethod{}. 
Evidently, we observe that the proposed two-level abstraction is not only performant but also essential as supplying raw commits to EoH adversely affect the performance. 
In addition, we observe that the weighted idea sampling strategy further improves the performance and reliability of \ourmethod{}, leading to improvements on all 66 neural network kernels. 



\begin{table}[ht]
\centering
\caption{Comparison of different settings of RAG on 66 neural network kernels. The final setting adopted in \ourmethod{} is highlighted in shades.}
\label{tab:results_abl_rag}
\begin{tabular}{@{\hspace{2mm}}l|c|c@{\hspace{2mm}}}
\toprule
\multirow{1}{*}{Variants of RAG Setting} & \multicolumn{1}{c|}{$\times$Speedup ($\uparrow$)} & \multicolumn{1}{c}{\#Success ($\uparrow$)} \\ \midrule
Without RAG & 1.24 & 40/66 \\
RAG w/ RISC-V technical documents & 1.39 & 62/66 \\
RAG w/ example codes (from Idea Pool) & 1.28 & 60/66 \\
\cellcolor{lightgray!60}RAG w/ RISC-V technical documents and example codes & \cellcolor{lightgray!60}\textbf{1.47} & \cellcolor{lightgray!60}\textbf{66/66} \\ \bottomrule
\end{tabular}%
\end{table}

\noindent\textbf{Impact of RAG.} 
As elaborated in Section~\S\ref{subsec:rag_eoh}, \ourmethod{} use retrieval augmented generation (RAG) to provide additional context to augment the prompt inputs to the LLM. 
In particular, we provide the official RISC-V ISA manual, the profiles of RISC-V extensions (i.e., RVV1.0 and RVA22), and the technical documents of Spacemit K1 as external reference material for RAG. 
To understand the effects of these external material, three variants of the RAG settings are studied, as follows.  


\begin{enumerate}[topsep=0pt, itemsep=3pt, parsep=2pt]
    \item \emph{Without RAG}: No context is provided to augment the prompts.
    \item \emph{RAG with RISC-V technical documents}: RISC-V ISA manual and the profiles of its extensions are supplied as external database for retrieval of relevant context. 
    \item \emph{RAG With example codes (from Idea Pool)}: Code examples from Idea Pool are supplied as external database for retrieval of relevant context. 
\end{enumerate}

Table~\ref{tab:results_abl_rag} summarizes the results comparing the above three variants along with the final setting of RAG adopted in \ourmethod{}. 
Evidently, we observe that supplying the RISC-V technical documents as the external database for RAG significantly improves the performance on RISC-V kernel optimization.  
In addition, we observe that supplying the example codes from Idea Pool further improves the performance and reliability of \ourmethod{}, leading to improvements on all 66 neural network kernels.

\section{Conclusion}

The rapid evolution of RISC-V hardware demands equally agile kernel optimization. 
This work addresses the absence of high-performance RISC-V kernels by introducing EoK, an LLM-driven evolutionary program search framework that automates kernel design for reference-scarce domains. 
By formalizing historical optimization knowledge and steering LLMs via RAG-enhanced RISC-V context, EoK achieves unprecedented results, outperforming human experts across 80 diverse kernel design tasks with a 1.27$\times$ median speedup. 
EoK provides the first scalable pathway to performant RISC-V kernels, bridging the gap between hardware innovation and deployable software.
While EoK significantly accelerates kernel development, we emphasize that human oversight remains indispensable—particularly for safety-critical systems.

\subsubsection*{Acknowledgments}
We would like extend our sincere gratitude to Dr. Fei Liu for the support on EoH and Dr. Ping Guo for the support on reproducing AI CUDA Engineer. 


{\small
\bibliographystyle{iclr2021_conference}
\bibliography{mybib}}

\clearpage
\appendix
\section{More Details of Idea Pool} \label{appx:idea-pool}

\subsection{Initial Set of Thoughts} \label{subappx:init-idea}
Table~\ref{appxtab:initial_thoughts} presents a set of widely-used kernel optimization techniques collected from the literature as our initial thoughts.
Commits from existing kernel library will be associated to one of these thoughts based on the similarity in the embedding space. 

\begin{table}[th]
\caption{Summary of twenty initial thoughts used for constructing the Idea Pool.}
\label{appxtab:initial_thoughts}
\centering
\begin{tabular}{@{\hspace{2mm}}ll@{\hspace{2mm}}}
\toprule
\textbf{Category}                 & \textbf{Initial Thought}  \\ \midrule
\multirow{6}{*}{Memory}  & Pinned memory                                           \\ 
                         & Asynchronous and overlapping transfers with computation \\ 
                         & Zero copy                                               \\ 
                         & Unified virtual addressing                              \\  
                         & Device memory spaces                                    \\  
                         & Device memory allocation                                \\  
                         & NUMA tuning                                             \\ \midrule
\multirow{5}{*}{\begin{tabular}[c]{@{}l@{}}Execution\\Configuration\end{tabular}}  & Occupancy \\  
                         & Hiding Register Dependencies                            \\  
                         & Thread and Block Heuristics                             \\  
                         & Effects of Shared Memory                                \\  
                         & Concurrent Kernel Execution                             \\  
                         & Multiple contexts                                       \\ \midrule
\multirow{2}{*}{\begin{tabular}[c]{@{}l@{}}Instruction\\Optimization\end{tabular}} & Arithmetic Instructions   \\ 
                         & Memory Instructions                                     \\ \midrule
\multirow{2}{*}{Control Flow} & Branching and Divergence                                \\ 
                         & Branch Predication                                      \\ \midrule
\multirow{3}{*}{Others}  & Maximizing parallel execution                           \\ 
                         & Maximizing memory bandwidth                             \\  
                         & Maximizing instruction throughput                       \\ \bottomrule
\end{tabular}%
\end{table}

\subsection{Refined Set of Ideas} \label{subappx:final-idea}
\begin{figure}[h!]
    \centering
    \begin{subfigure}[b]{0.98\textwidth}
         \centering
         \includegraphics[width=.95\textwidth]{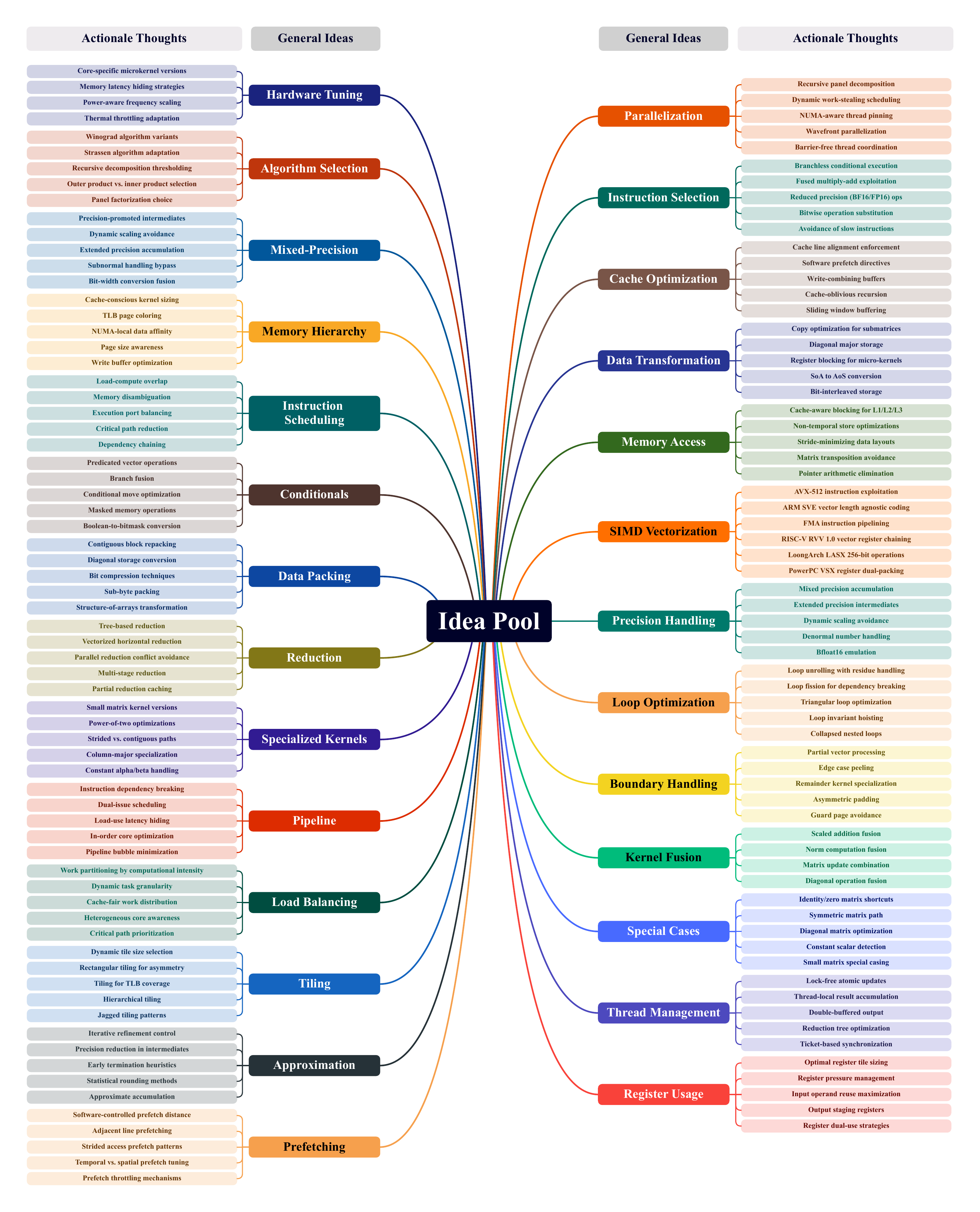}
     \end{subfigure}
     \caption{Visualization of the general ideas with their actionable thoughts from the learned Idea Pool.}
     \label{appxfig:final-idea}
\end{figure}

Two-level of abstraction is applied to transform raw commits into general kernel design ideas with actionable thoughts to form the Idea Pool.
A pictorial illustration of the learned Idea Pool is shown in Figure~\ref{appxfig:final-idea}.

\section{Prompts} \label{appx:prompts}
LLMs are used to \ding{172} summarize the common technique or principle behind a set of commits; 
\ding{173} create an initial population of individuals for EoH by applying a sampled thought to a reference kernel; 
\ding{174} create new candidate kernel within each iteration of EoH. 
The corresponding prompts are provided in the following subsections, respectively.

\subsection{Idea Summarization Prompts} \label{subappx:prompt_idea}
\vspace{-.5em}
\begin{tcolorbox}[colback=blue!5,colframe=blue!75!black,title=Summarization Prompts]
{\footnotesize\textcolor{black!80}{=== System Prompt ===}}
\begin{lstlisting}[language=C++, escapeinside={(*}{*)}, directivestyle={\color{black}}, emph={int,char,double,float,unsigned}, emphstyle={\color{blue}},]
You are an expert in high-performance computing kernel optimization and try to learn the existing kernel optimization methods.
\end{lstlisting}
{\footnotesize\textcolor{black!80}{=== Base Prompt ===}}
\begin{lstlisting}[language=C++, escapeinside={(*}{*)}, directivestyle={\color{black}}, emph={int,char,double,float,unsigned}, emphstyle={\color{blue}},]
Based on the given materials from a well-maintained code repository, please summarize the key idea of the commit messages and code diff records.

Your summarization should only contain the information and be no longer than 20 words and begin with an action verb (e.g., "Apply", "Utilize")
    {commit messages and code diff records}
\end{lstlisting}
\end{tcolorbox} 

\vspace{-1em}
\subsection{Initialization Prompts} \label{subappx:prompt_init}
\vspace{-.5em}
\begin{tcolorbox}[colback=blue!5,colframe=blue!75!black,title=Initialization Prompts]
{\footnotesize\textcolor{black!80}{=== System Prompt ===}}
\begin{lstlisting}[language=C++, escapeinside={(*}{*)}, directivestyle={\color{black}}, emph={int,char,double,float,unsigned}, emphstyle={\color{blue}},]
You are an expert in high-performance computing kernel optimization, trying to reduce the runtime of a {operation} kernel in RISC-V. Make sure the kernel returns the correct result. The kernel will be run on {hardware_type} with {extensions}. 

Here is a reference implementation of the kernel:
    {code_of_reference_implementation}
\end{lstlisting}
{\footnotesize\textcolor{black!80}{=== Base Prompt ===}}
\begin{lstlisting}[language=C++, escapeinside={(*}{*)}, directivestyle={\color{black}}, emph={int,char,double,float,unsigned}, emphstyle={\color{blue}},]
Please modify the code by the given thought and its code examples.  // sampled from Idea Pool
    {thought}
    {code_examples}
\end{lstlisting}
\end{tcolorbox} 

\vspace{-1em}
\subsection{EoH Prompts} \label{subappx:prompt_eoh}
\vspace{-.5em}
\begin{tcolorbox}[colback=blue!5,colframe=blue!75!black,title=RAG-augmented EoH Prompts]
{\footnotesize\textcolor{black!80}{=== System Prompt ===}}
\begin{lstlisting}[language=C++, escapeinside={(*}{*)}, directivestyle={\color{black}}, emph={int,char,double,float,unsigned}, emphstyle={\color{blue}},]
You are an expert in high-performance computing kernel optimization, trying to reduce the runtime of a {operation} kernel in RISC-V. Make sure the kernel returns the correct result. The kernel will be run on {hardware_type} with {extensions}. 

Here is a reference implementation of the kernel:
    {code_of_reference_implementation}

\end{lstlisting}
{\footnotesize\textcolor{black!80}{=== Base Prompt ===}}
\begin{lstlisting}[language=C++, escapeinside={(*}{*)}, directivestyle={\color{black}}, emph={int,char,double,float,unsigned}, emphstyle={\color{blue}},]
1. First, describe your new thought and main steps in one sentence. The description must be inside within boxed {{}}. 

2. Next, optimize the following kernel by your new thought with RISC-V extensions:
    {kernel code}
    
Do not give additional explanations. 
\end{lstlisting}
\end{tcolorbox} 

\section{Benchmark Details} \label{appx:benchmark}
In this work, we consider 14 general-purpose and 66 neural network computation related kernels. 
The detailed list of the kernels and their brief descriptions are provided in Table~\ref{appxtab:general_kernel}, \ref{appxtab:grouped_nn_kernel}, and \ref{appxtab:nn_kernel}. 

\begin{table}[h]
\centering
\caption{Summary of 14 general-purpose kernels evaluated in this work.}
\label{appxtab:general_kernel}
\begin{tabular}{@{\hspace{2mm}}lcl@{\hspace{2mm}}}
\toprule
\textbf{ID} & \textbf{Kernel}            & \multicolumn{1}{l}{\textbf{Description}}                                                             \\ \midrule
1                & Reduce and scan          & Core primitives of parallel computing.                                                               \\ 
2                & Bulk remove              & \begin{tabular}[c]{@{}l@{}}Compact an array by removing all elements\\specified by index.\end{tabular}                                        \\ 
3                & Bulk insert              & Merge two arrays in parallel by insertion index.                                                     \\ 
4                & Merge                    & Merge two sorted sequences in parallel.                                                              \\ 
5                & Merge sort               & \begin{tabular}[c]{@{}l@{}}Sort data by recursively dividing and\\ merging subarrays.\end{tabular}                                             \\ 
6                & Segmented sort           & Sort many variable-length arrays in parallel.                                                        \\ 
7                & Locality sort            & \begin{tabular}[c]{@{}l@{}}Detect regions of approximate sortedness\\without requiring annotations.\end{tabular}                              \\ 
8                & Vectorized sorted search & \begin{tabular}[c]{@{}l@{}}Run many concurrent searches where\\both the needles and haystack arrays are sorted.\end{tabular}                   \\ 
9                & Load-balancing search & A specialization of vectorized sorted search.                                                        \\ 
10               & Interval expand           & \begin{tabular}[c]{@{}l@{}}Expand a memory interval (e.g., array or buffer)\\by reallocating or extending its size.\end{tabular}               \\ 
11               & Interval move             & Relocates data between memory intervals.                                                             \\ 
12               & Relational joins         & \begin{tabular}[c]{@{}l@{}}Sort-merge joins supporting inner, left, right,\\and outer variants.\end{tabular}                                  \\ 
13               & Multisets                & \begin{tabular}[c]{@{}l@{}}Replace the Merge Path partitioning with the \\ Balanced Path to search for key-rank matches.\end{tabular}  \\ 
14               & Segmented reduction & \begin{tabular}[c]{@{}l@{}}A parallel reduction over many irregular-\\ length segments.\end{tabular}                                            \\ \bottomrule
\end{tabular}
\end{table}

\begin{table}[h]
\vspace{-.4em}
\centering
\caption{Summary of three types of neural network related kernels evaluated in this work. We also consider the FP16 version of kernels marked with $\dagger$. Details of each kernel is provided in Table~\ref{appxtab:nn_kernel}.}
\label{appxtab:grouped_nn_kernel}
\begin{tabular}{@{\hspace{2mm}}lcl@{\hspace{2mm}}}
\toprule
\textbf{ID} & \textbf{Type} & \textbf{Kernels} \\ \midrule
1 & \begin{tabular}[c]{@{}c@{}}Feature Extraction and \\ Transformation\end{tabular} & \begin{tabular}[c]{@{}l@{}}Conv1d$^{\dagger}$, Convdepth$^{\dagger}$, Conv$^{\dagger}$, Deconvdepth$^{\dagger}$, Deconv$^{\dagger}$, Gemm, \\ Innerproduct$^{\dagger}$, Pooling$^{\dagger}$, GRU$^{\dagger}$\end{tabular} \\
2 & \begin{tabular}[c]{@{}c@{}}Activation, Normalization,\\and Regularization\end{tabular} & \begin{tabular}[c]{@{}l@{}}Bnll$^{\dagger}$, Celu$^{\dagger}$, Gelu, Hardsigmoid$^{\dagger}$, Hardswish$^{\dagger}$, Mish$^{\dagger}$, PReLU$^{\dagger}$, \\ ReLU$^{\dagger}$, SELU, Sigmoid$^{\dagger}$, Softmax, Swish$^{\dagger}$, Tanh$^{\dagger}$, BatchNorm$^{\dagger}$, \\ InstanceNorm$^{\dagger}$, Dropout\end{tabular} \\
3 & \begin{tabular}[c]{@{}c@{}}Tensor Operations and\\Manipulation\end{tabular} & \begin{tabular}[c]{@{}l@{}}Absval$^{\dagger}$, Bias$^{\dagger}$, Binaryop$^{\dagger}$, Eltwise$^{\dagger}$, Unaryop$^{\dagger}$, Cast$^{\dagger}$, Clip$^{\dagger}$, \\ Concat, Crop, Flatten, Interp$^{\dagger}$, Packing, Padding\end{tabular} \\ \bottomrule
\end{tabular}%
\end{table}

\begin{table}[t]
\centering
\caption{Summary of neural network related kernels evaluated in this work. We also consider the FP16 version of kernels marked with $\dagger$, resulting in a total of 66 kernels.}
\label{appxtab:nn_kernel}
\begin{tabular}{@{\hspace{2mm}}lcl@{\hspace{2mm}}}
\toprule
\textbf{ID} & \textbf{Kernel} & \textbf{Description} \\ \midrule
1 & Absval$^{\dagger}$ & Absolute value computation. \\
2 & BatchNorm$^{\dagger}$ & Batch normalization layer. \\
3 & Bias$^{\dagger}$ & Element-wise addition layer. \\
4 & Binaryop$^{\dagger}$ & Element-wise binary operations. \\
5 & Bnll$^{\dagger}$ & Binomial normal log likelihood (BNLL) activation function. \\
6 & Cast$^{\dagger}$ & Data type casting operations \\
7 & Celu$^{\dagger}$ & \begin{tabular}[c]{@{}l@{}}Continuously differentiable exponential linear unit (CELU)\\activation function.\end{tabular} \\
8 & Clip$^{\dagger}$ & Clips tensor values between a minimum and maximum threshold. \\
9 & Concat & Tensor concatenation along a specified dimension. \\
10 & Conv1d$^{\dagger}$ & 1D convolution layer. \\
11 & Convdepth$^{\dagger}$ & Depthwise convolution layer. \\
12 & Conv$^{\dagger}$ & Standard 2D/3D convolution layer. \\
13 & Crop & Cropping of input tensors along spatial or channel dimensions. \\
14 & Deconvdepth$^{\dagger}$ & Depthwise deconvolution (transposed convolution) layer. \\
15 & Deconv$^{\dagger}$ & Standard deconvolution (transposed convolution) layer. \\
16 & Dropout & Dropout regularization layer. \\
17 & Eltwise$^{\dagger}$ & Element-wise operations on multiple input tensors. \\
18 & Flatten & Flattening of input tensors into a 1D vector. \\
19 & Gelu & Gaussian error linear unit (GELU) activation function. \\
20 & Gemm & General matrix multiply operations. \\
21 & GRU$^{\dagger}$ & Gated recurrent unit (GRU) recurrent layer. \\
22 & Hardsigmoid$^{\dagger}$ & Hard sigmoid activation function. \\
23 & Hardswish$^{\dagger}$ & Hard swish activation function. \\
24 & Innerproduct$^{\dagger}$ & Inner product (fully connected) layer. \\
25 & Instancenorm$^{\dagger}$ & Instance normalization layer. \\
26 & Interp$^{\dagger}$ & Image resizing layers (nearest, bilinear, bicubic). \\
27 & Mish$^{\dagger}$ & Mish activation function. \\
28 & Packing & Tensor packing/unpacking operations. \\
29 & padding & Padding operations around tensor edges. \\
30 & Pooling$^{\dagger}$ & Pooling operations (max, average). \\
31 & PReLU$^{\dagger}$ & Parametric rectified linear unit activation function. \\
32 & ReLU$^{\dagger}$ & Rectified linear unit activation function. \\
33 & SELU & Scaled exponential linear unit activation. \\
34 & Sigmoid$^{\dagger}$ & Sigmoid activation function. \\
35 & Softmax & Softmax normalization. \\
36 & Swish$^{\dagger}$ & Swish activation function. \\
37 & Tanh$^{\dagger}$ & Hyperbolic tangent activation function. \\
38 & Unaryop$^{\dagger}$ & Various unary operations. \\ \bottomrule
\end{tabular}%
\end{table}

\clearpage
\section{Additional Results} \label{appx:results}
\begin{figure}[h]
    \captionsetup[subfigure]{justification=centering}
     \begin{subfigure}[t]{0.14\textwidth}
         \centering
         \includegraphics[width=.95\textwidth]{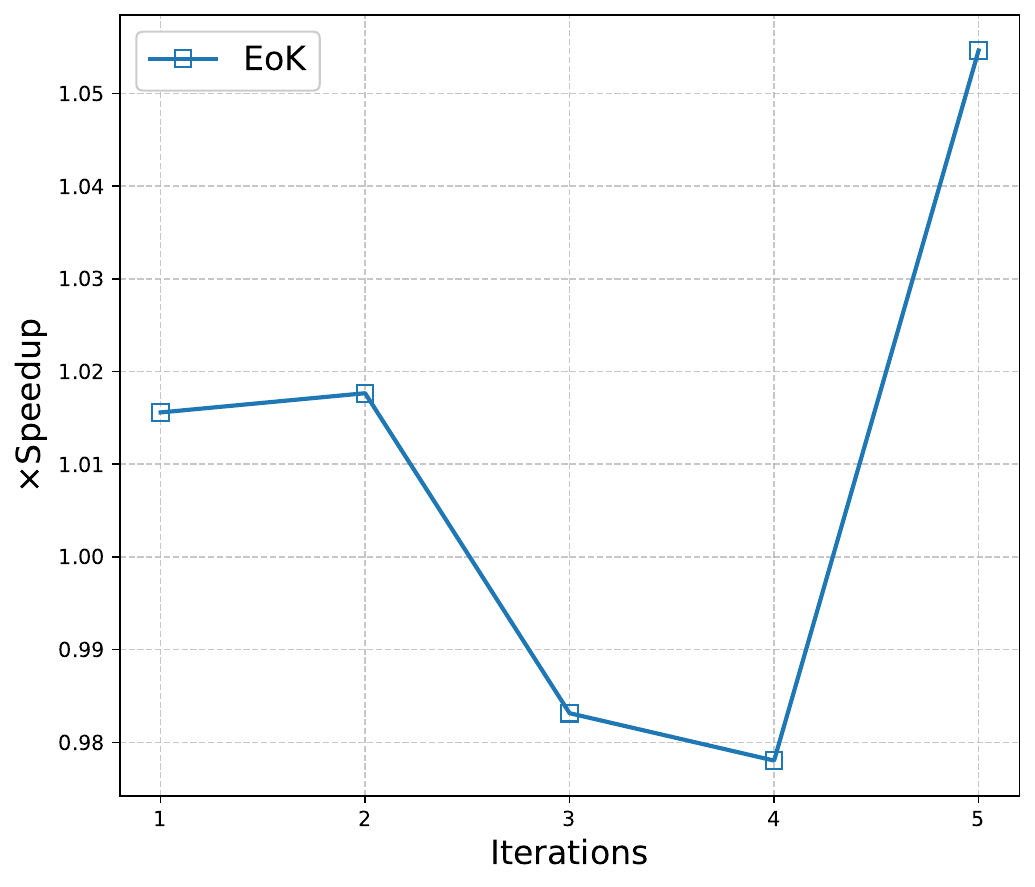}
         \caption*{Interval Move}         
     \end{subfigure}\hfill
     \begin{subfigure}[t]{0.14\textwidth}
         \centering
         \includegraphics[width=.95\textwidth]{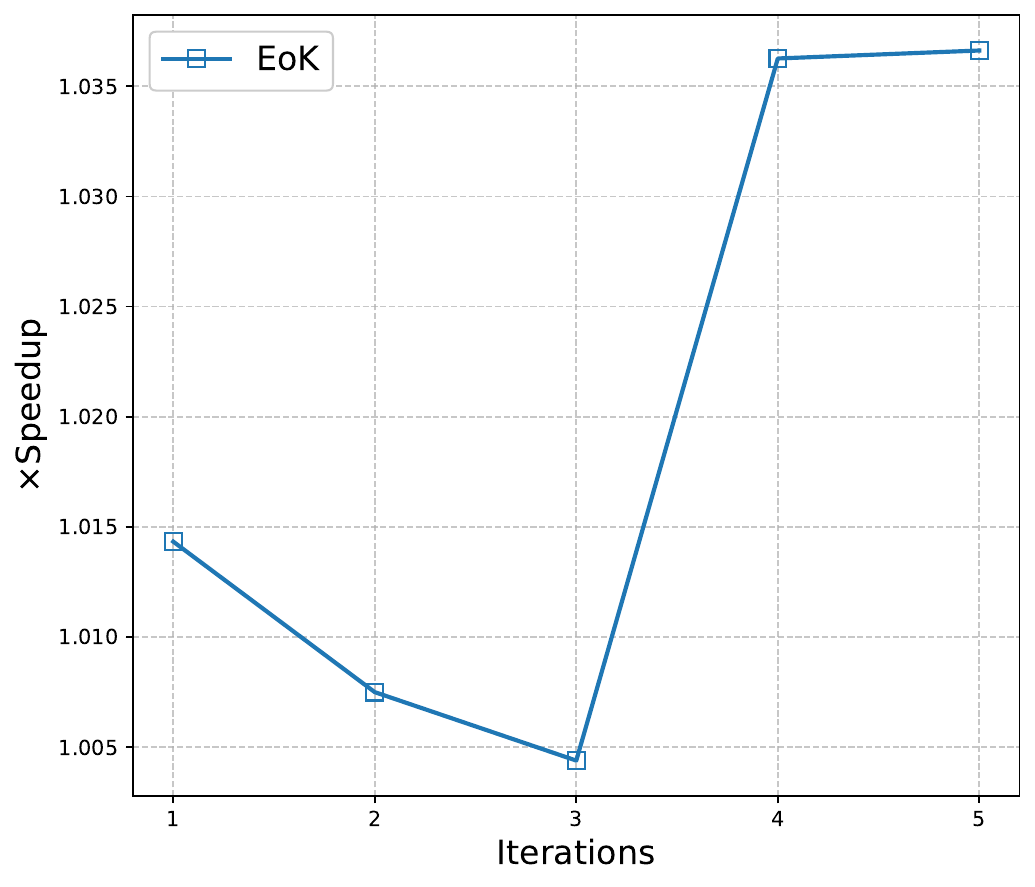}
         \caption*{Bulk Remove}
     \end{subfigure}\hfill
     \begin{subfigure}[t]{0.14\textwidth}
         \centering
         \includegraphics[width=.95\textwidth]{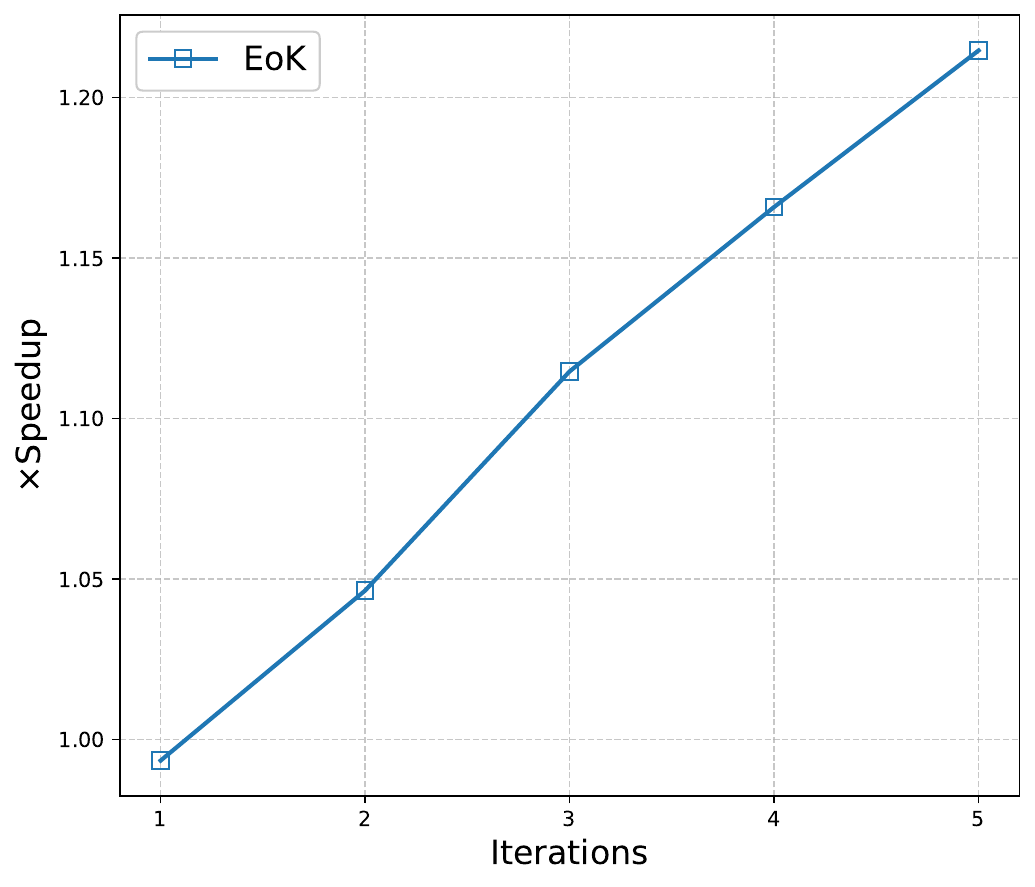}
         \caption*{Bulk Insert}  
     \end{subfigure}\hfill
     \begin{subfigure}[t]{0.14\textwidth}
         \centering
         \includegraphics[width=.95\textwidth]{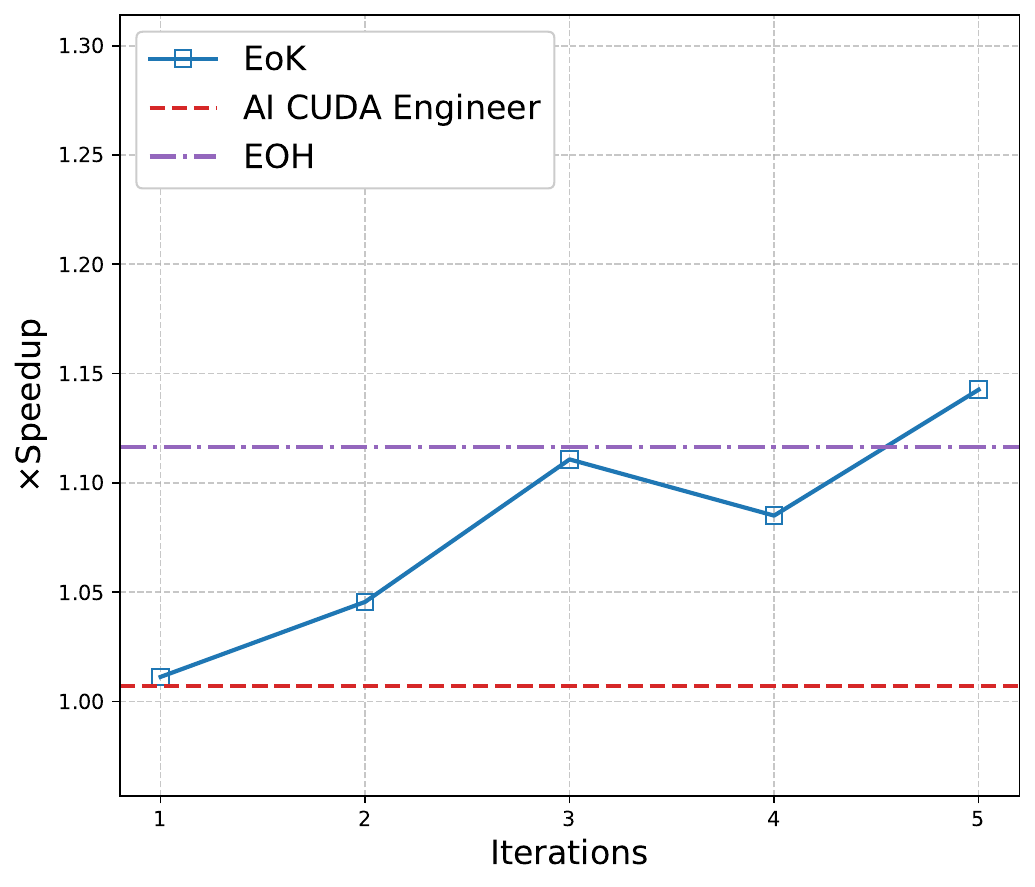}
         \caption*{Merge}         
     \end{subfigure}\hfill
     \begin{subfigure}[t]{0.14\textwidth}
         \centering
         \includegraphics[width=.95\textwidth]{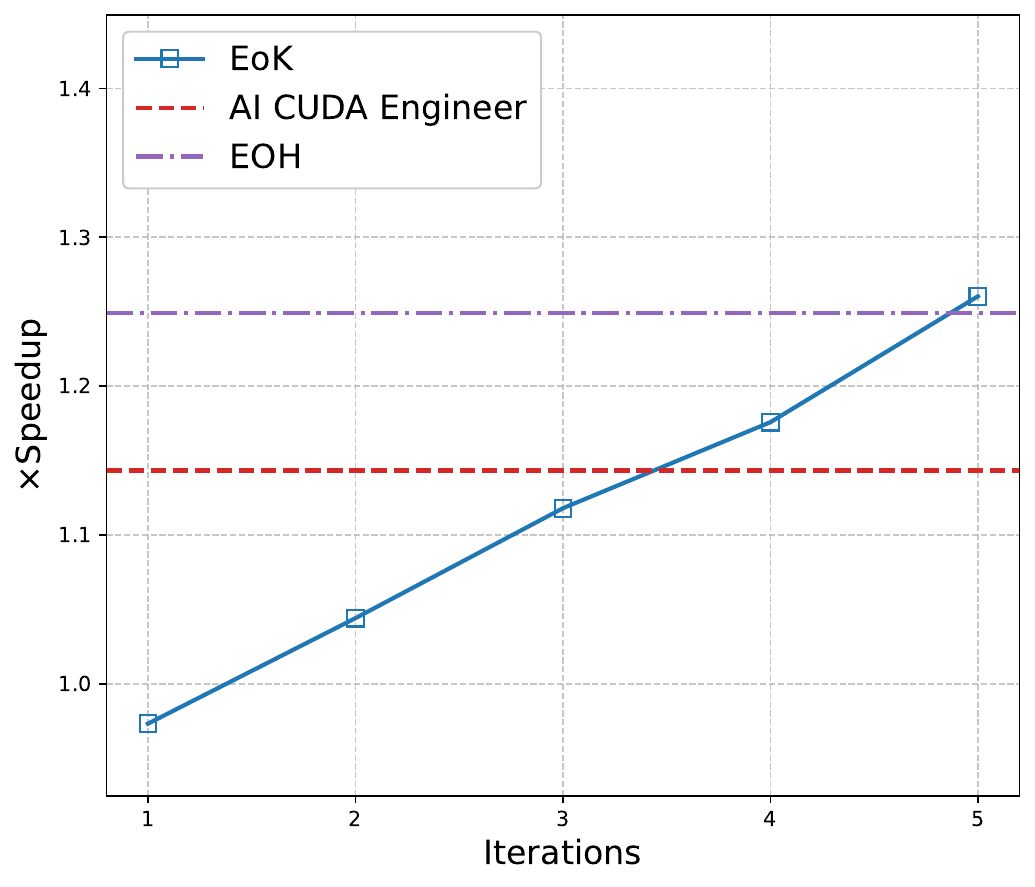}
         \caption*{Merge Sort}     
     \end{subfigure}\hfill
     \begin{subfigure}[t]{0.14\textwidth}
         \centering
         \includegraphics[width=.95\textwidth]{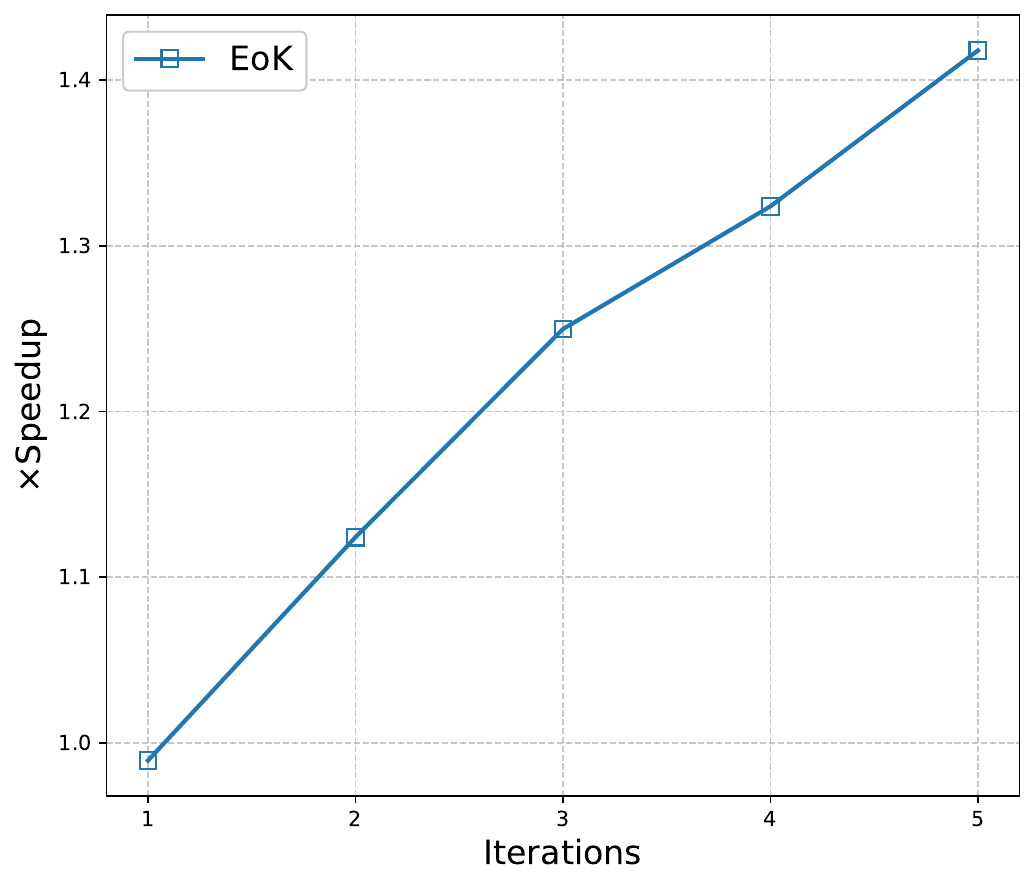}
         \caption*{Multisets}         
     \end{subfigure}\hfill
     \begin{subfigure}[t]{0.14\textwidth}
         \centering
         \includegraphics[width=.95\textwidth]{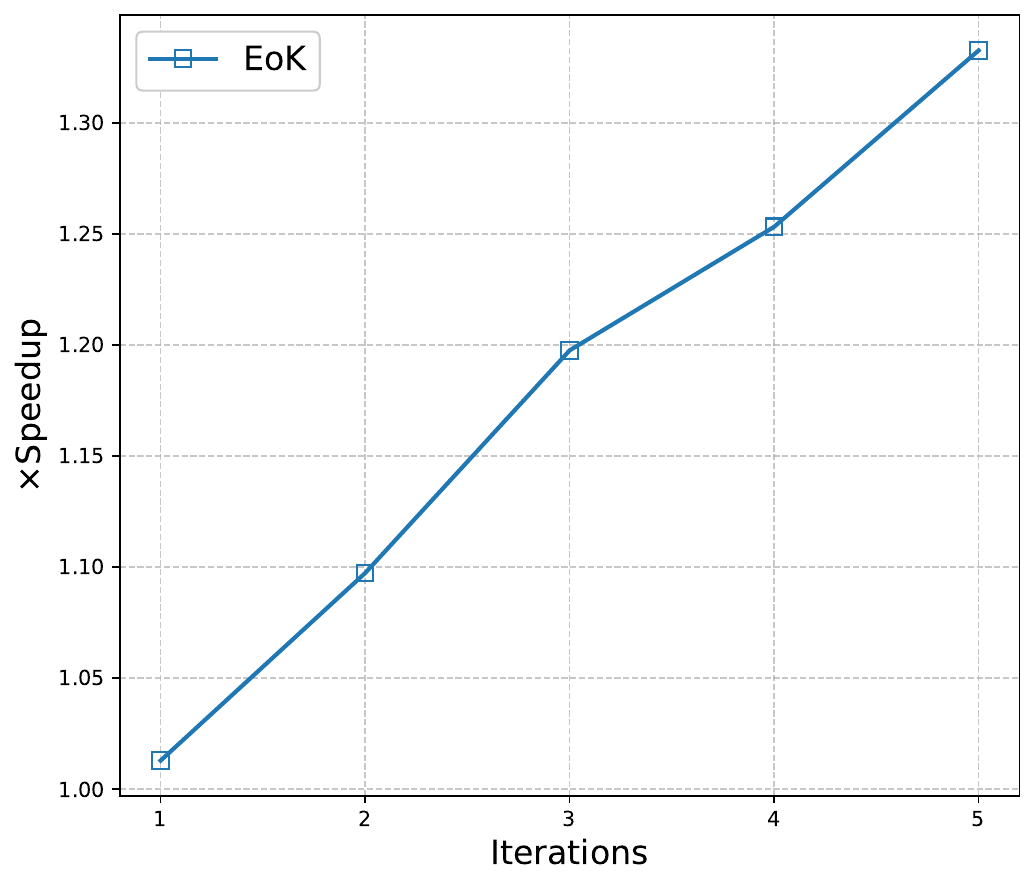}
         \caption*{Locality Sort}         
     \end{subfigure}\\
     \begin{subfigure}[t]{0.14\textwidth}
         \centering
         \includegraphics[width=.95\textwidth]{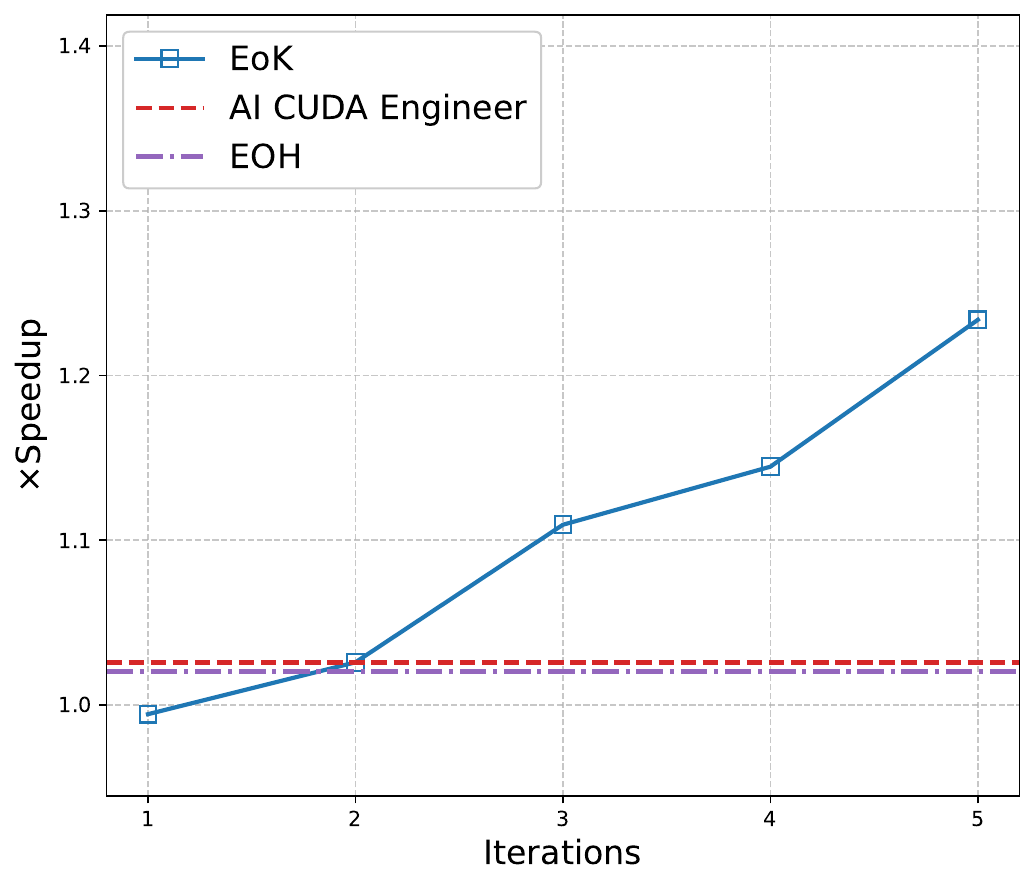}
         \caption*{Reduce and Scan}         
     \end{subfigure}\hfill
     \begin{subfigure}[t]{0.14\textwidth}
         \centering
         \includegraphics[width=.95\textwidth]{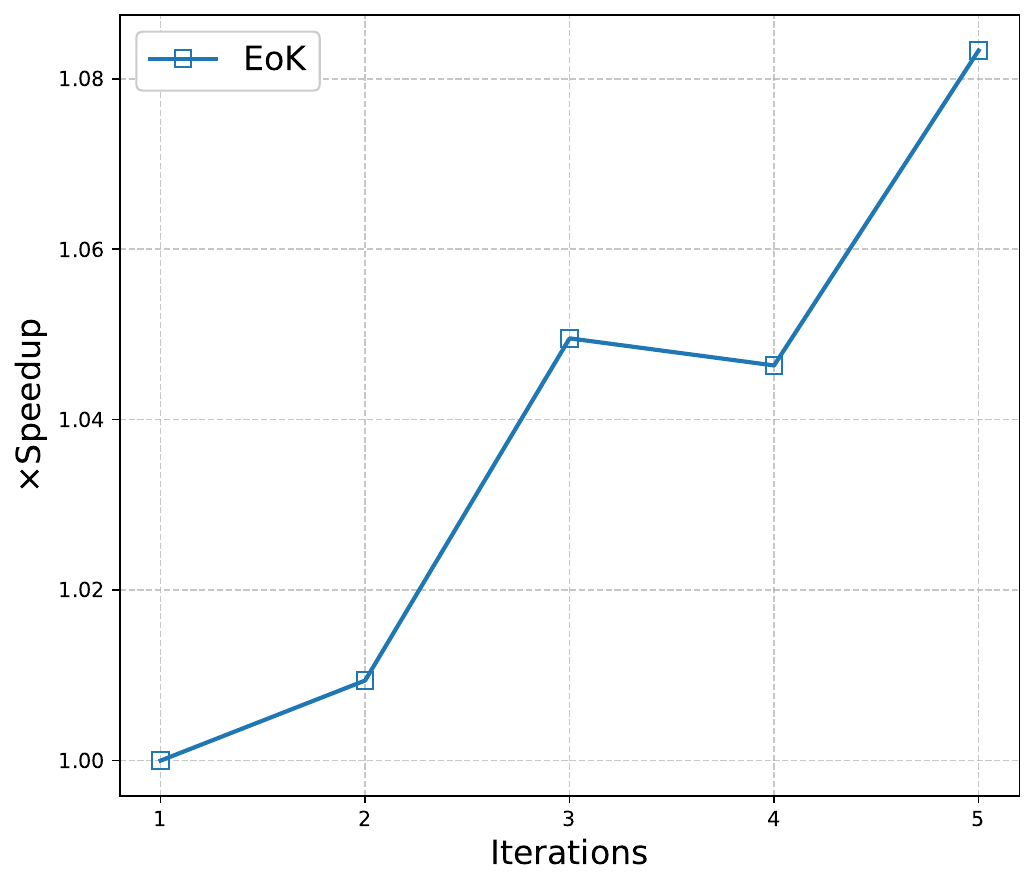}
         \caption*{Vectorized Sorted Search}         
     \end{subfigure}\hfill
     \begin{subfigure}[t]{0.14\textwidth}
         \centering
         \includegraphics[width=.95\textwidth]{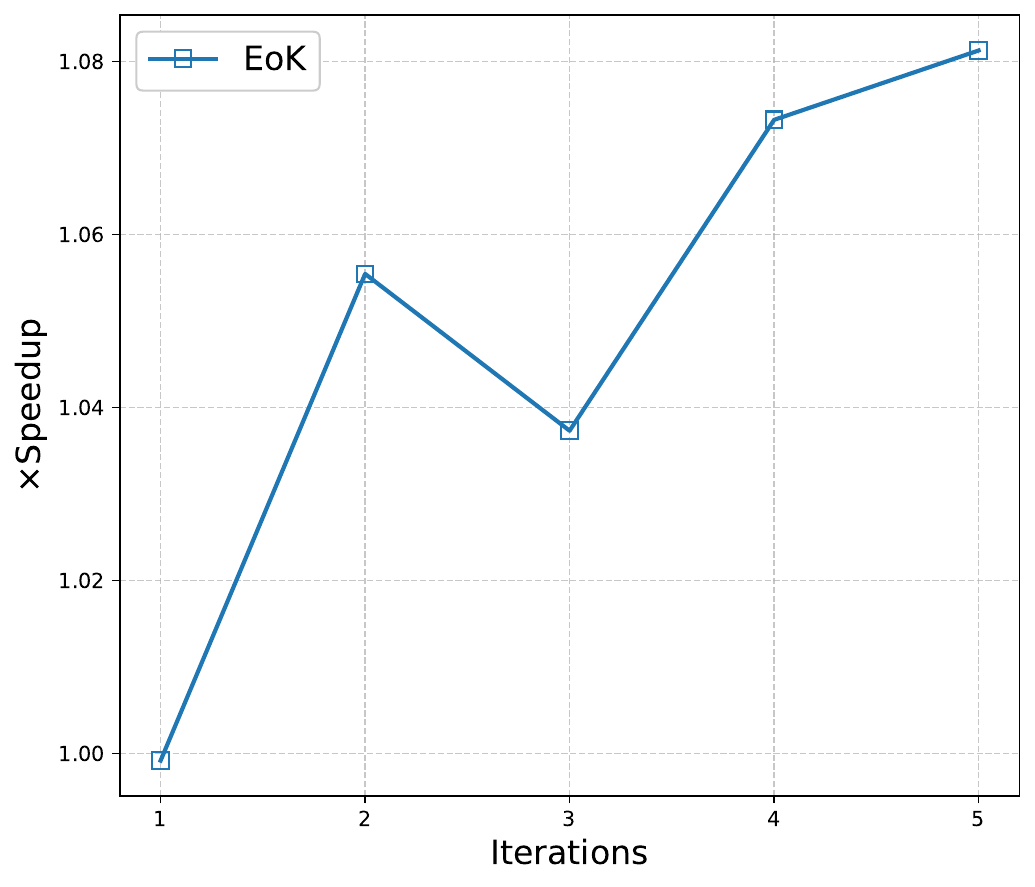}
         \caption*{Load Balancing Search}        
     \end{subfigure}\hfill
     \begin{subfigure}[t]{0.14\textwidth}
         \centering
         \includegraphics[width=.95\textwidth]{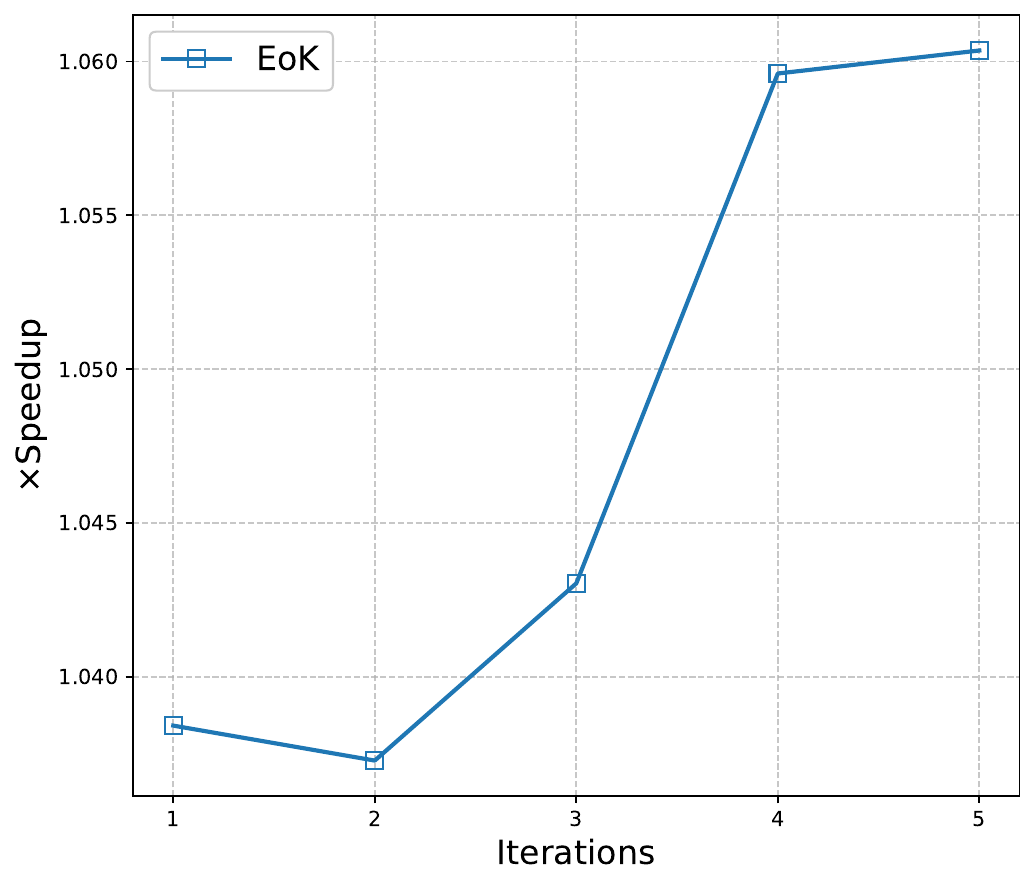}
         \caption*{Interval Expand}         
     \end{subfigure}\hfill
     \begin{subfigure}[t]{0.14\textwidth}
         \centering
         \includegraphics[width=.95\textwidth]{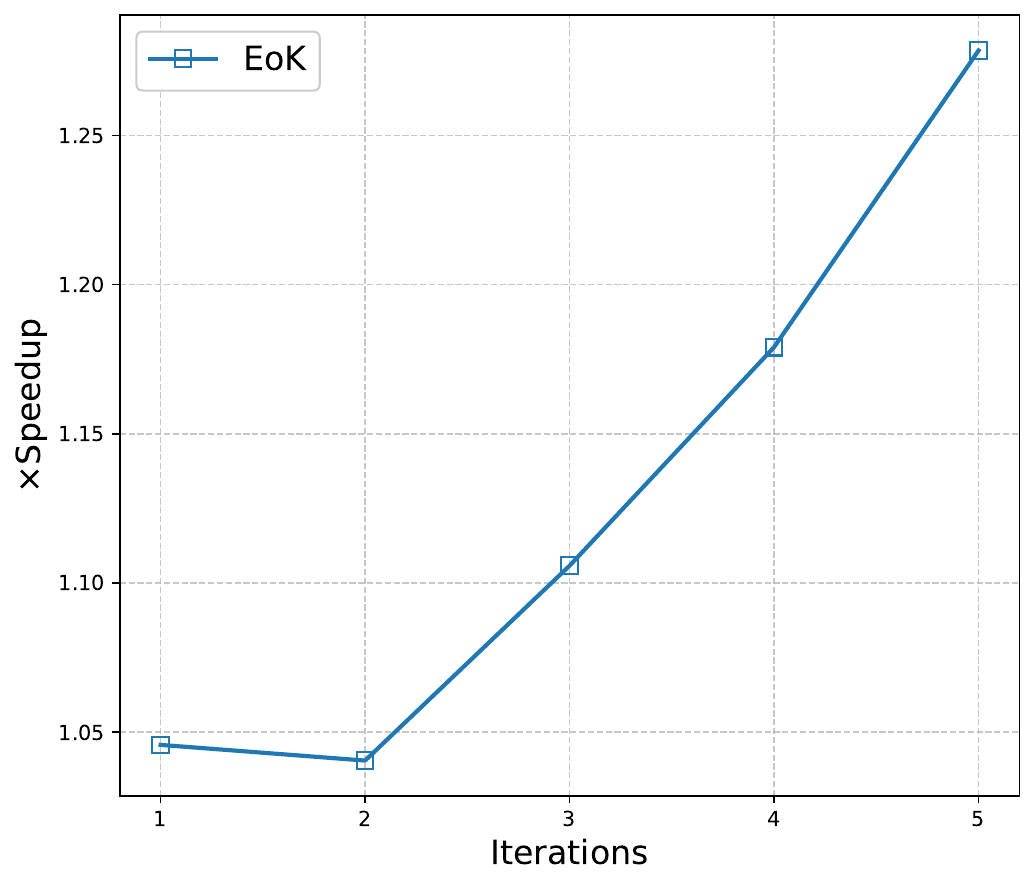}
         \caption*{Relational Joins}         
     \end{subfigure}\hfill
     \begin{subfigure}[t]{0.14\textwidth}
         \centering
         \includegraphics[width=.95\textwidth]{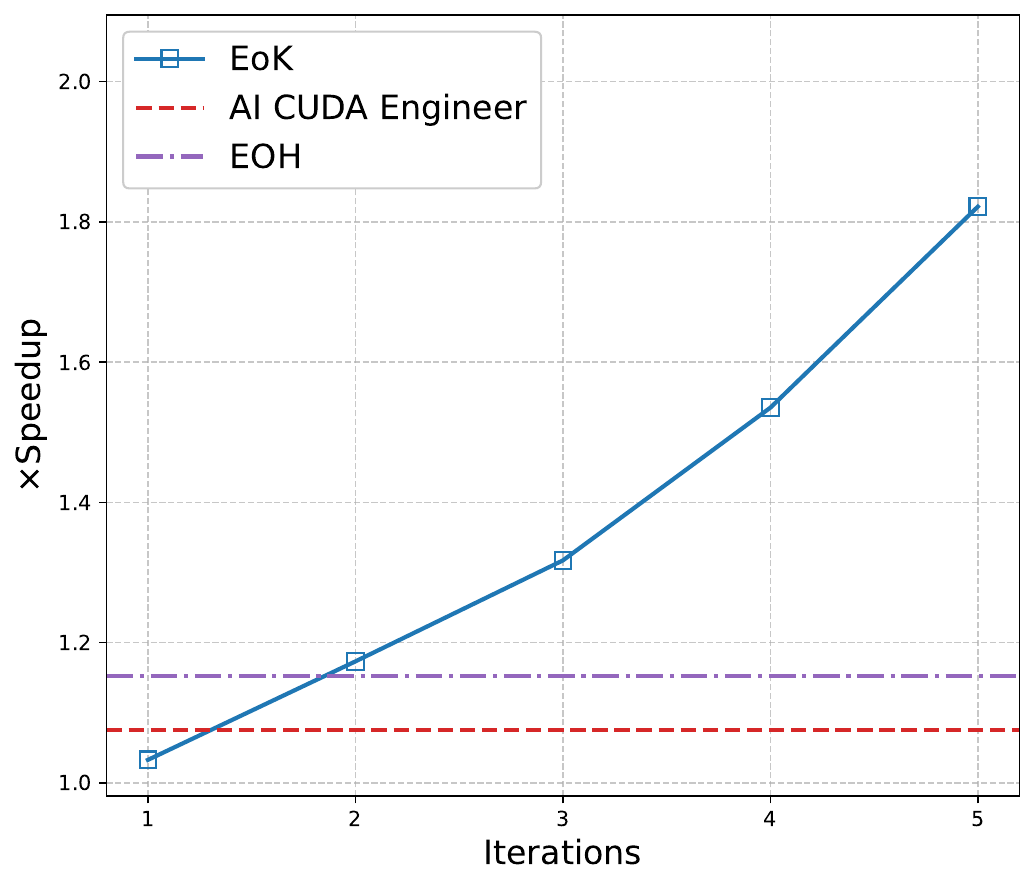}
         \caption*{Segmented Sort}        
     \end{subfigure}\hfill
     \begin{subfigure}[t]{0.14\textwidth}
         \centering
         \includegraphics[width=.95\textwidth]{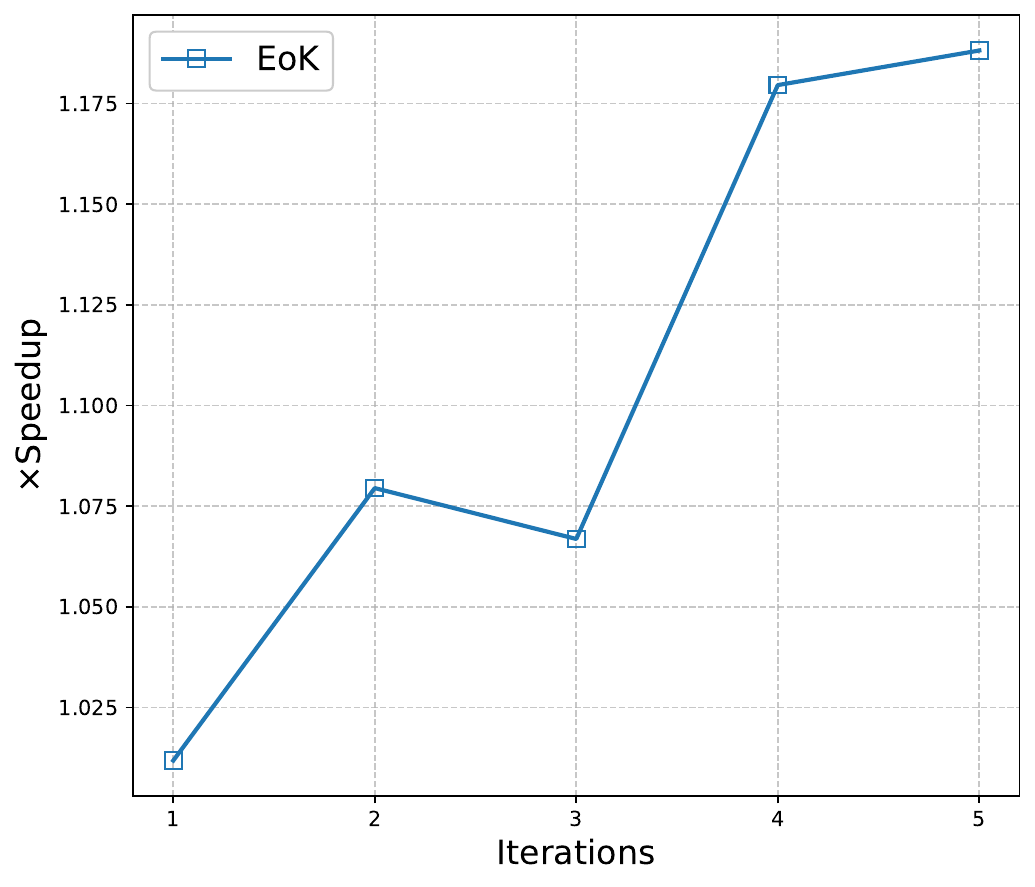}
         \caption*{Segmented Reduction}       
     \end{subfigure}
     \caption{Plots of iteration versus speedup for \ourmethod{} across 14 general kernels. Speedups from other baseline methods are shown for a subset of kernels where these methods achieved at least one successful implementation.}
     \label{appxfig:gen _kernel_convergence_individual}
\end{figure}

\begin{figure}[b]
    \vspace{-20em}
    \captionsetup[subfigure]{justification=centering}
    \begin{subfigure}[b]{0.14\textwidth}
         \centering
         \includegraphics[width=.95\textwidth]{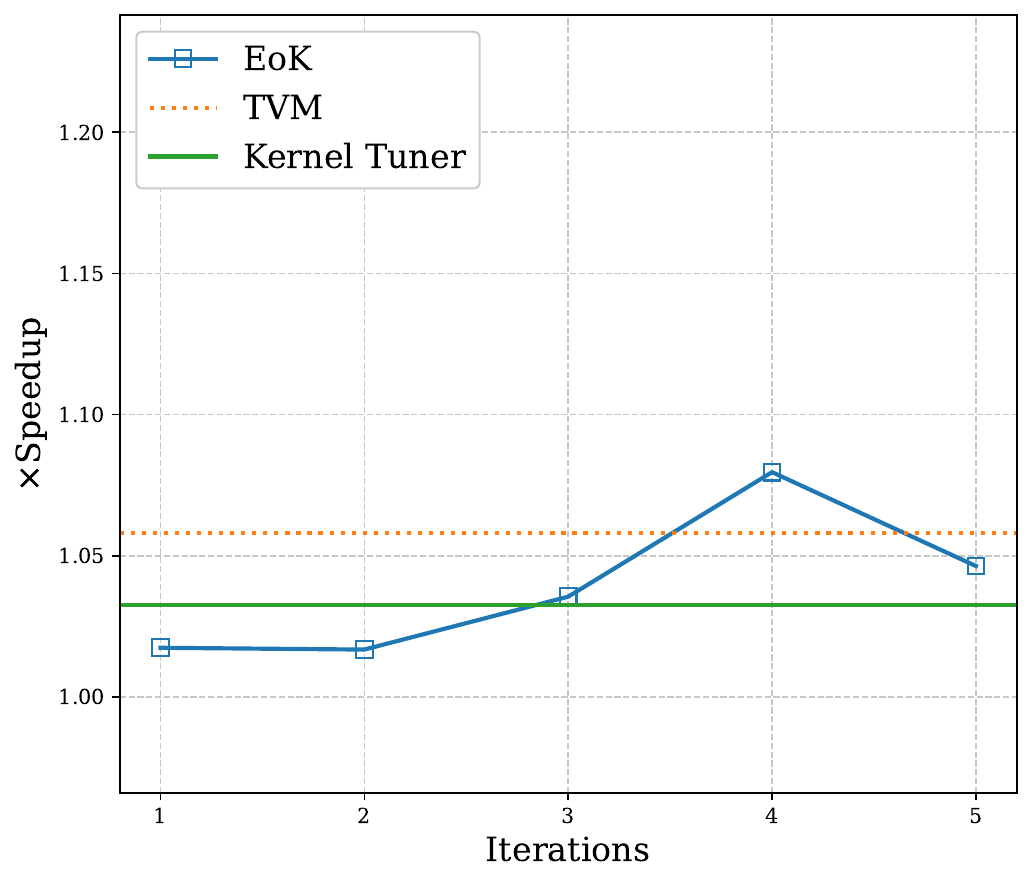}
         \caption*{Abs value}         
     \end{subfigure}\hfill
     \begin{subfigure}[b]{0.14\textwidth}
         \centering
         \includegraphics[width=.95\textwidth]{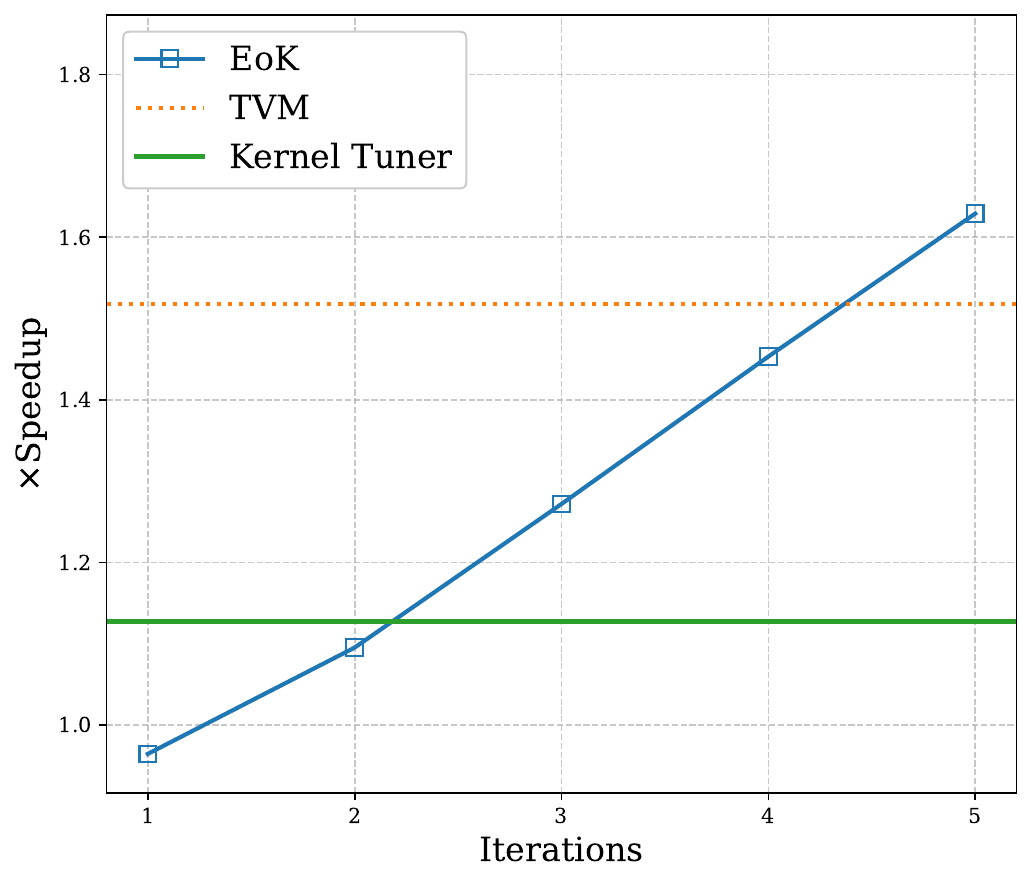}
         \caption*{BatchNorm}
     \end{subfigure}\hfill
     \begin{subfigure}[b]{0.14\textwidth}
         \centering
         \includegraphics[width=.95\textwidth]{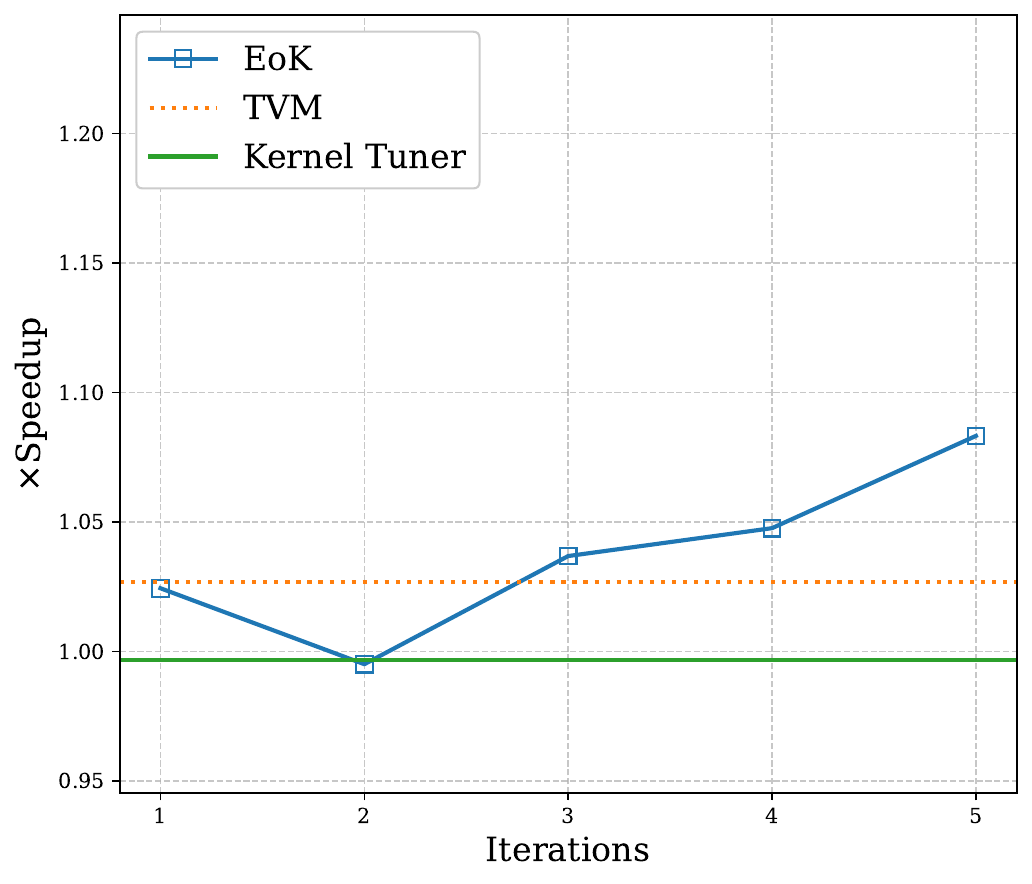}
         \caption*{Bias}  
     \end{subfigure}\hfill
     \begin{subfigure}[b]{0.14\textwidth}
         \centering
         \includegraphics[width=.95\textwidth]{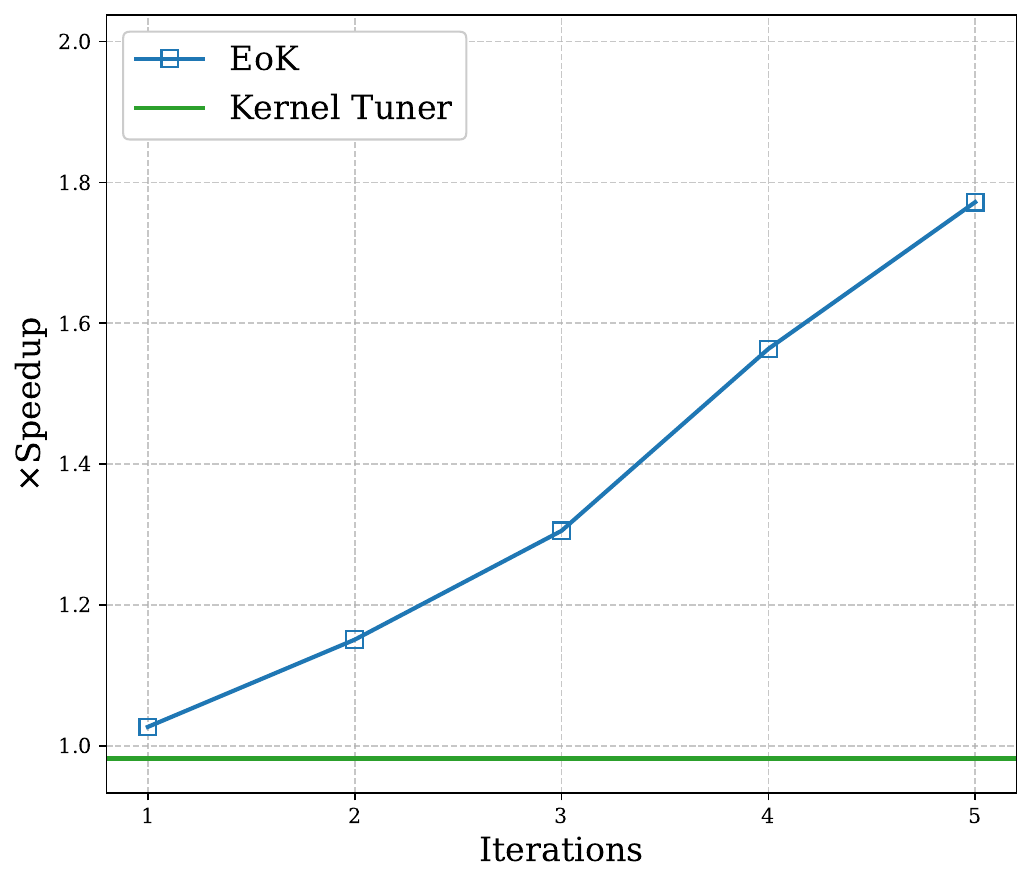}
         \caption*{Binary op}         
     \end{subfigure}\hfill
     \begin{subfigure}[b]{0.14\textwidth}
         \centering
         \includegraphics[width=.95\textwidth]{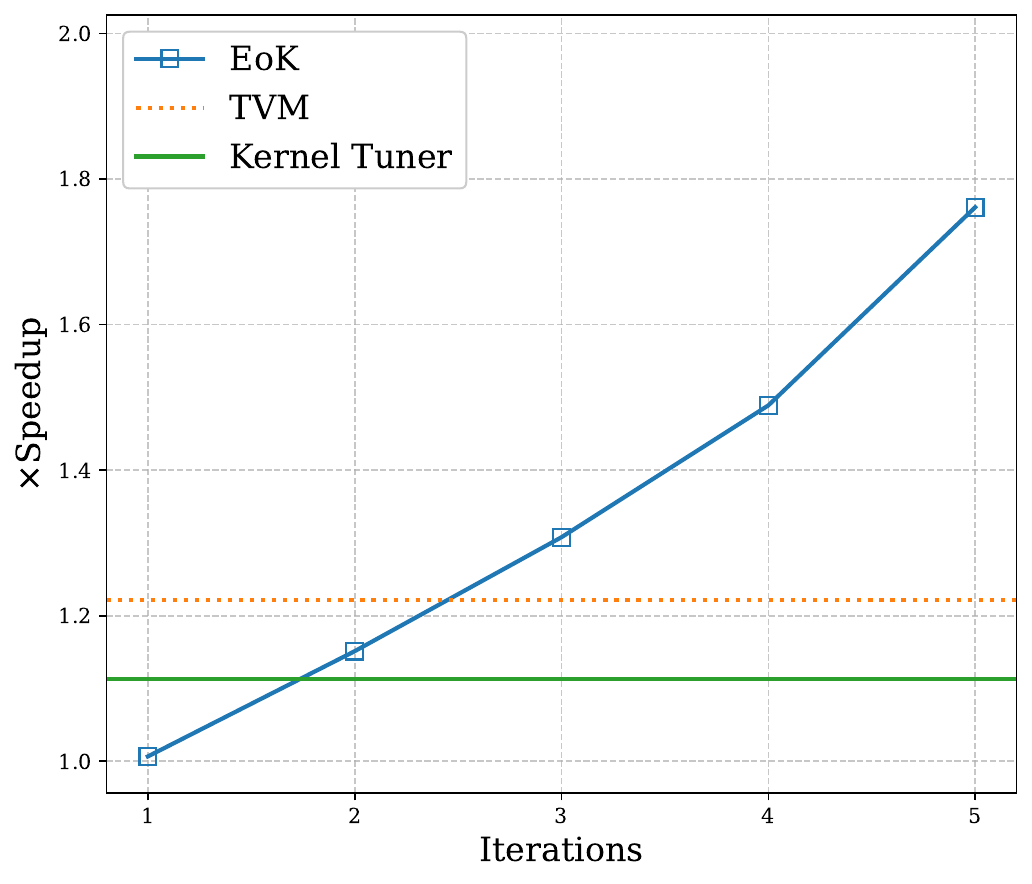}
         \caption*{BNLL}     
     \end{subfigure}\hfill
     \begin{subfigure}[b]{0.14\textwidth}
         \centering
         \includegraphics[width=.95\textwidth]{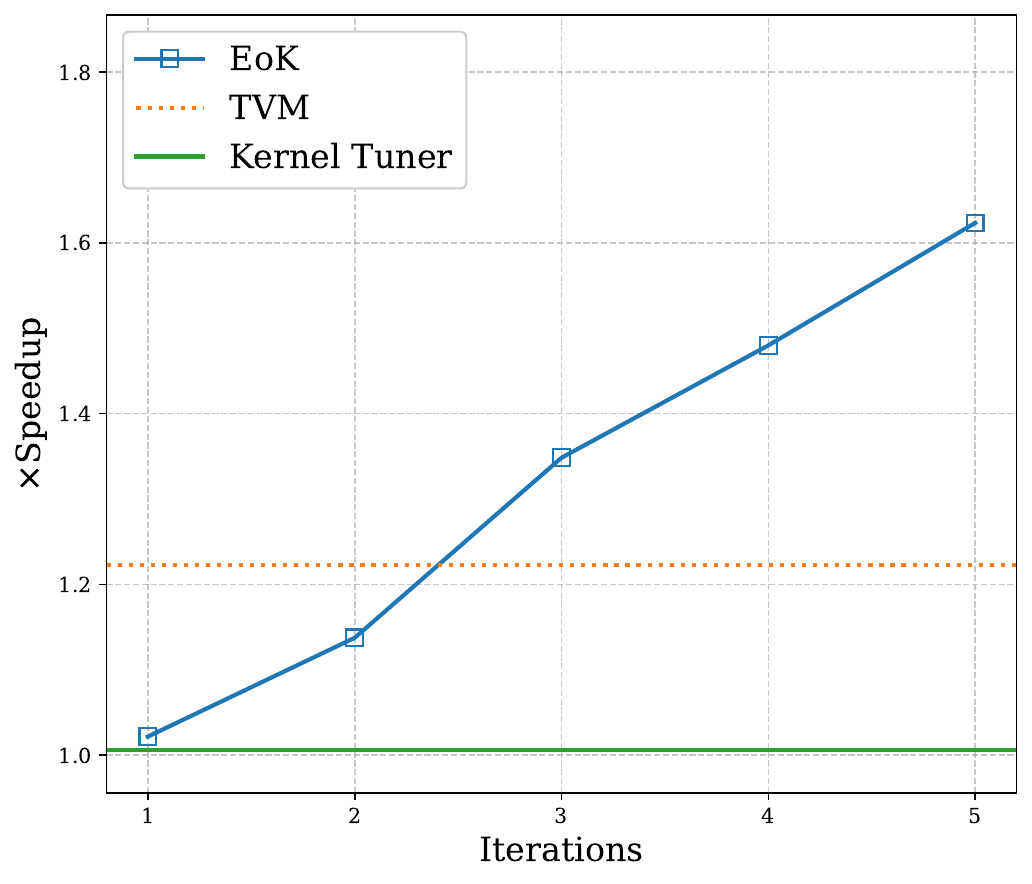}
         \caption*{Cast}        
     \end{subfigure}\hfill
     \begin{subfigure}[b]{0.14\textwidth}
         \centering
         \includegraphics[width=.95\textwidth]{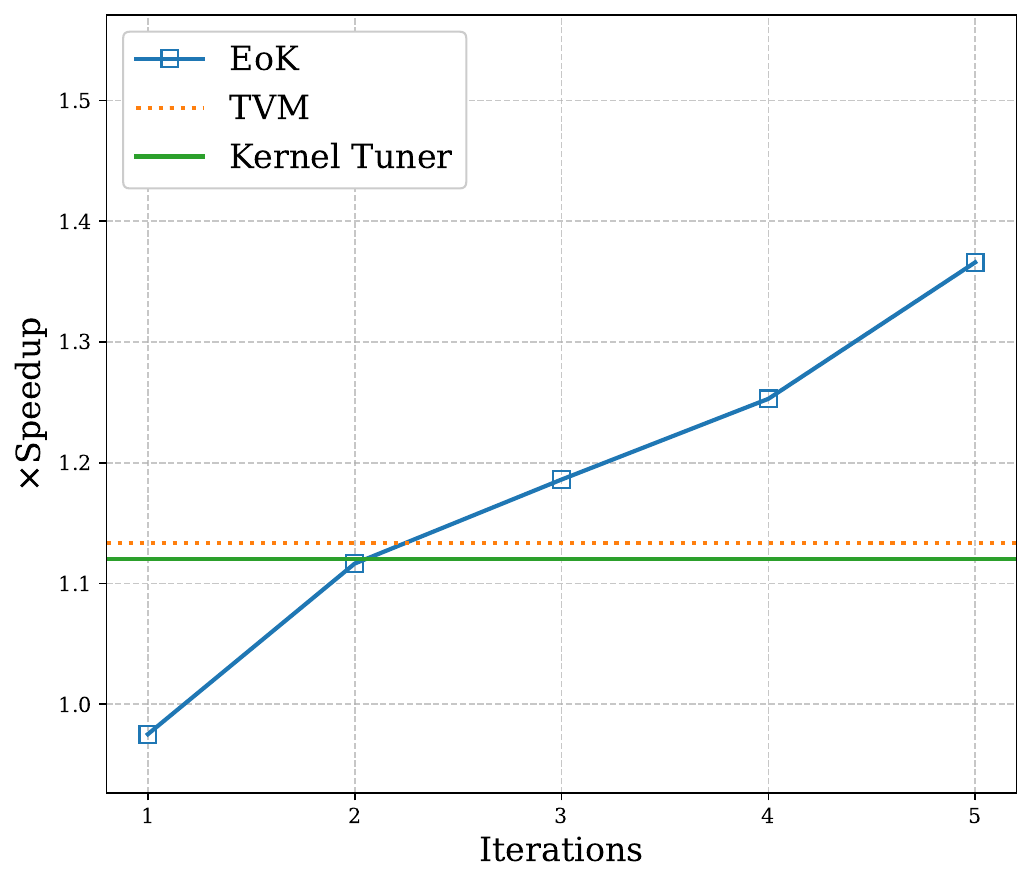}
         \caption*{CELU}         
     \end{subfigure}\\
     \begin{subfigure}[b]{0.14\textwidth}
         \centering
         \includegraphics[width=.95\textwidth]{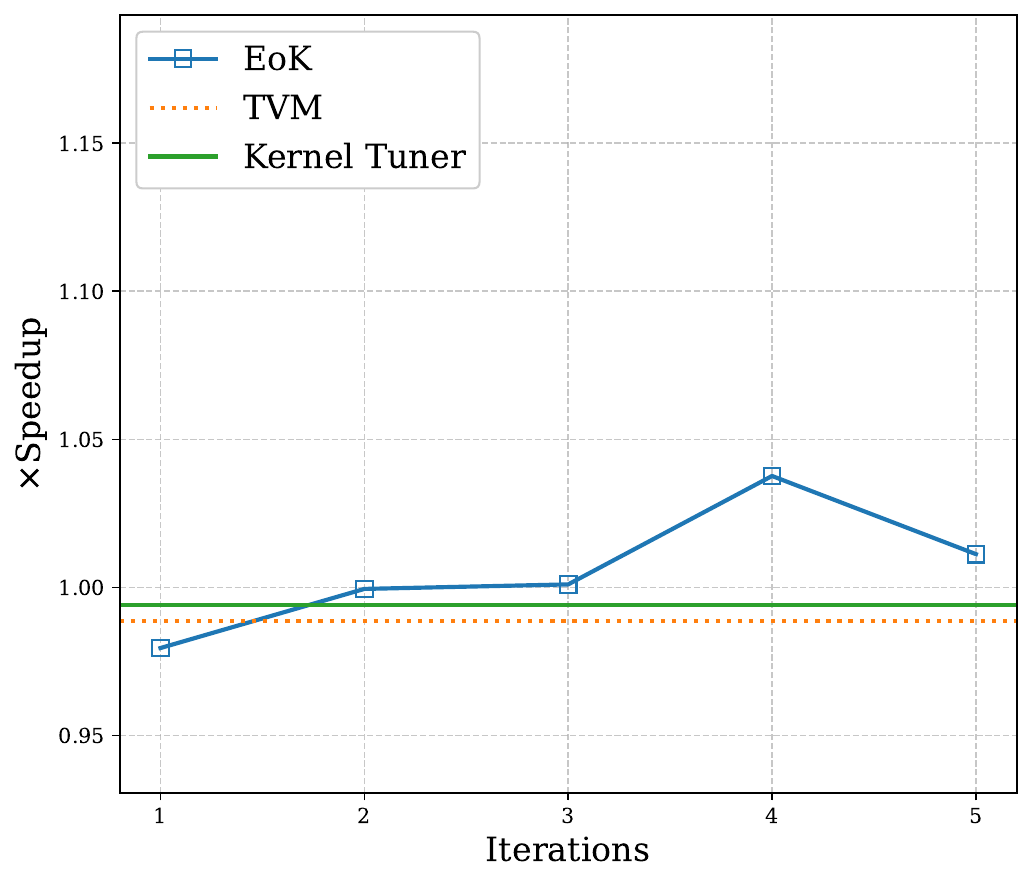}
         \caption*{Clip}         
     \end{subfigure}\hfill
     \begin{subfigure}[b]{0.14\textwidth}
         \centering
         \includegraphics[width=.95\textwidth]{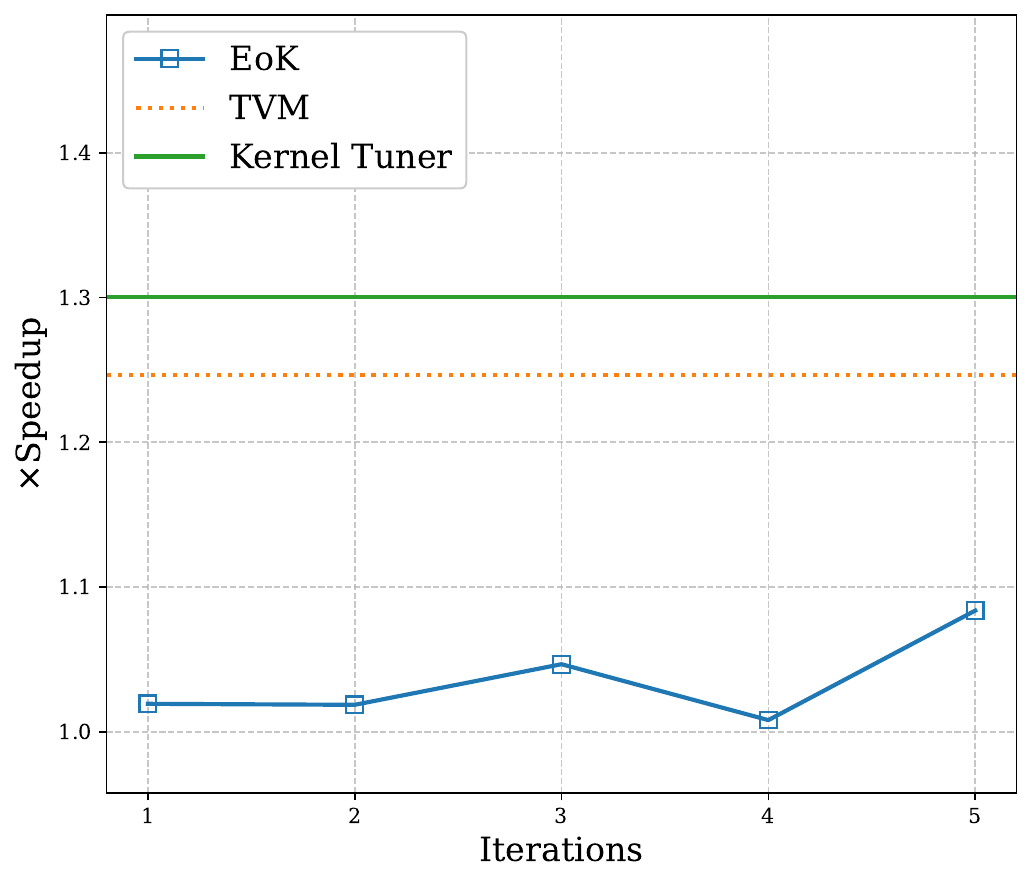}
         \caption*{Concat}        
     \end{subfigure}\hfill
     \begin{subfigure}[b]{0.14\textwidth}
         \centering
         \includegraphics[width=.95\textwidth]{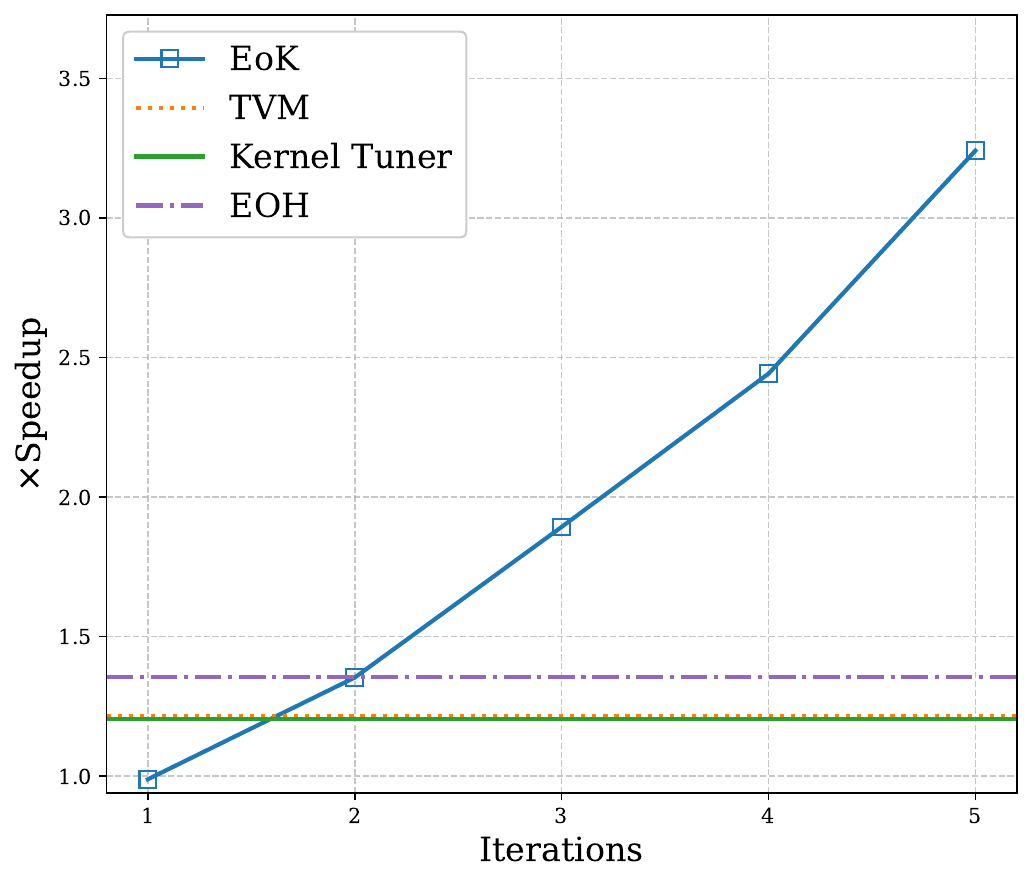}
         \caption*{Conv 1D}         
     \end{subfigure}\hfill
     \begin{subfigure}[b]{0.14\textwidth}
         \centering
         \includegraphics[width=.95\textwidth]{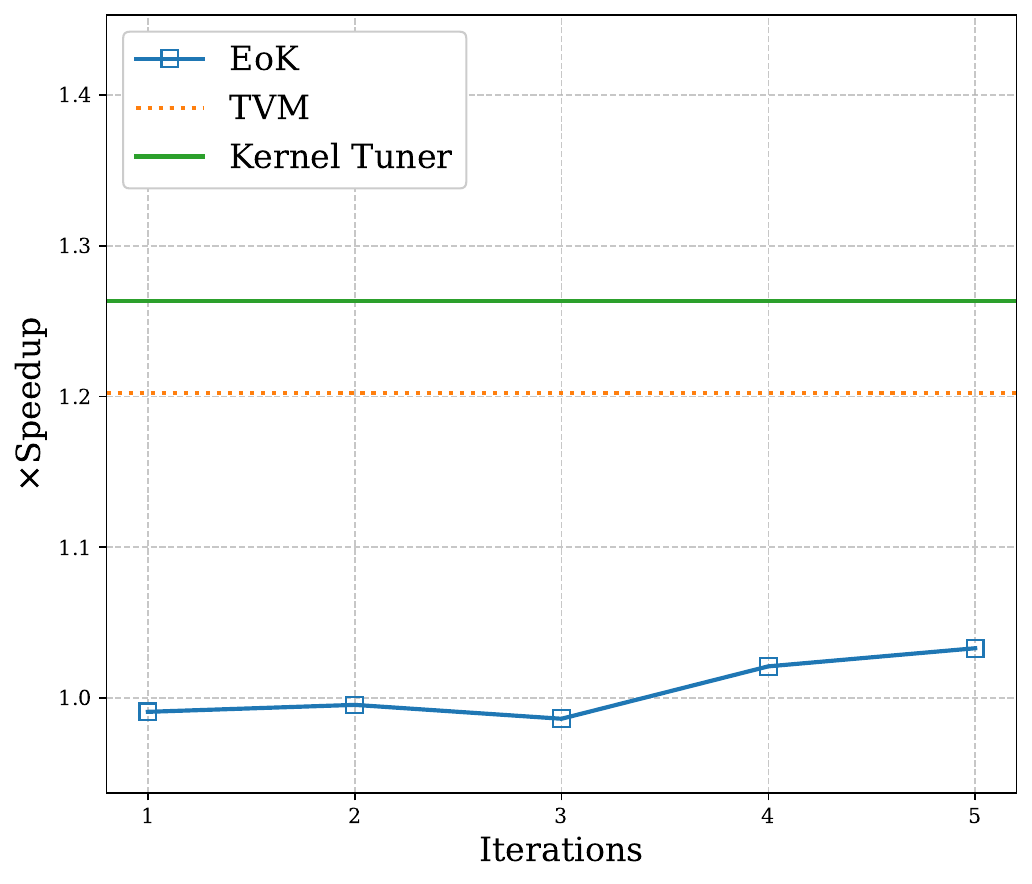}
         \caption*{DeConv}         
     \end{subfigure}\hfill
     \begin{subfigure}[b]{0.14\textwidth}
         \centering
         \includegraphics[width=.95\textwidth]{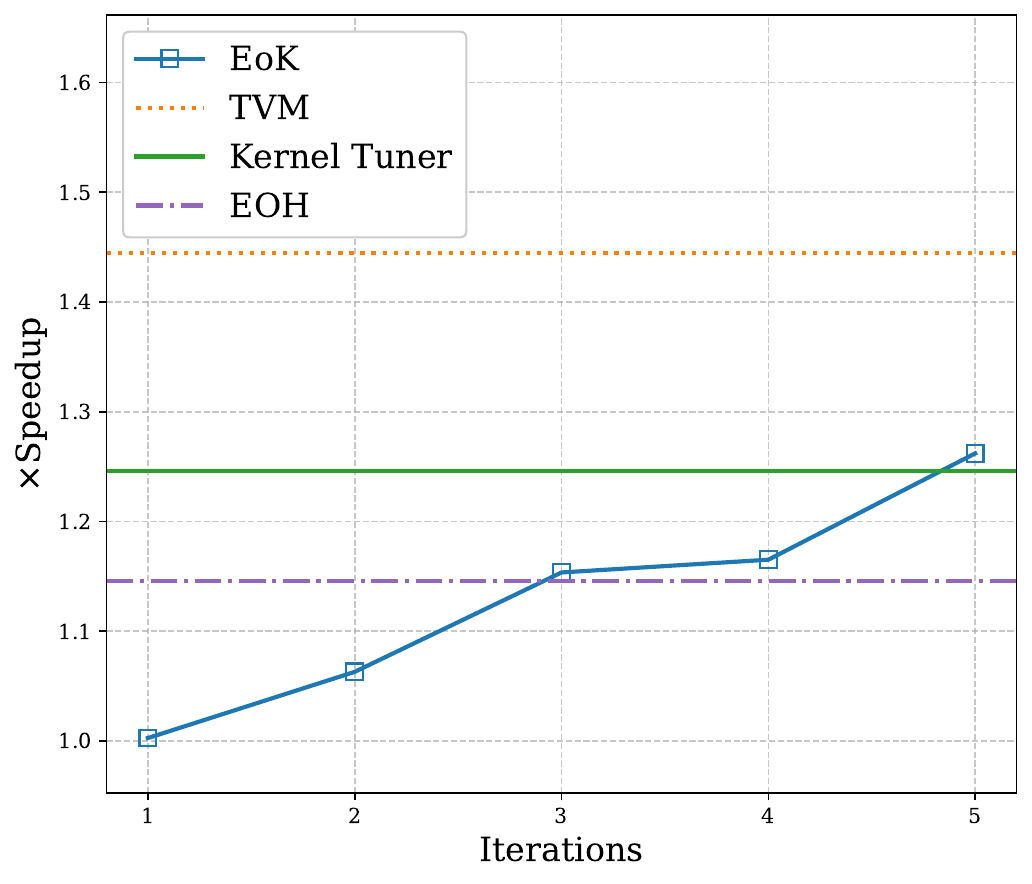}
         \caption*{Dep Conv}         
     \end{subfigure}\hfill
     \begin{subfigure}[b]{0.14\textwidth}
         \centering
         \includegraphics[width=.95\textwidth]{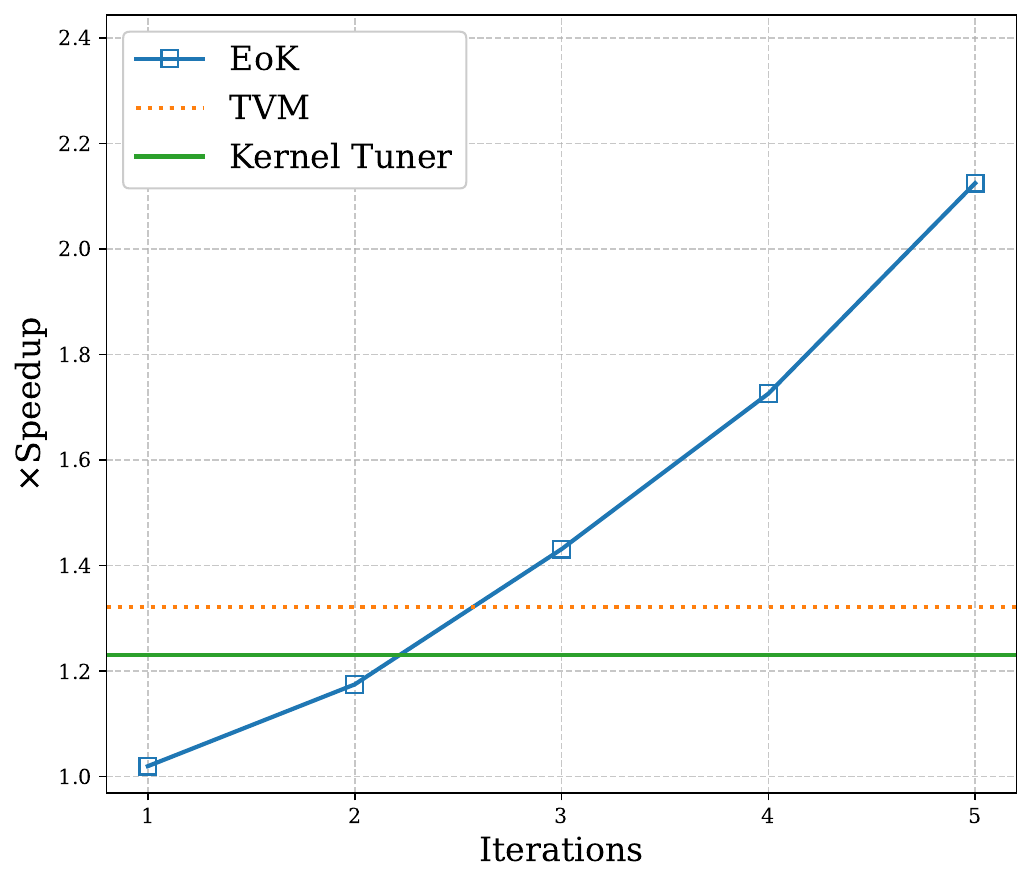}
         \caption*{Dep DeConv}         
     \end{subfigure}\hfill
     \begin{subfigure}[b]{0.14\textwidth}
         \centering
         \includegraphics[width=.95\textwidth]{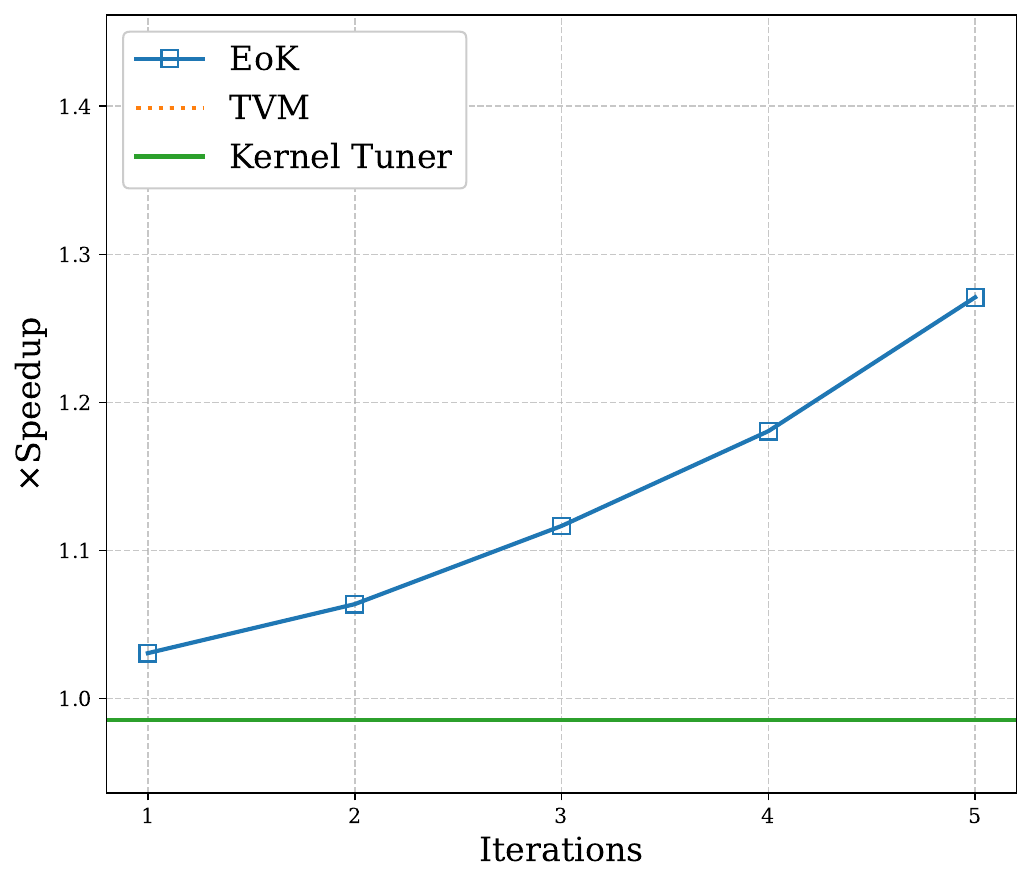}
         \caption*{Eltwise}       
     \end{subfigure}\\
     \begin{subfigure}[b]{0.14\textwidth}
         \centering
         \includegraphics[width=.95\textwidth]{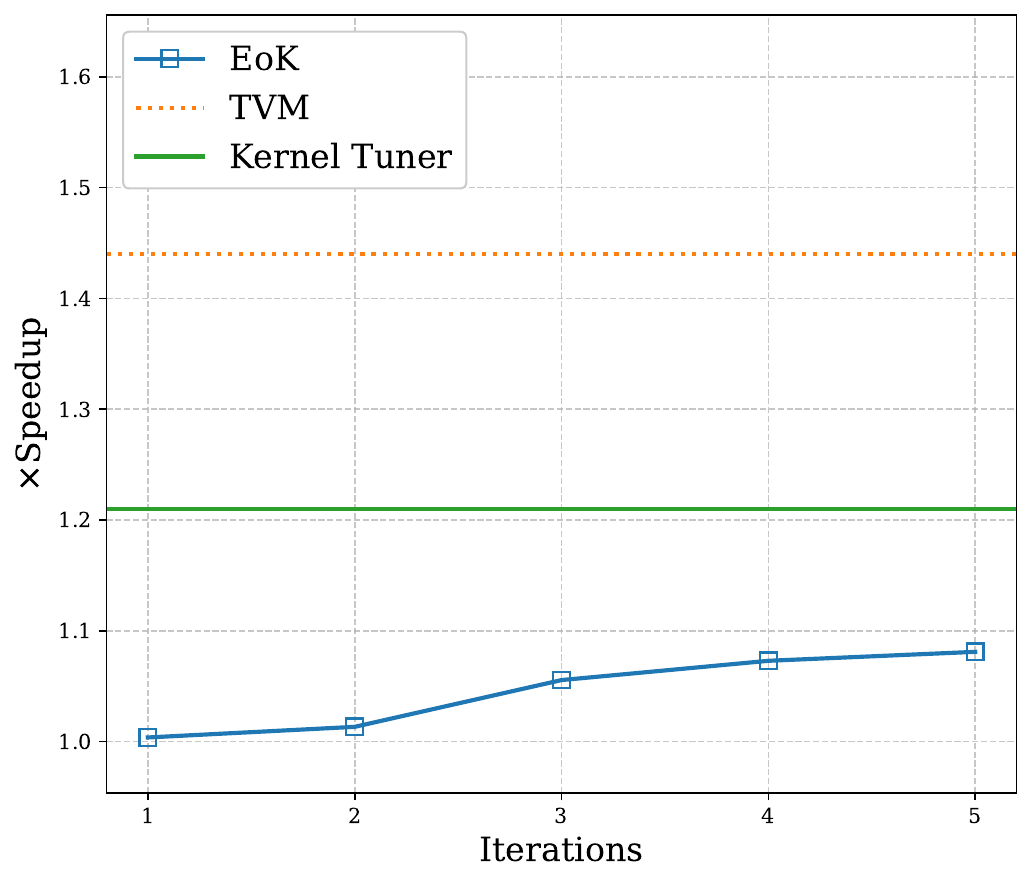}
         \caption*{GRU}         
     \end{subfigure}\hfill
     \begin{subfigure}[b]{0.14\textwidth}
         \centering
         \includegraphics[width=.95\textwidth]{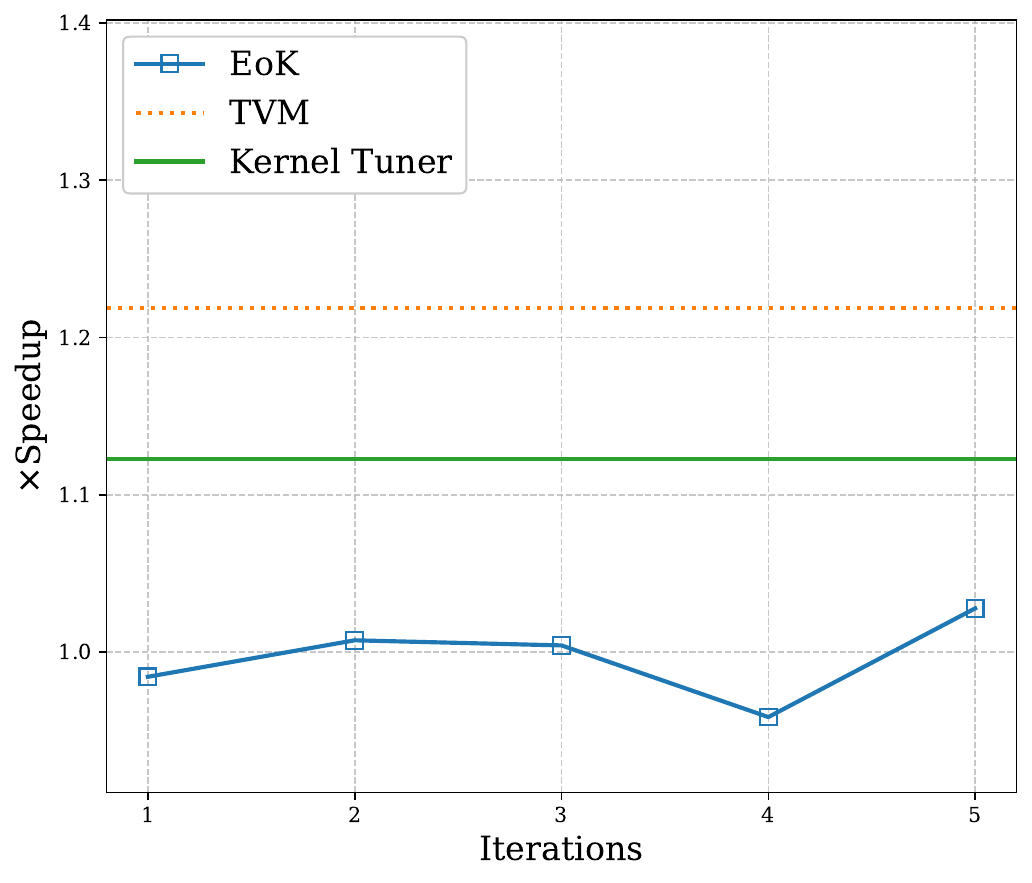}
         \caption*{Hard Sigmoid}         
     \end{subfigure}\hfill
     \begin{subfigure}[b]{0.14\textwidth}
         \centering
         \includegraphics[width=.95\textwidth]{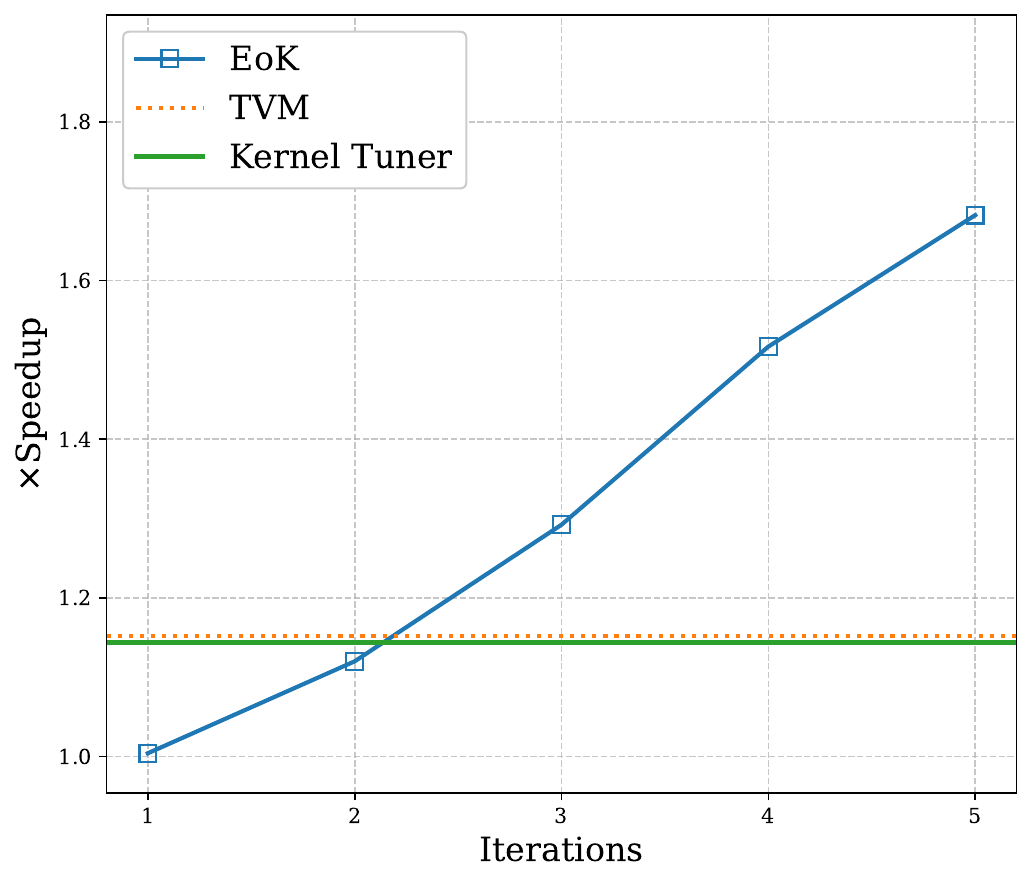}
         \caption*{Hard Swish}       
     \end{subfigure}\hfill
     \begin{subfigure}[b]{0.14\textwidth}
         \centering
         \includegraphics[width=.95\textwidth]{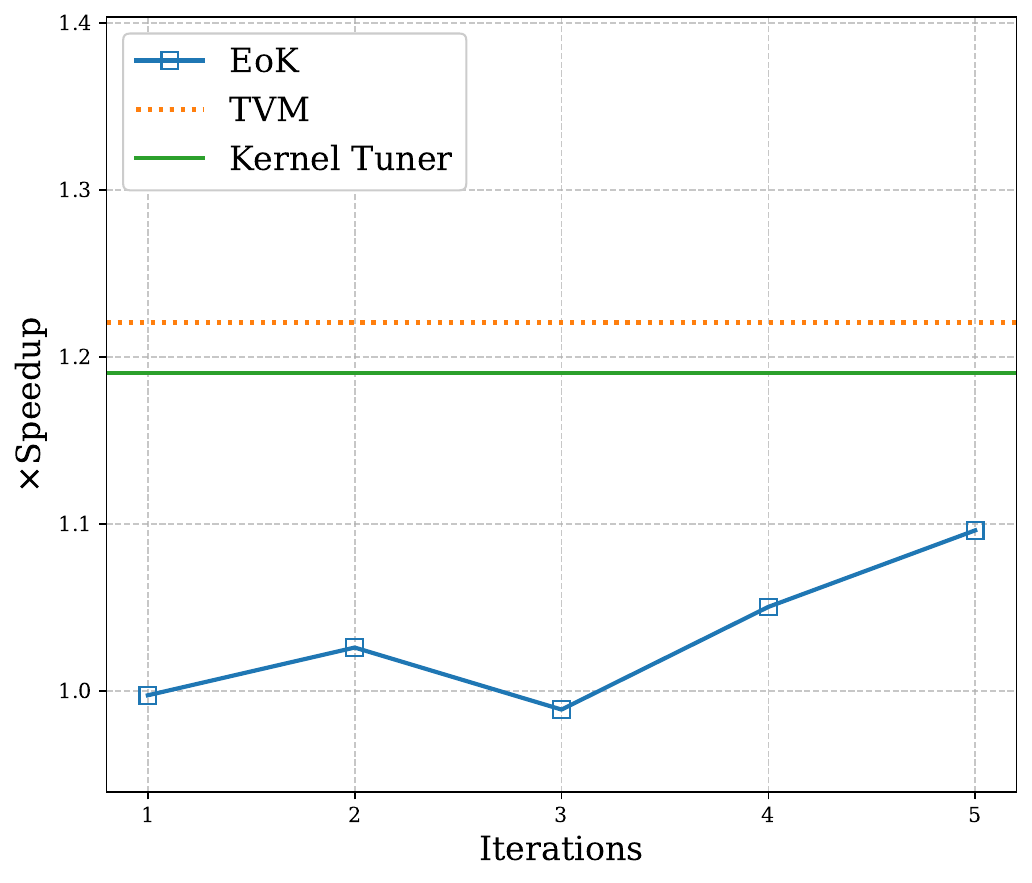}
         \caption*{Inner Product}         
     \end{subfigure}\hfill
     \begin{subfigure}[b]{0.14\textwidth}
         \centering
         \includegraphics[width=.95\textwidth]{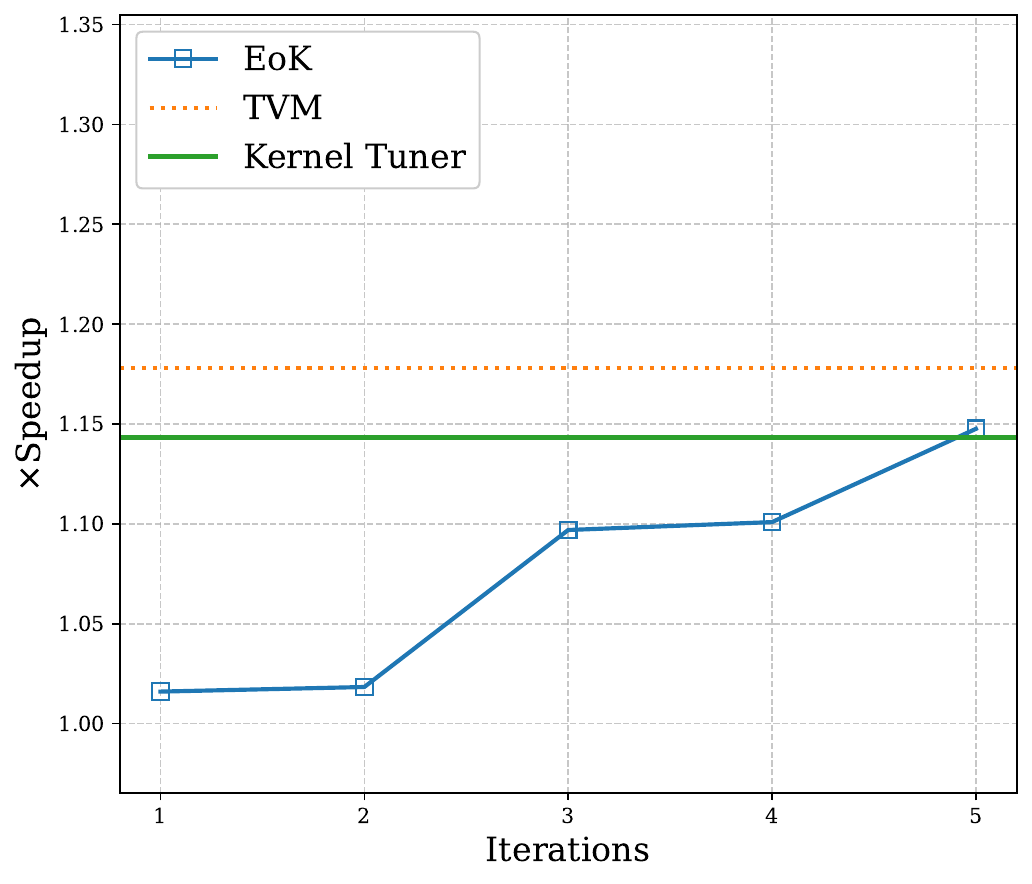}
         \caption*{InstanceNorm}         
     \end{subfigure}\hfill
     \begin{subfigure}[b]{0.14\textwidth}
         \centering
         \includegraphics[width=.95\textwidth]{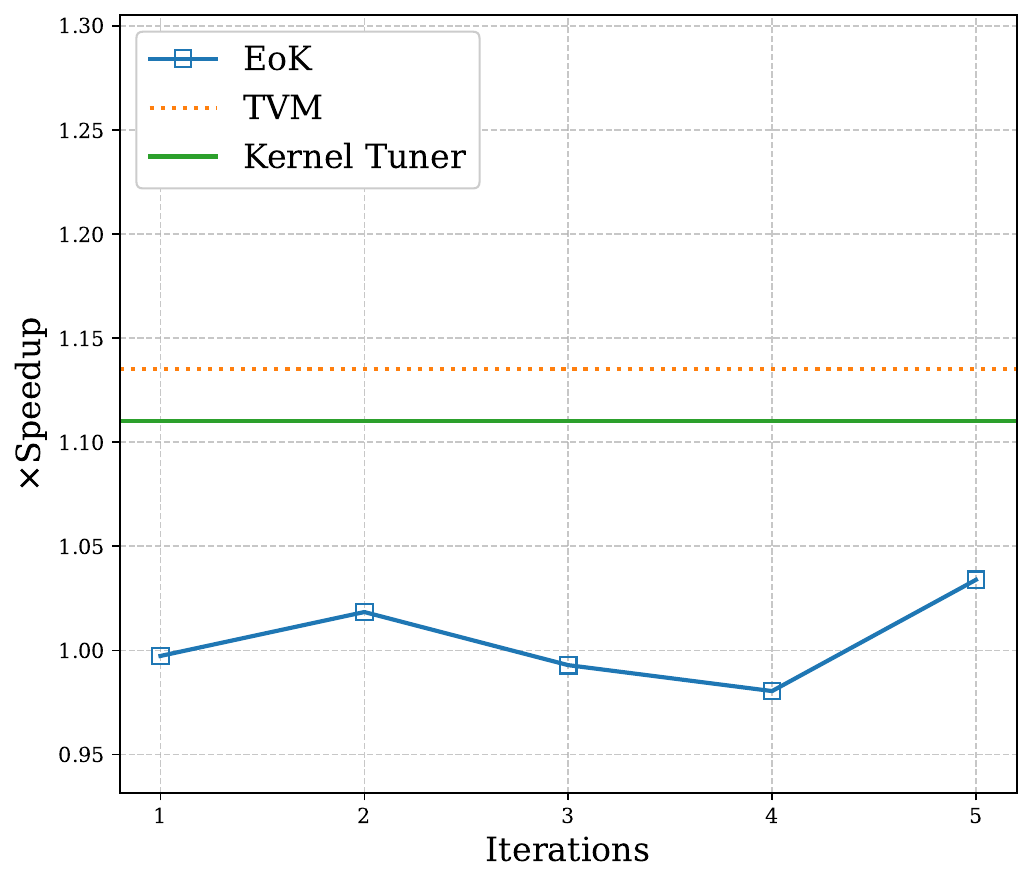}
         \caption*{Interp}         
     \end{subfigure}\hfill
     \begin{subfigure}[b]{0.14\textwidth}
         \centering
         \includegraphics[width=.95\textwidth]{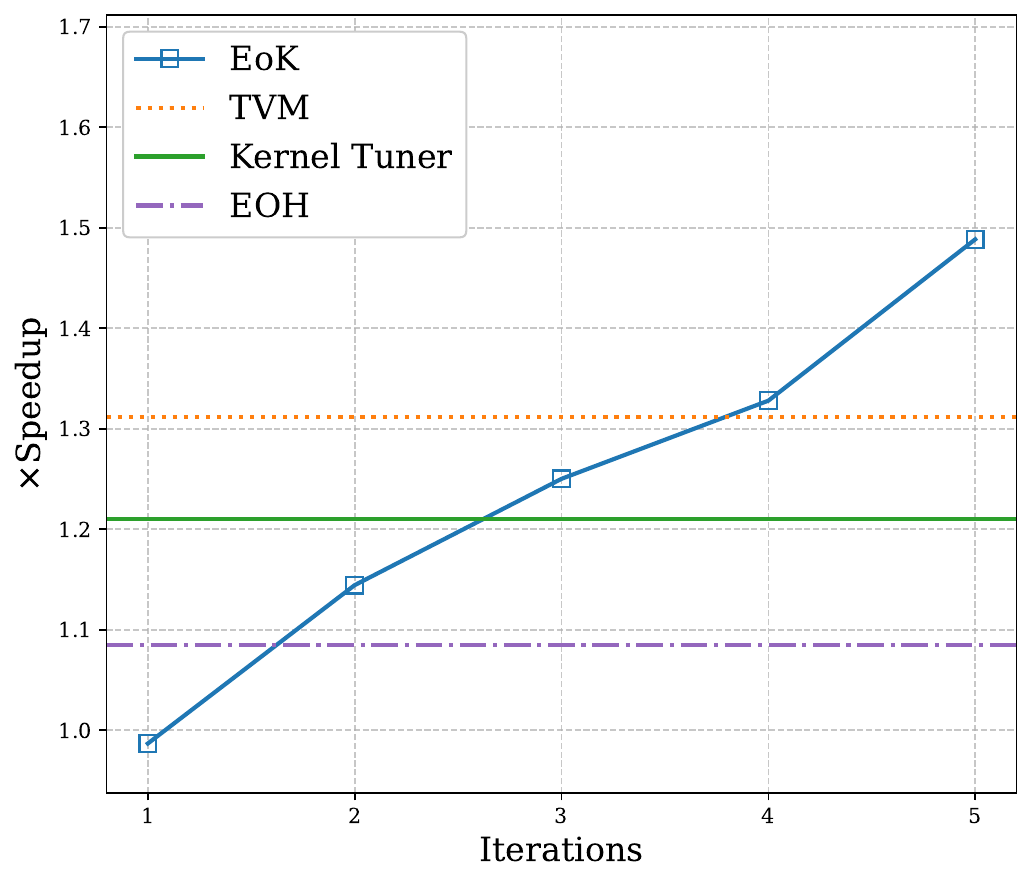}
         \caption*{Mish}        
     \end{subfigure}\\
     \begin{subfigure}[b]{0.14\textwidth}
         \centering
         \includegraphics[width=.95\textwidth]{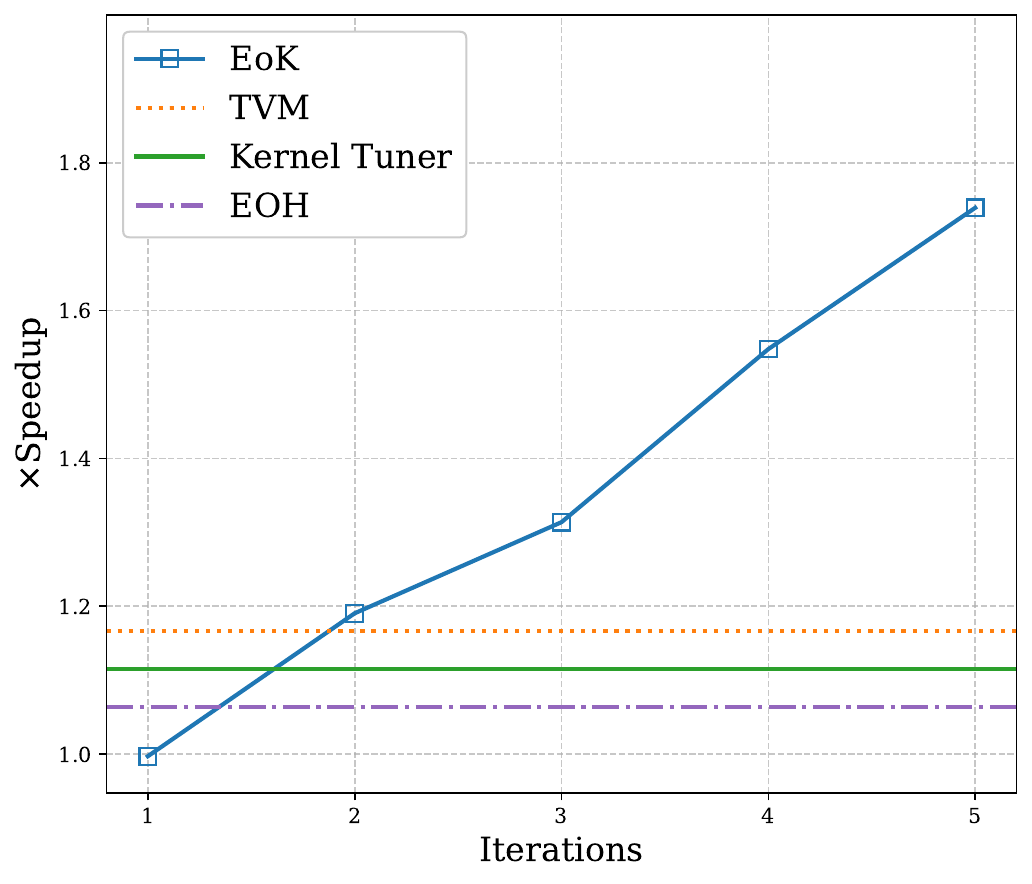}
         \caption*{Pooling}         
     \end{subfigure}\hfill
     \begin{subfigure}[b]{0.14\textwidth}
         \centering
         \includegraphics[width=.95\textwidth]{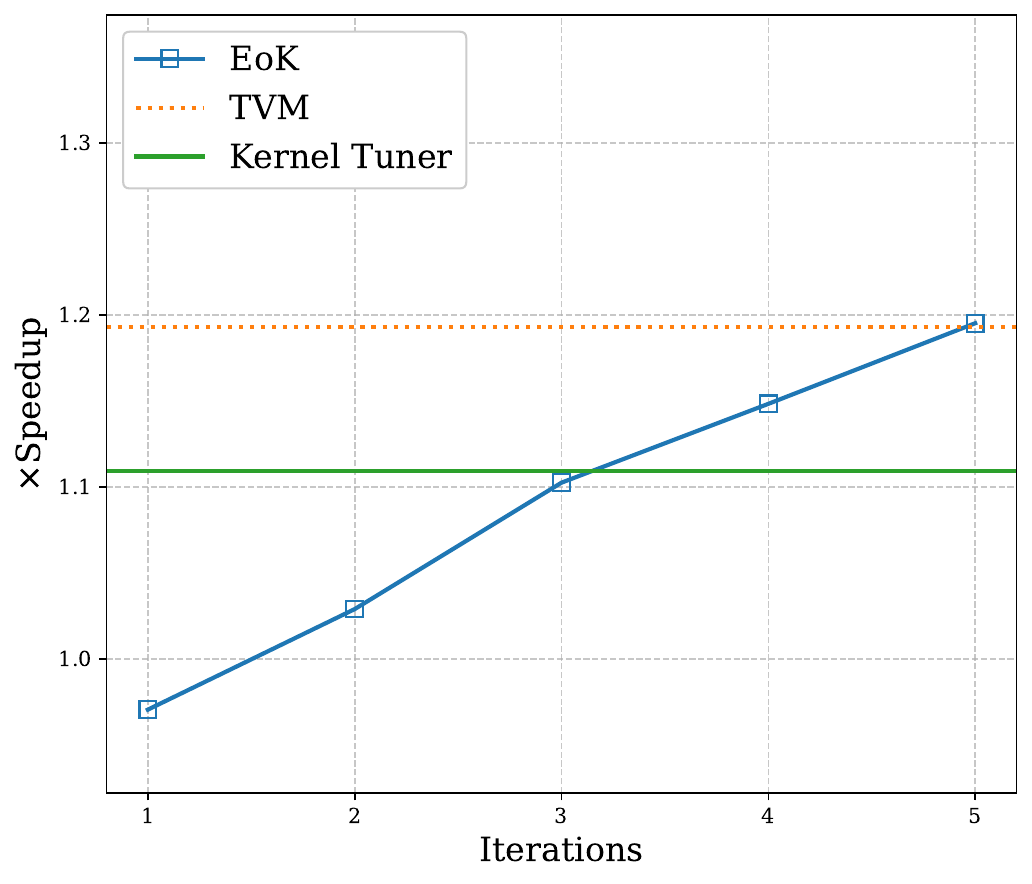}
         \caption*{PReLU}        
     \end{subfigure}\hfill
     \begin{subfigure}[b]{0.14\textwidth}
         \centering
         \includegraphics[width=.95\textwidth]{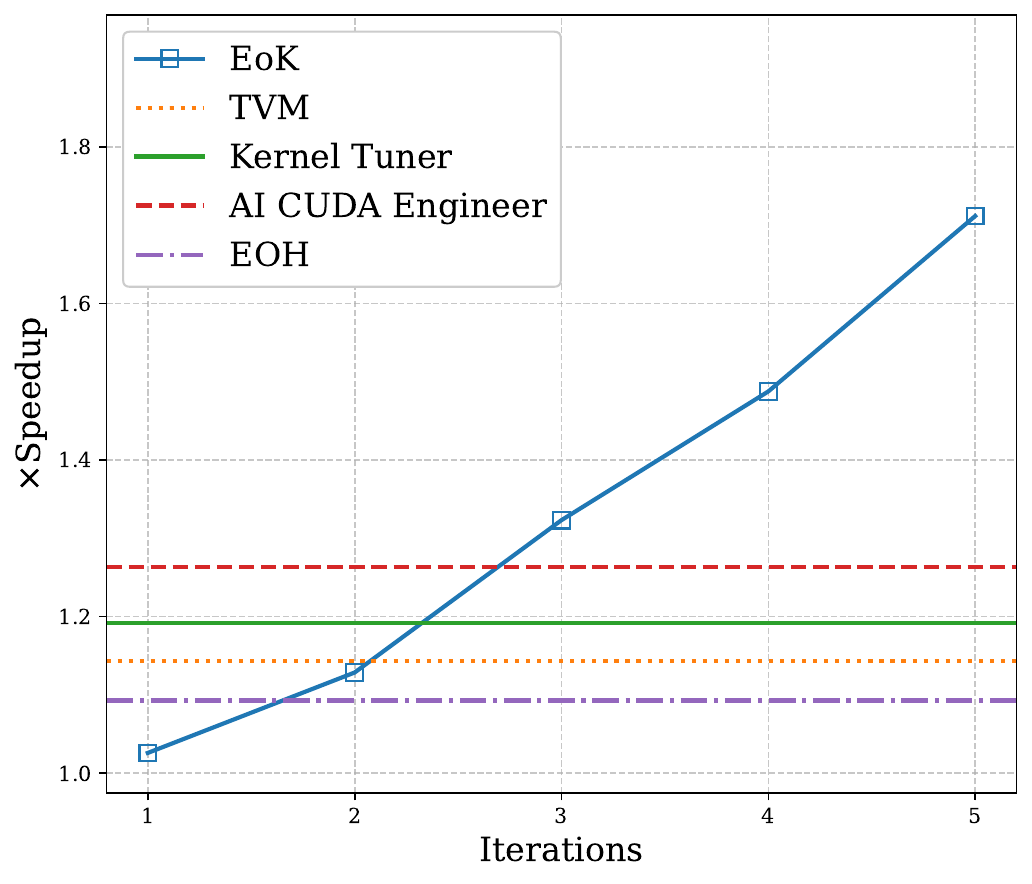}
         \caption*{ReLU}         
     \end{subfigure}\hfill
     \begin{subfigure}[b]{0.14\textwidth}
         \centering
         \includegraphics[width=.95\textwidth]{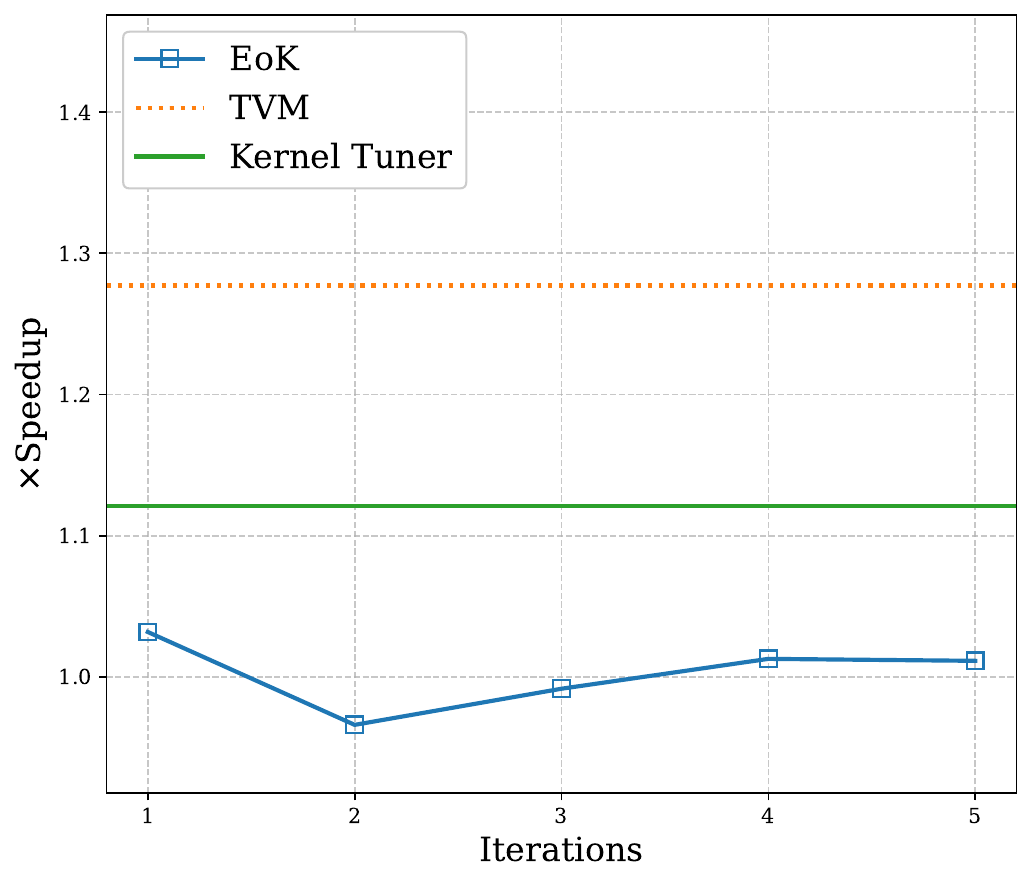}
         \caption*{Sigmoid}         
     \end{subfigure}\hfill
     \begin{subfigure}[b]{0.14\textwidth}
         \centering
         \includegraphics[width=.95\textwidth]{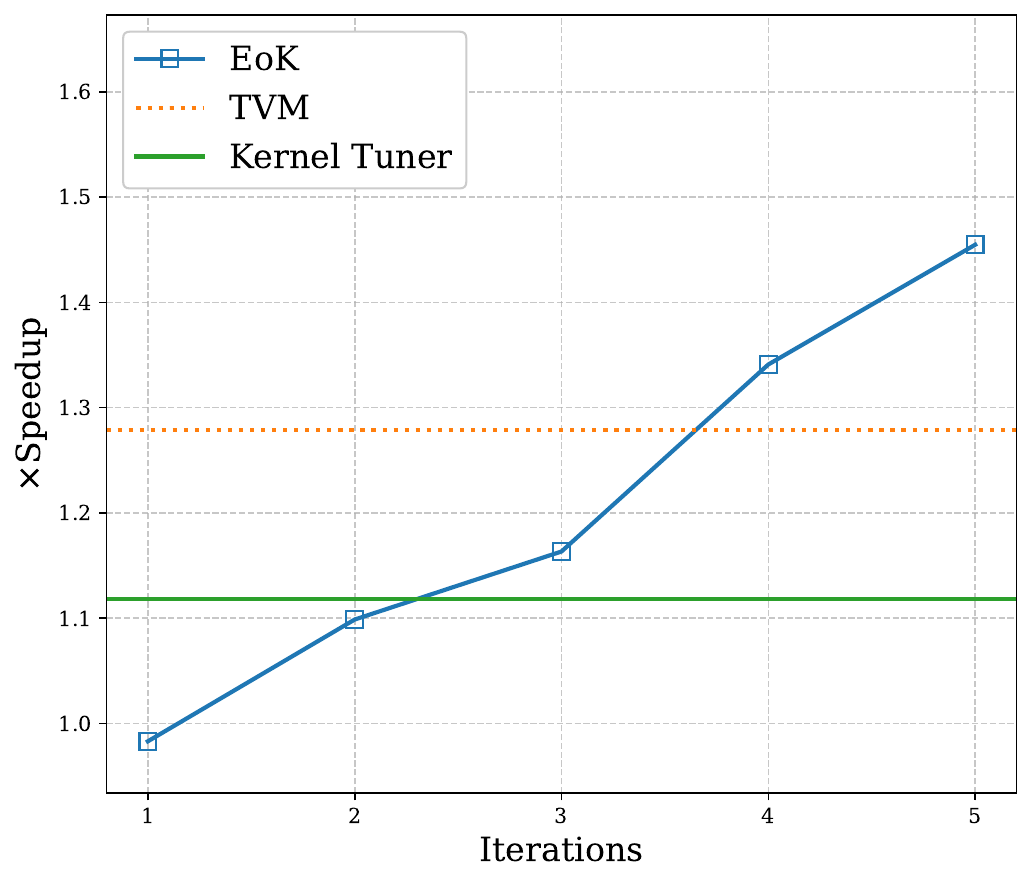}
         \caption*{Swish}        
     \end{subfigure}\hfill
     \begin{subfigure}[b]{0.14\textwidth}
         \centering
         \includegraphics[width=.95\textwidth]{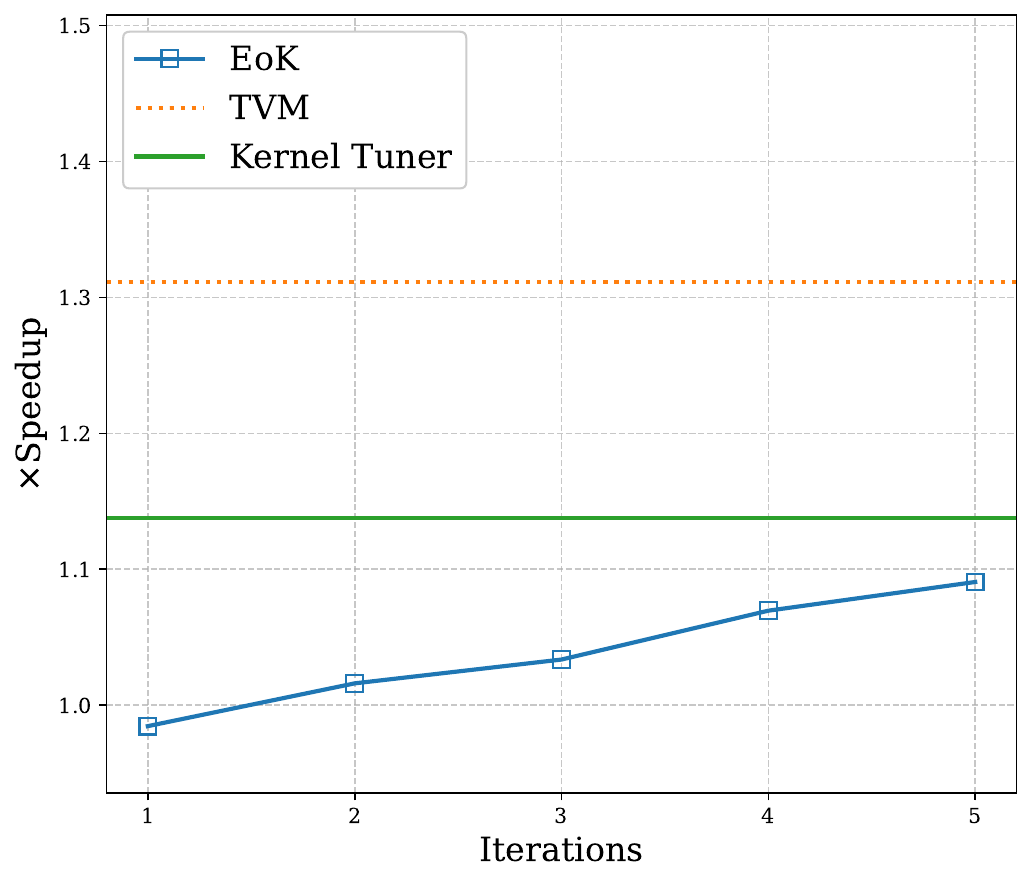}
         \caption*{Tanh}         
     \end{subfigure}\hfill
     \begin{subfigure}[b]{0.14\textwidth}
         \centering
         \includegraphics[width=.95\textwidth]{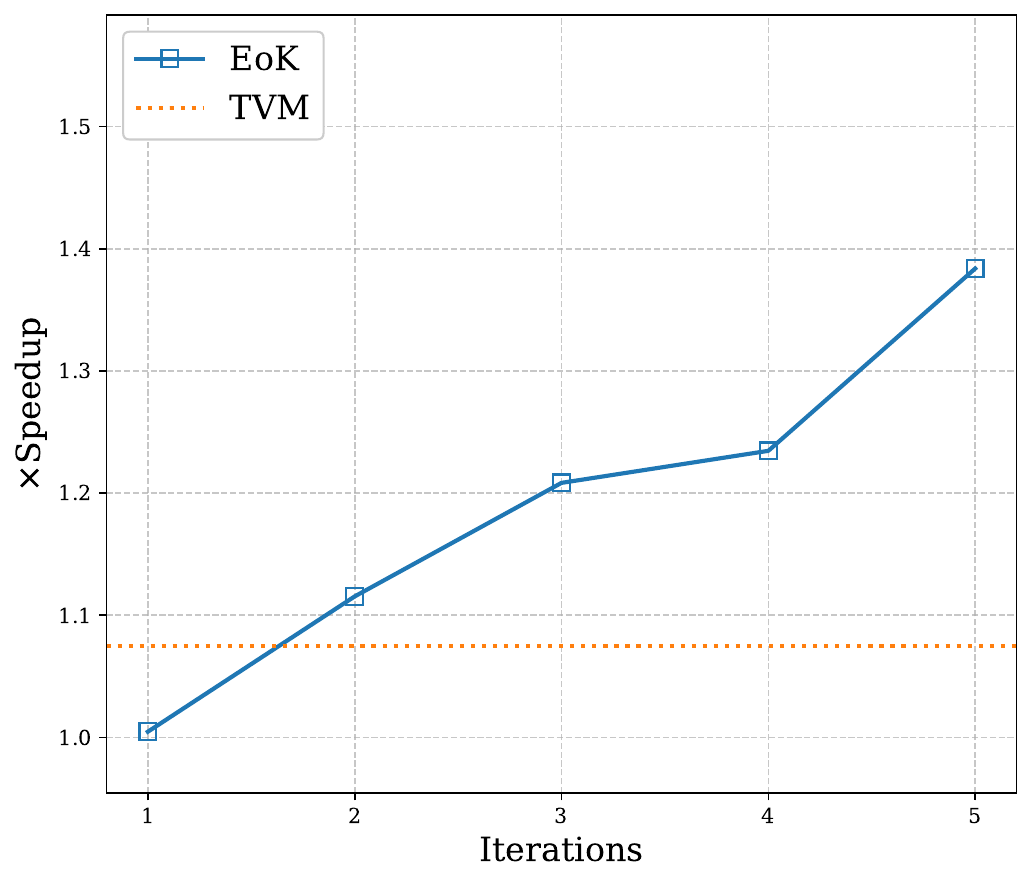}
         \caption*{Unary Op}         
     \end{subfigure}
     \caption{Plots of iteration versus speedup for \ourmethod{} across 28 Half-precision Neural Network Kernels with Zfh extension. Speedups from other baseline methods are shown for a subset of kernels where these methods achieved at least one successful implementation.}
     \label{appxfig:nn_zfh_kernel_convergence_individual}
\end{figure}

\begin{figure}[t]
    \captionsetup[subfigure]{justification=centering}
    \begin{subfigure}[b]{0.14\textwidth}
         \centering
         \includegraphics[width=.95\textwidth]{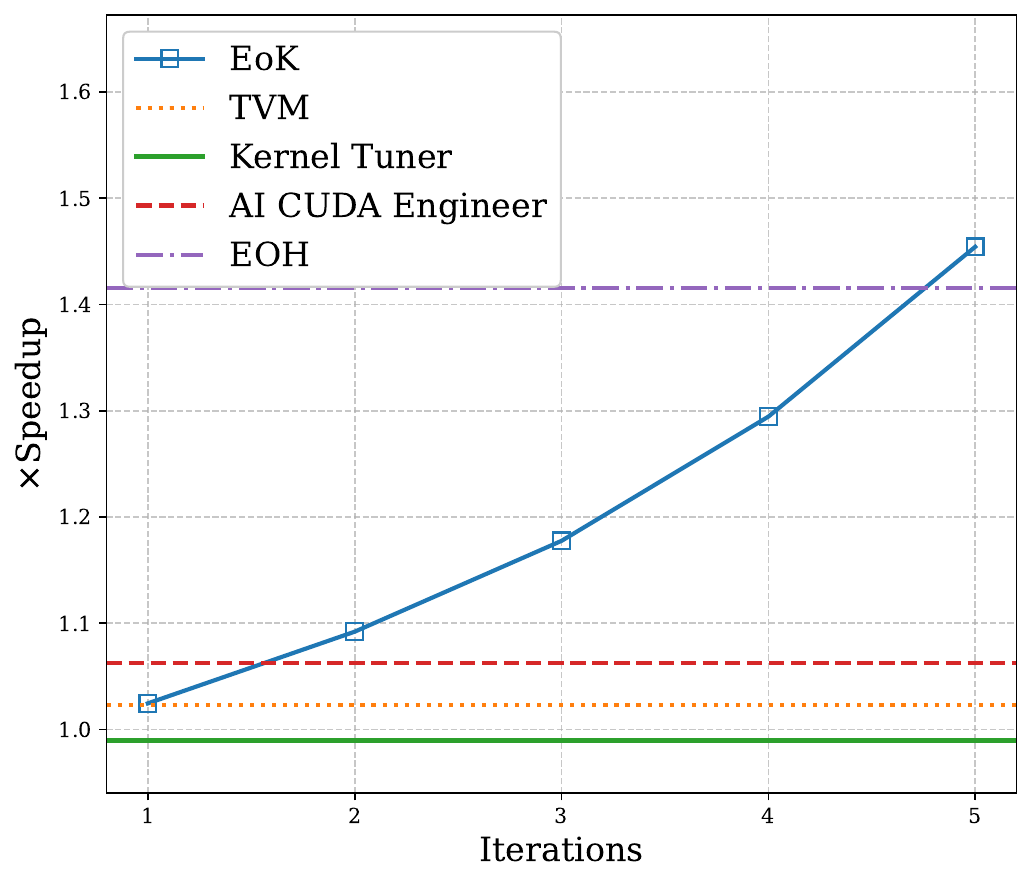}
         \caption*{Abs value}         
     \end{subfigure}\hfill
     \begin{subfigure}[b]{0.14\textwidth}
         \centering
         \includegraphics[width=.95\textwidth]{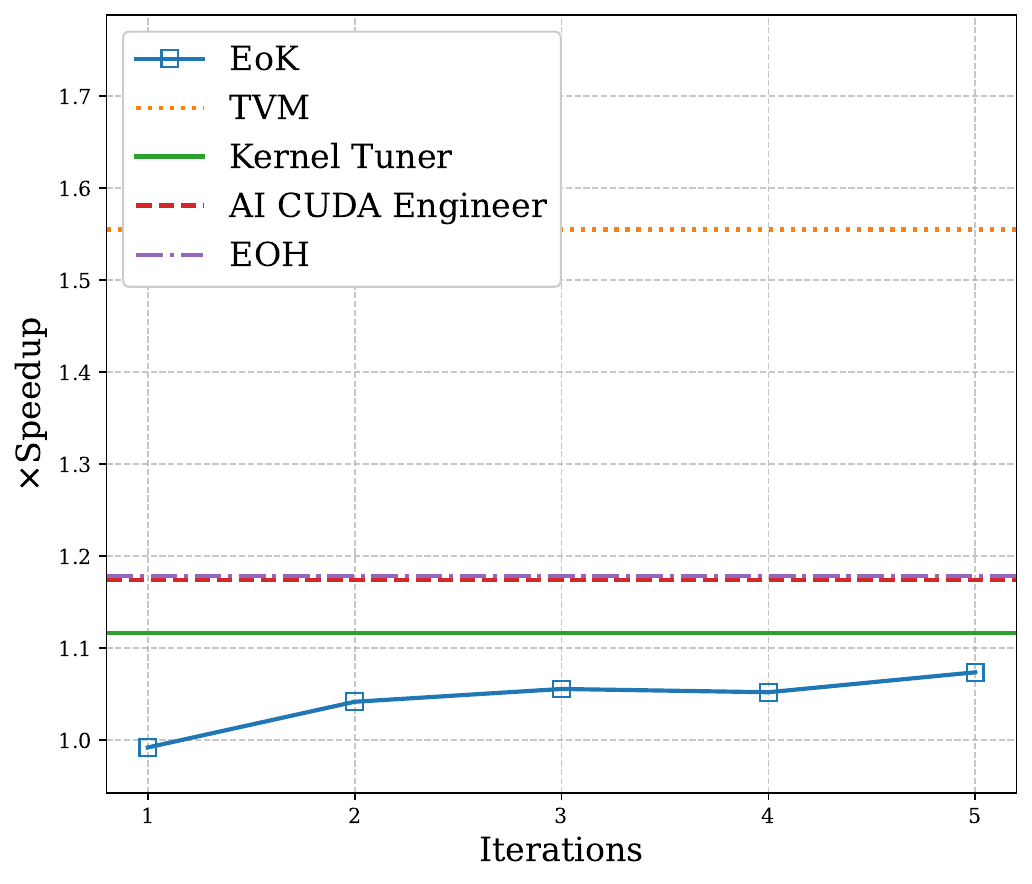}
         \caption*{BatchNorm}
     \end{subfigure}\hfill
     \begin{subfigure}[b]{0.14\textwidth}
         \centering
         \includegraphics[width=.95\textwidth]{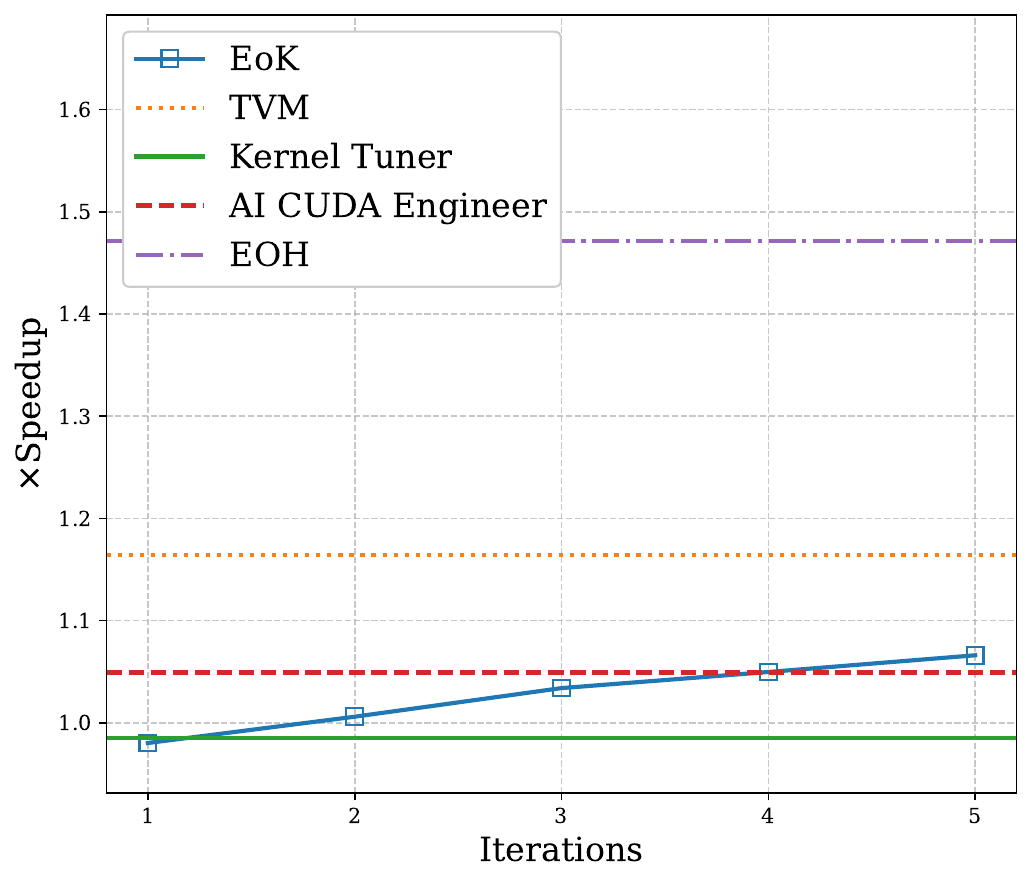}
         \caption*{Bias}  
     \end{subfigure}\hfill
     \begin{subfigure}[b]{0.14\textwidth}
         \centering
         \includegraphics[width=.95\textwidth]{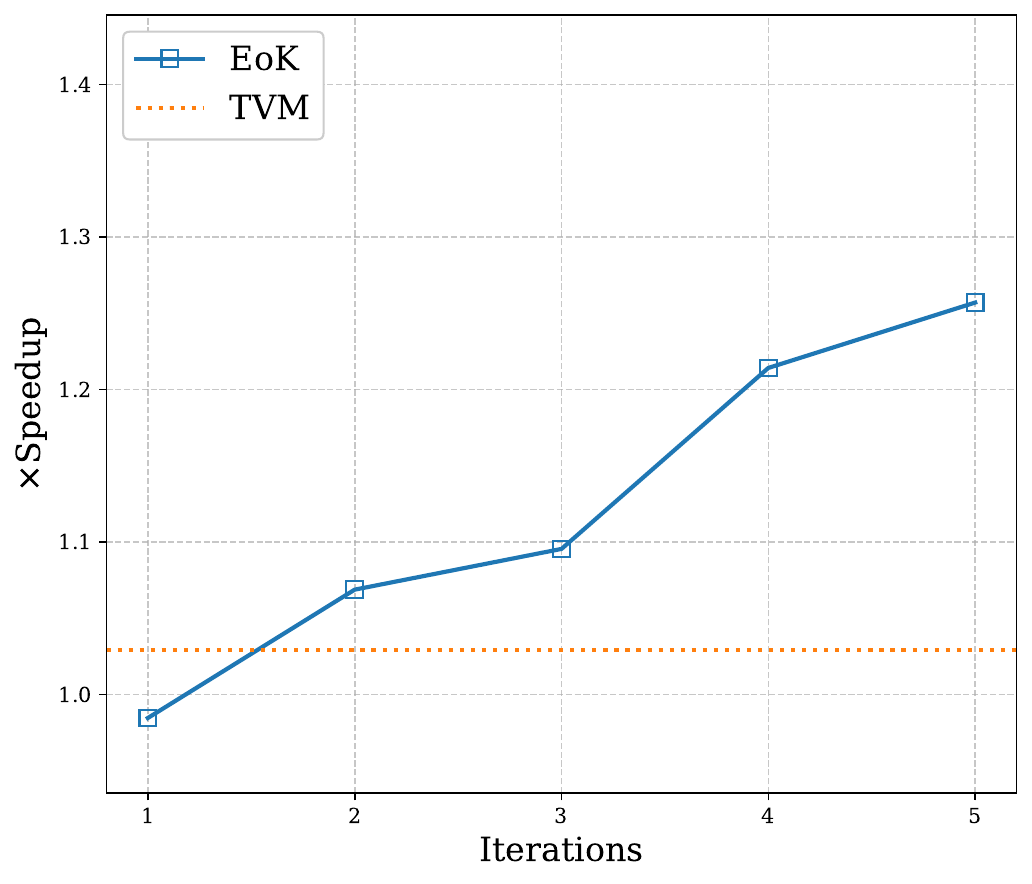}
         \caption*{Binary op}         
     \end{subfigure}\hfill
     \begin{subfigure}[b]{0.14\textwidth}
         \centering
         \includegraphics[width=.95\textwidth]{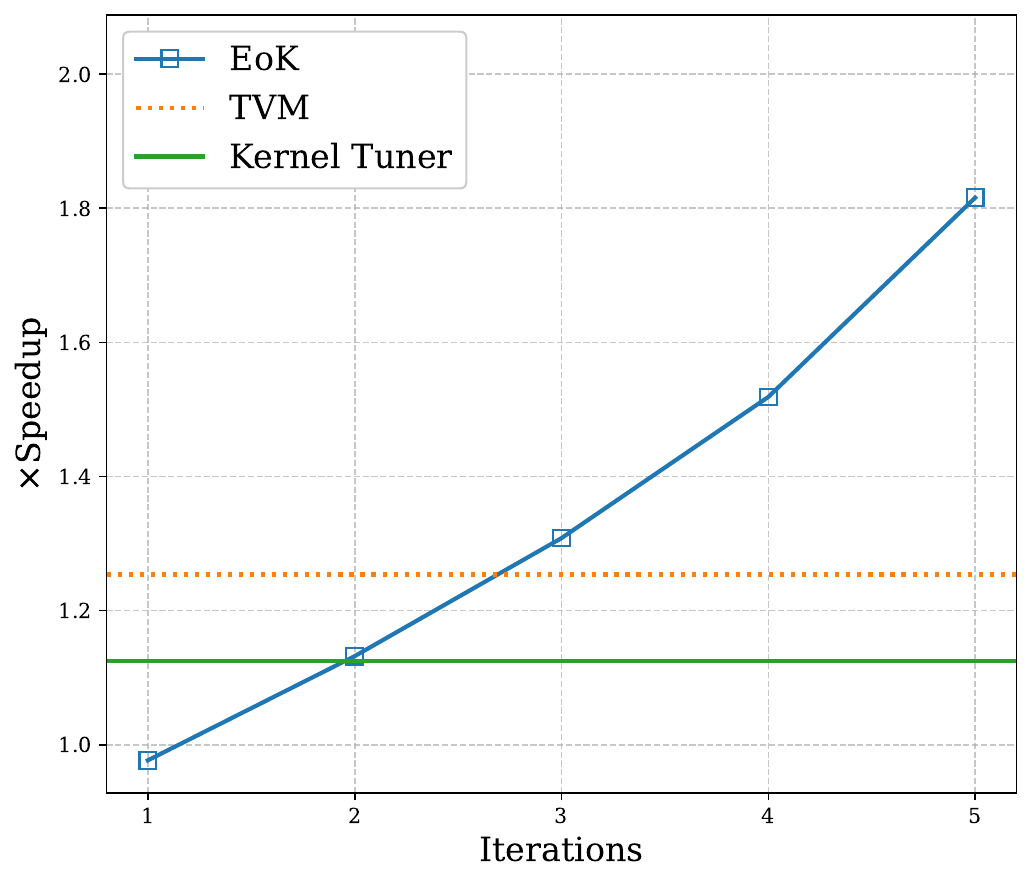}
         \caption*{BNLL}     
     \end{subfigure}\hfill
     \begin{subfigure}[b]{0.14\textwidth}
         \centering
         \includegraphics[width=.95\textwidth]{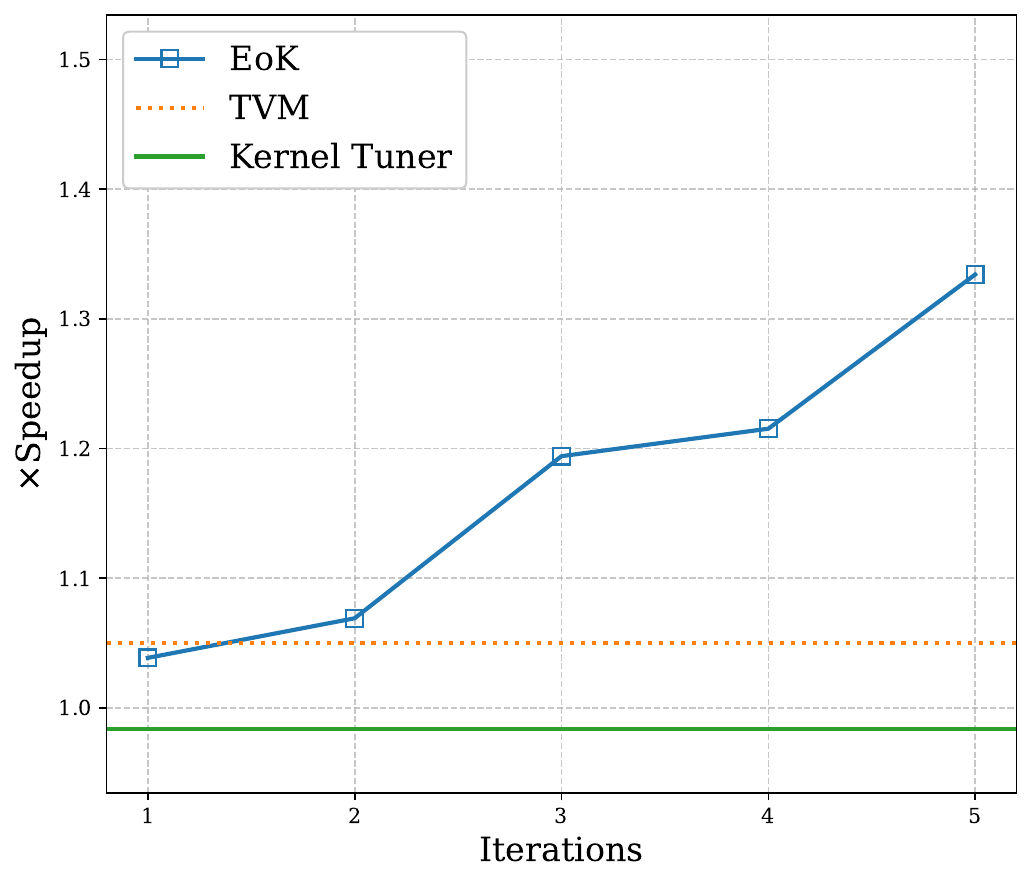}
         \caption*{Cast}        
     \end{subfigure}\hfill
     \begin{subfigure}[b]{0.14\textwidth}
         \centering
         \includegraphics[width=.95\textwidth]{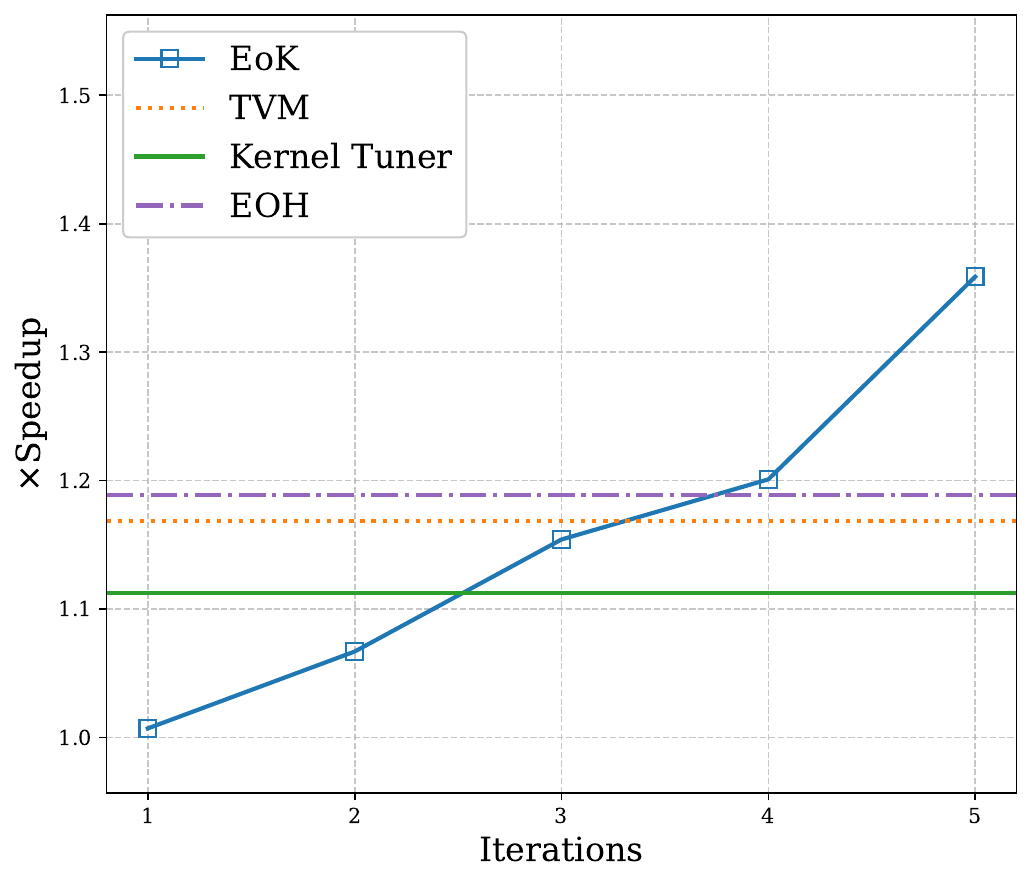}
         \caption*{CELU}         
     \end{subfigure}\\
     \begin{subfigure}[b]{0.14\textwidth}
         \centering
         \includegraphics[width=.95\textwidth]{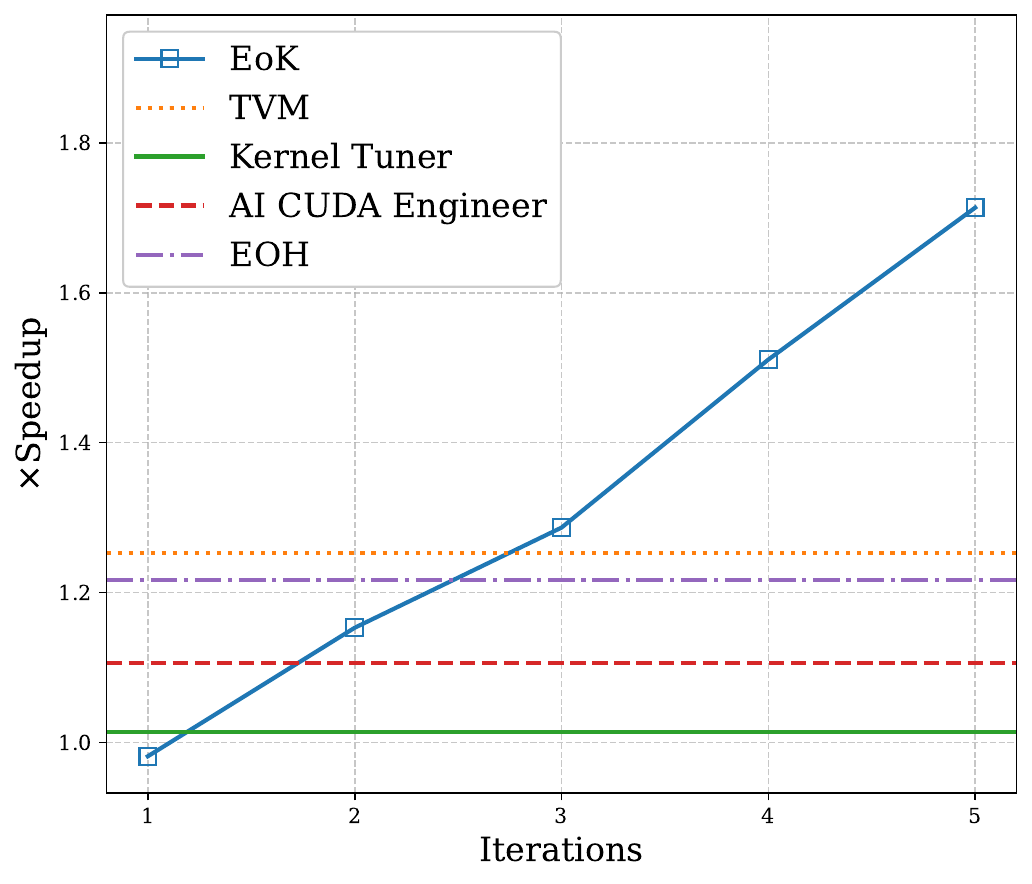}
         \caption*{Clip}         
     \end{subfigure}\hfill
     \begin{subfigure}[b]{0.14\textwidth}
         \centering
         \includegraphics[width=.95\textwidth]{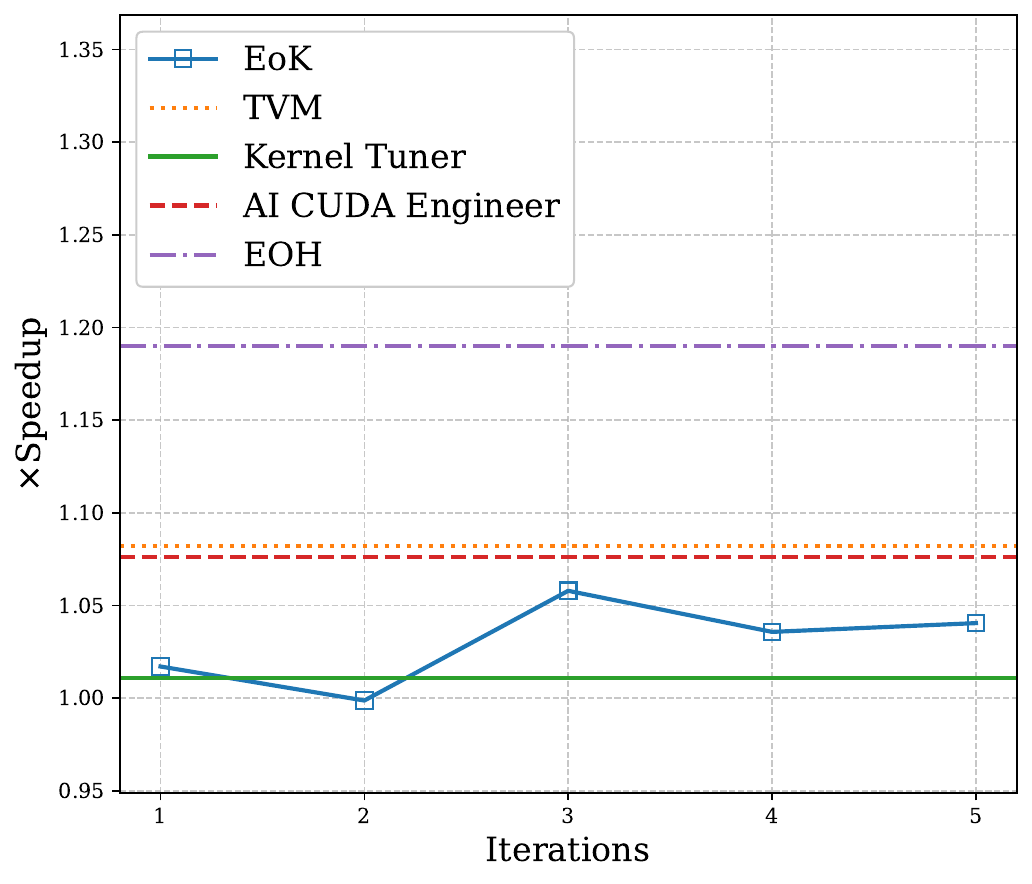}
         \caption*{Concat}        
     \end{subfigure}\hfill
     \begin{subfigure}[b]{0.14\textwidth}
         \centering
         \includegraphics[width=.95\textwidth]{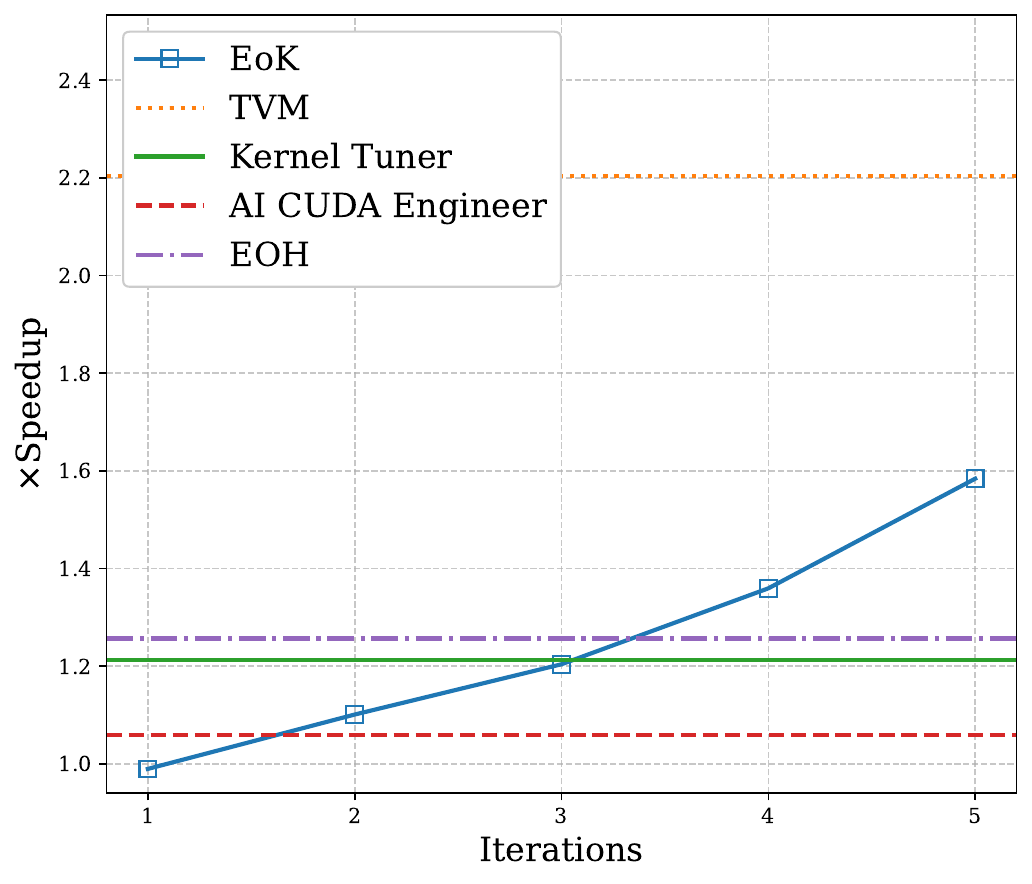}
         \caption*{Conv}         
     \end{subfigure}\hfill
     \begin{subfigure}[b]{0.14\textwidth}
         \centering
         \includegraphics[width=.95\textwidth]{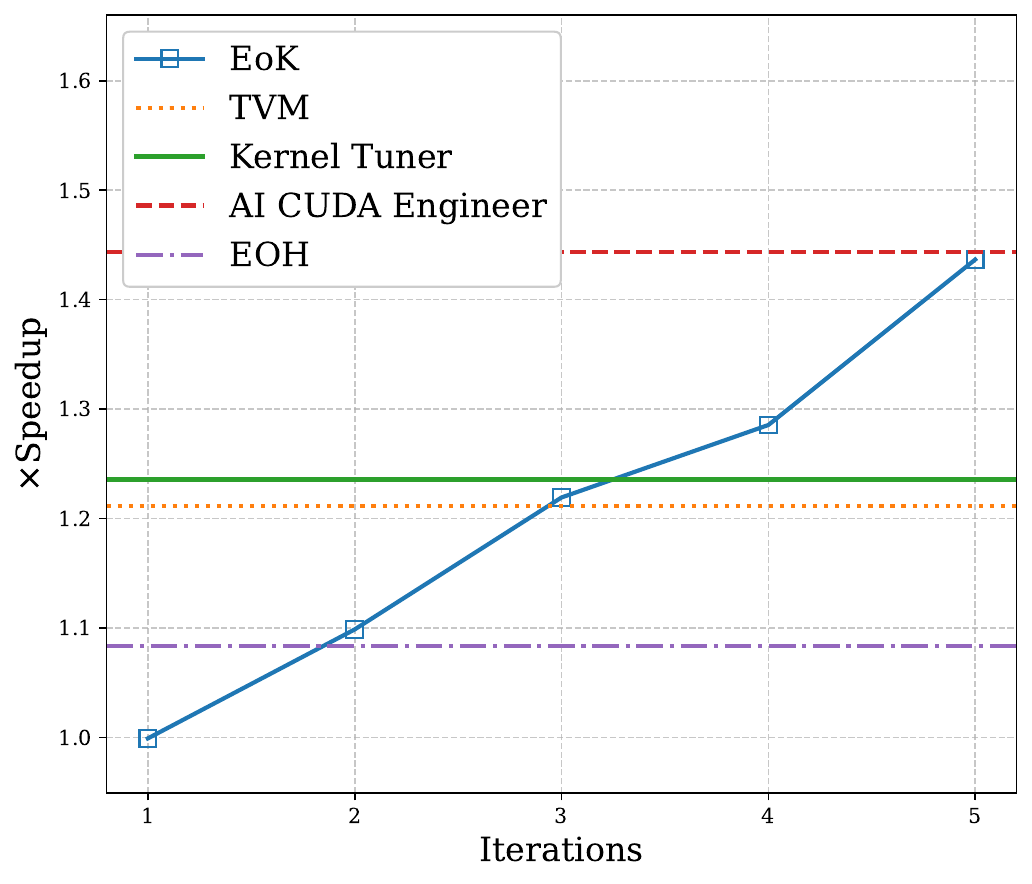}
         \caption*{Conv 1D}         
     \end{subfigure}\hfill
     \begin{subfigure}[b]{0.14\textwidth}
         \centering
         \includegraphics[width=.95\textwidth]{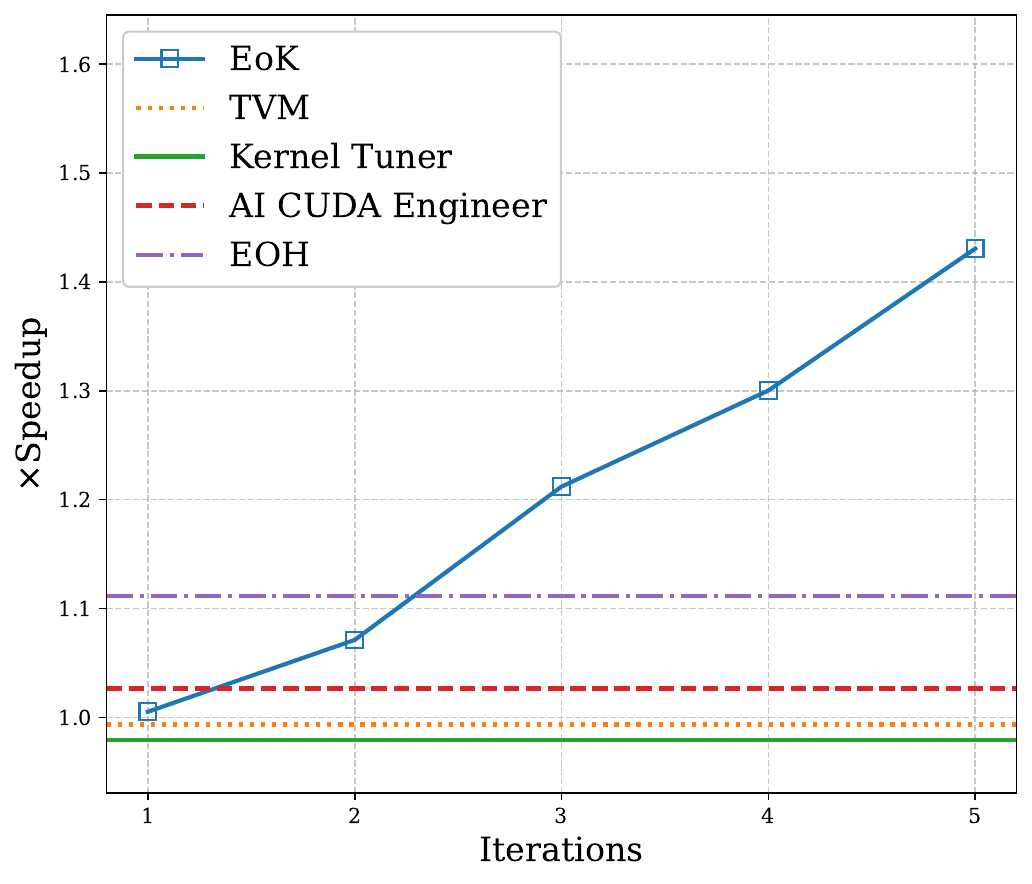}
         \caption*{Crop}         
     \end{subfigure}\hfill
     \begin{subfigure}[b]{0.14\textwidth}
         \centering
         \includegraphics[width=.95\textwidth]{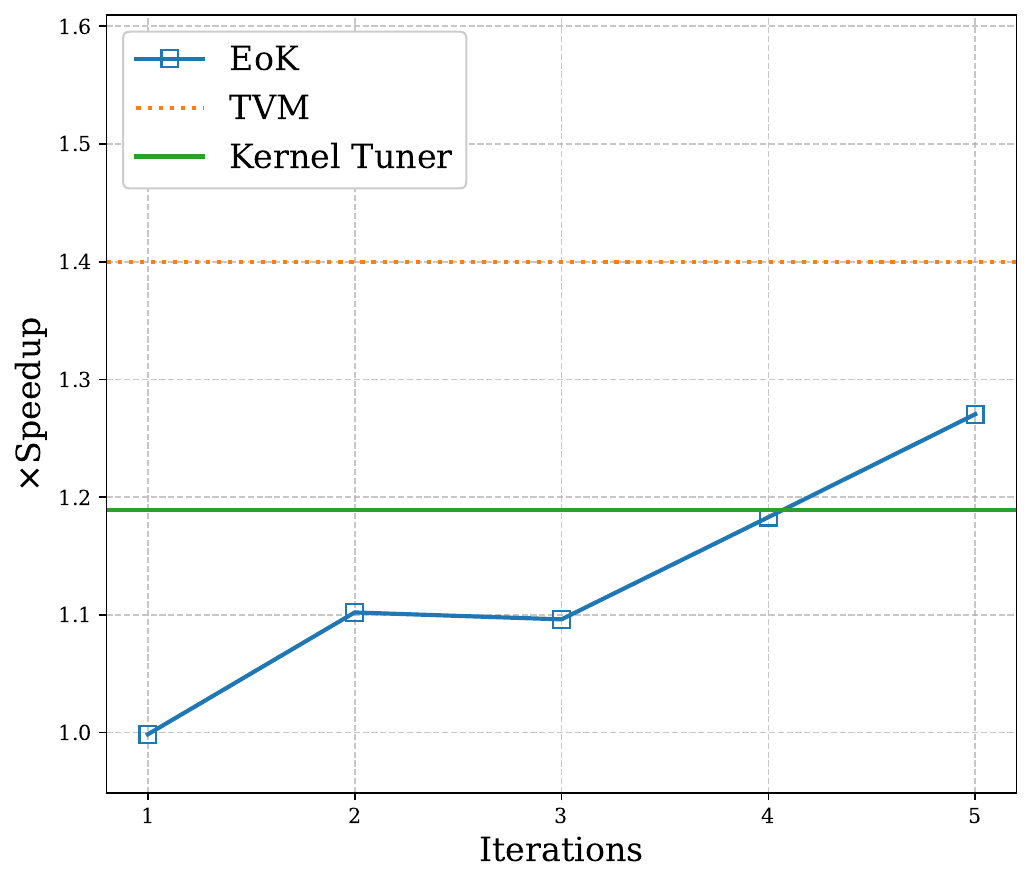}
         \caption*{DeConv}         
     \end{subfigure}\hfill
     \begin{subfigure}[b]{0.14\textwidth}
         \centering
         \includegraphics[width=.95\textwidth]{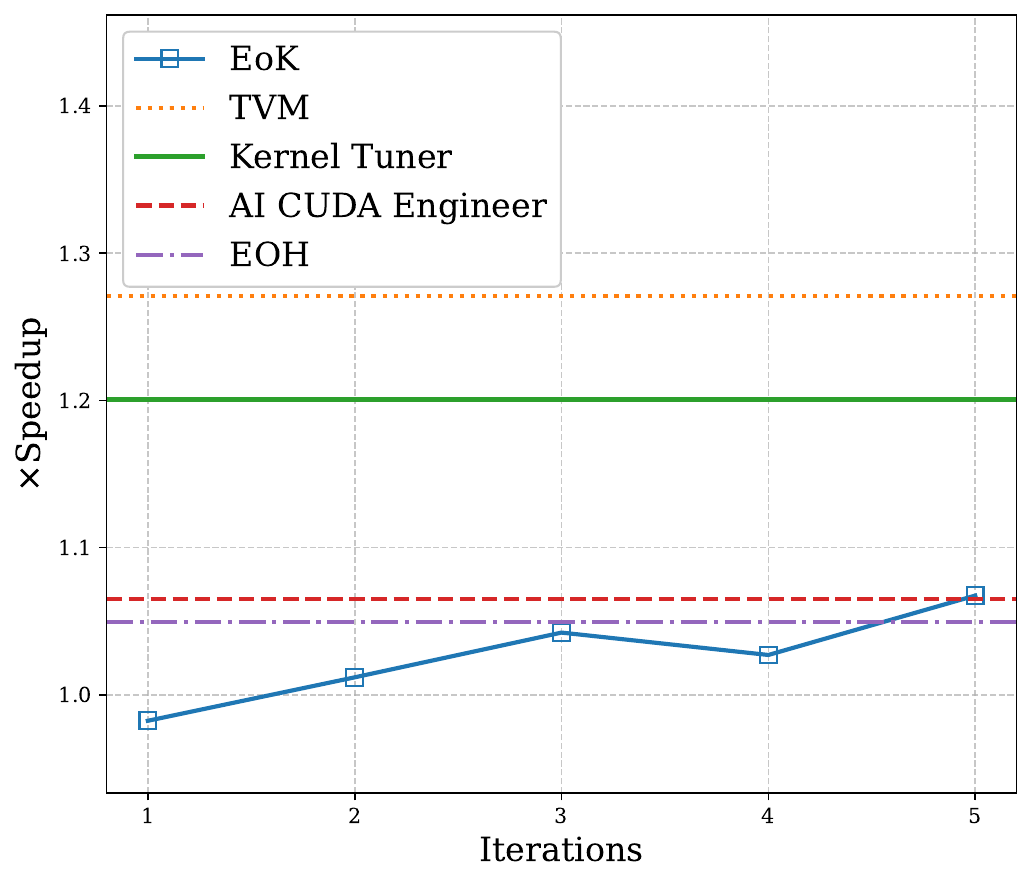}
         \caption*{Dep Conv}       
     \end{subfigure}\\
     \begin{subfigure}[b]{0.14\textwidth}
         \centering
         \includegraphics[width=.95\textwidth]{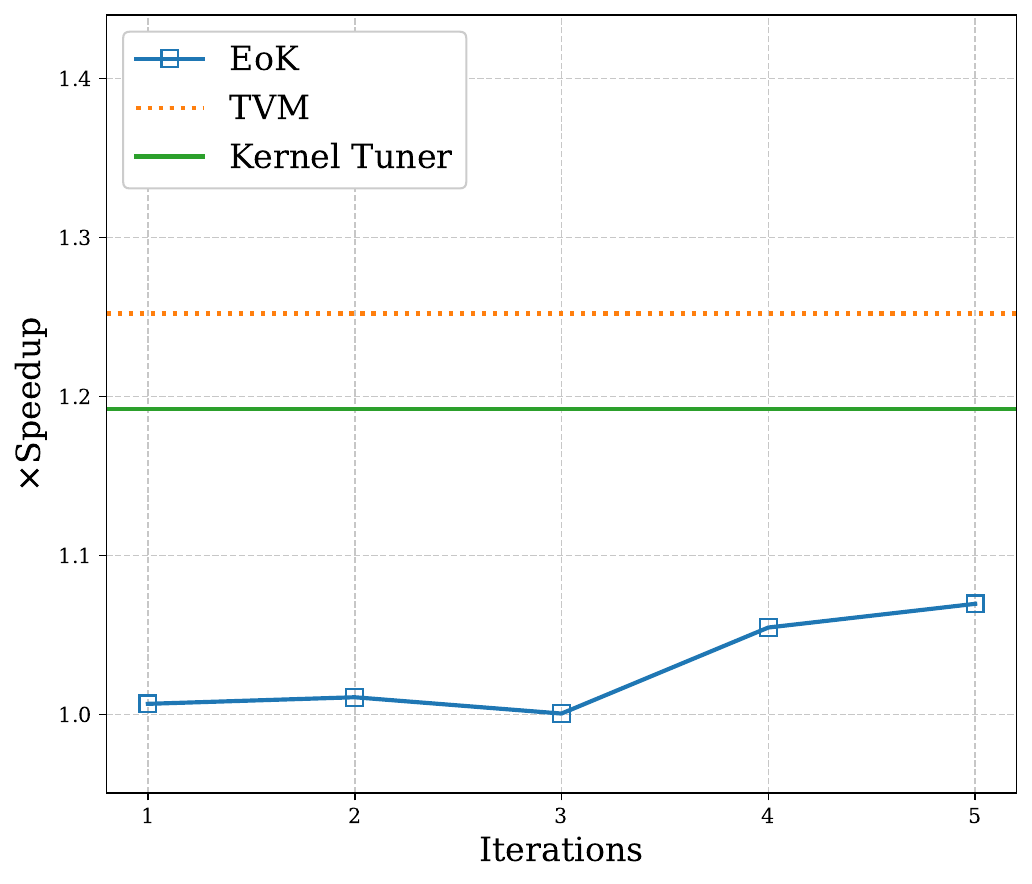}
         \caption*{Dep DeConv}         
     \end{subfigure}\hfill
     \begin{subfigure}[b]{0.14\textwidth}
         \centering
         \includegraphics[width=.95\textwidth]{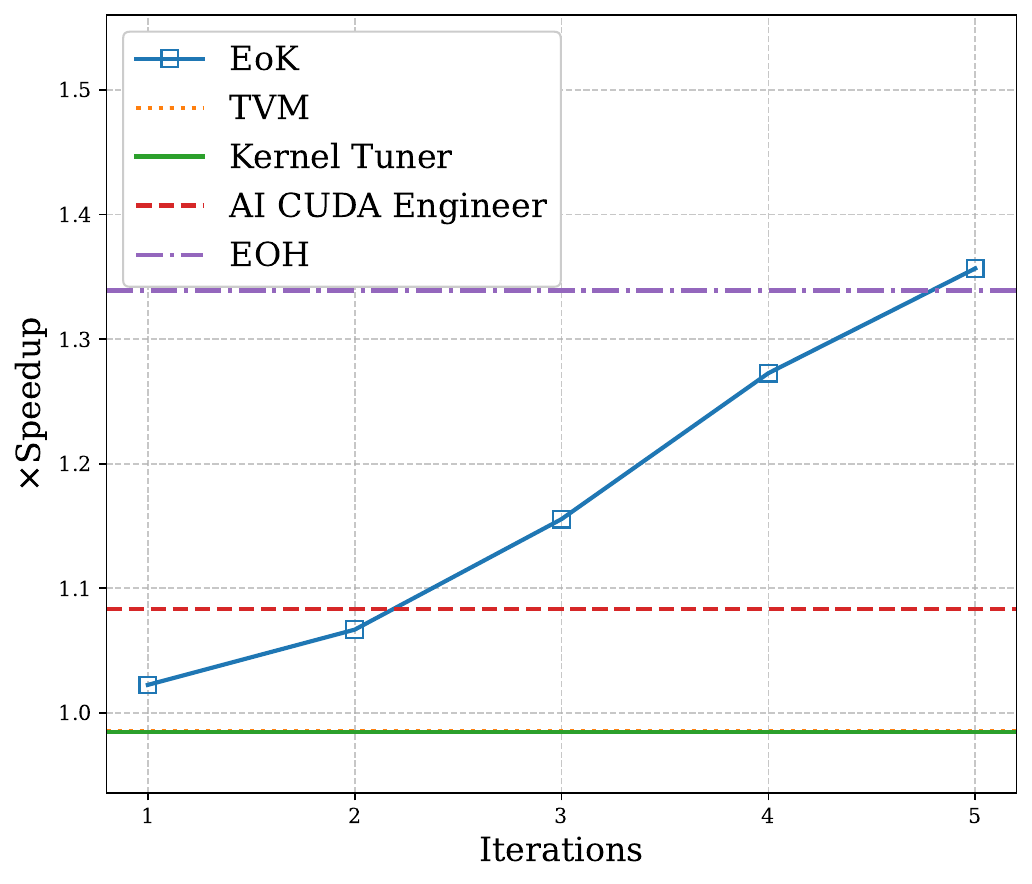}
         \caption*{Dropout}         
     \end{subfigure}\hfill
     \begin{subfigure}[b]{0.14\textwidth}
         \centering
         \includegraphics[width=.95\textwidth]{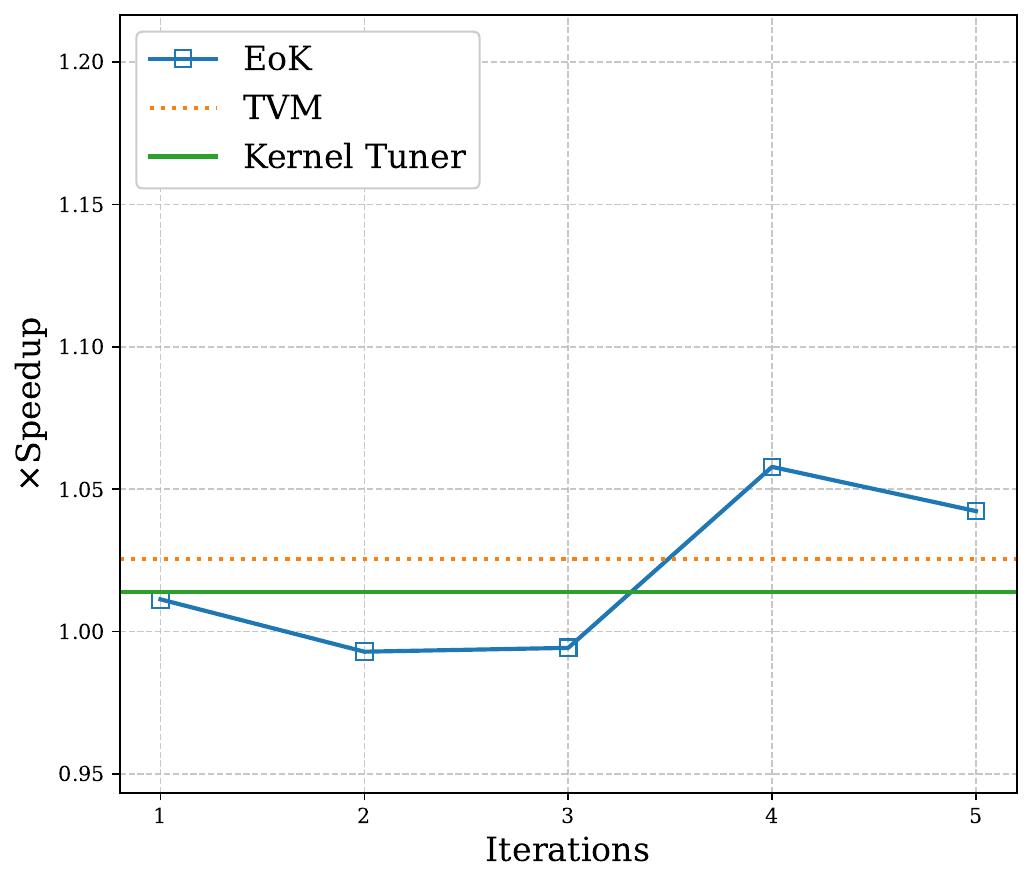}
         \caption*{Eltwise}       
     \end{subfigure}\hfill
     \begin{subfigure}[b]{0.14\textwidth}
         \centering
         \includegraphics[width=.95\textwidth]{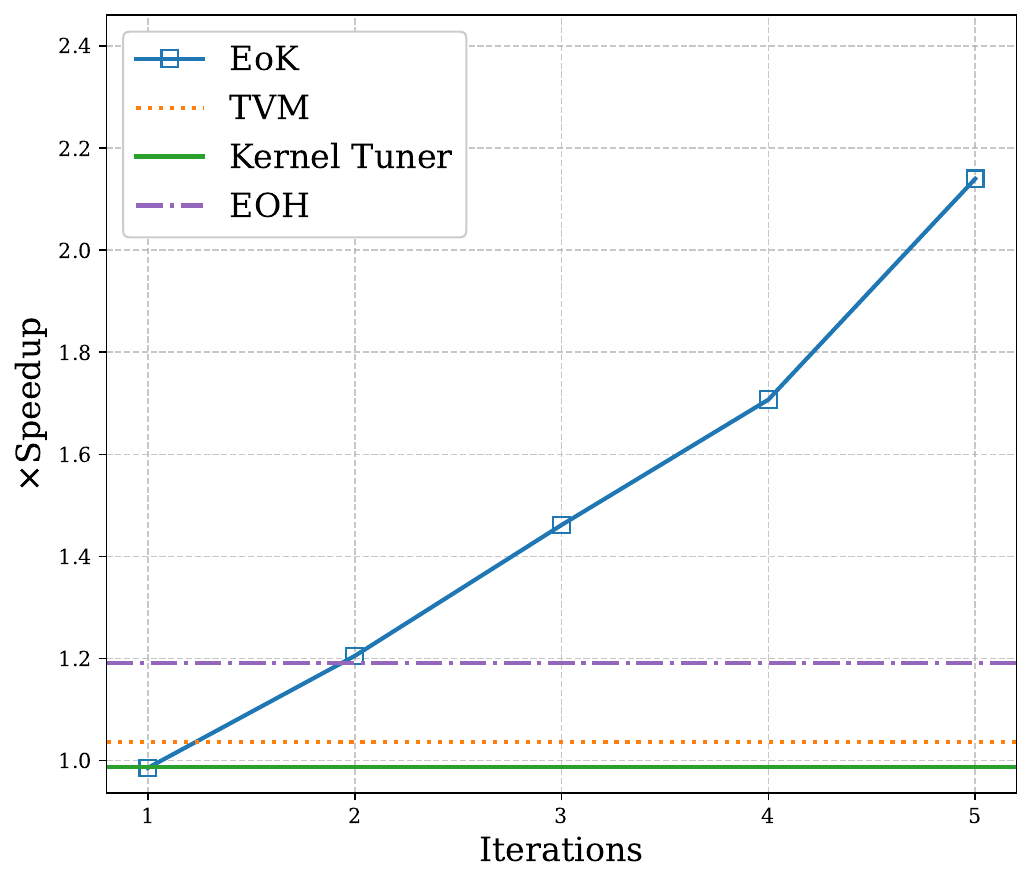}
         \caption*{Flatten}         
     \end{subfigure}\hfill
     \begin{subfigure}[b]{0.14\textwidth}
         \centering
         \includegraphics[width=.95\textwidth]{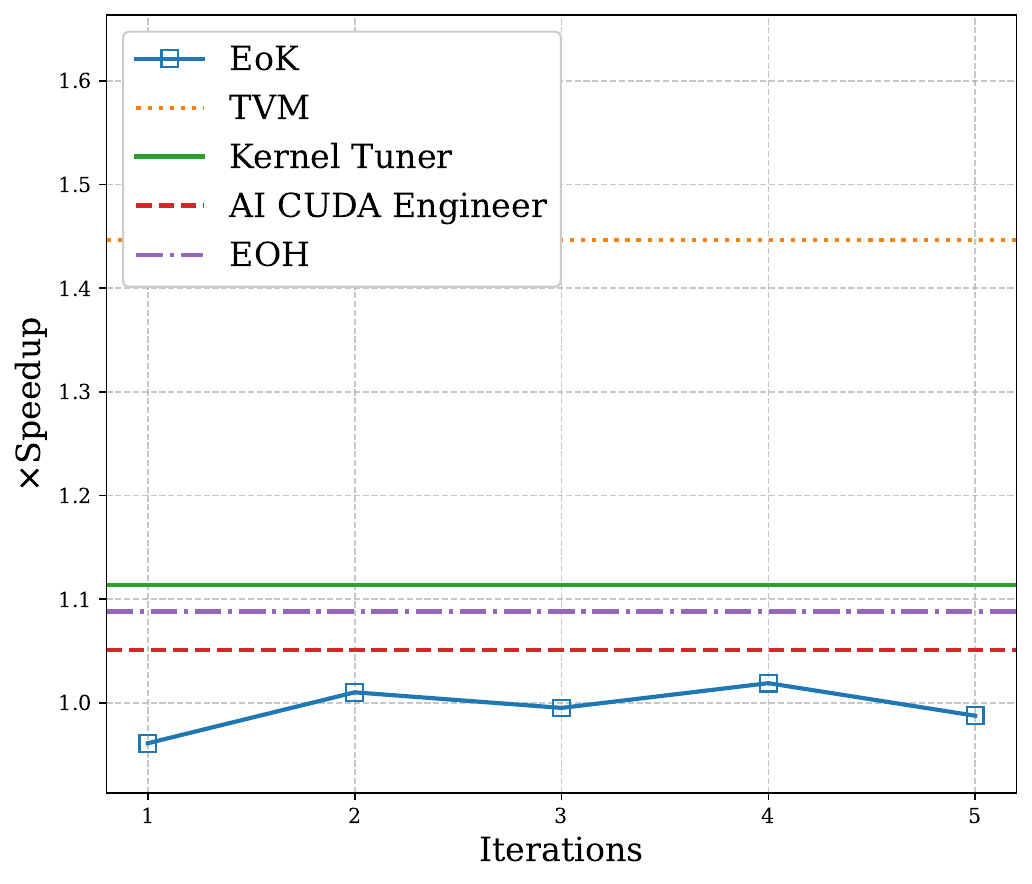}
         \caption*{GELU}         
     \end{subfigure}\hfill
     \begin{subfigure}[b]{0.14\textwidth}
         \centering
         \includegraphics[width=.95\textwidth]{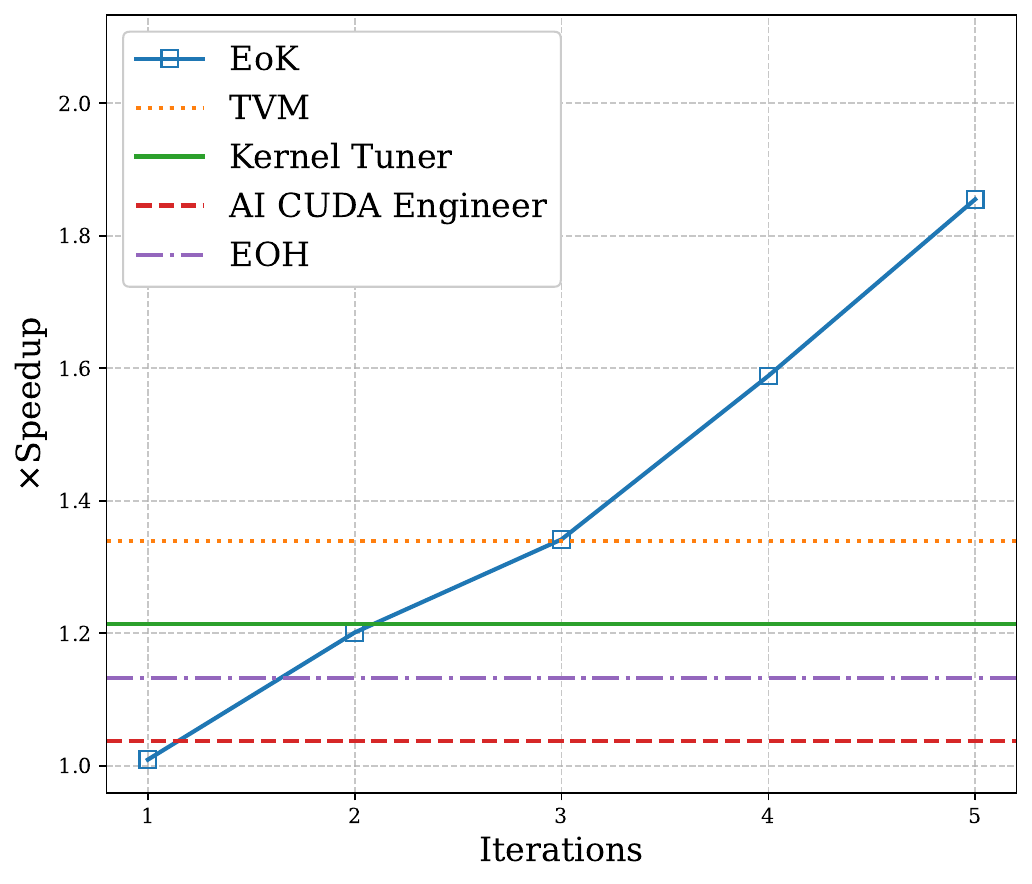}
         \caption*{GEMM}         
     \end{subfigure}\hfill
     \begin{subfigure}[b]{0.14\textwidth}
         \centering
         \includegraphics[width=.95\textwidth]{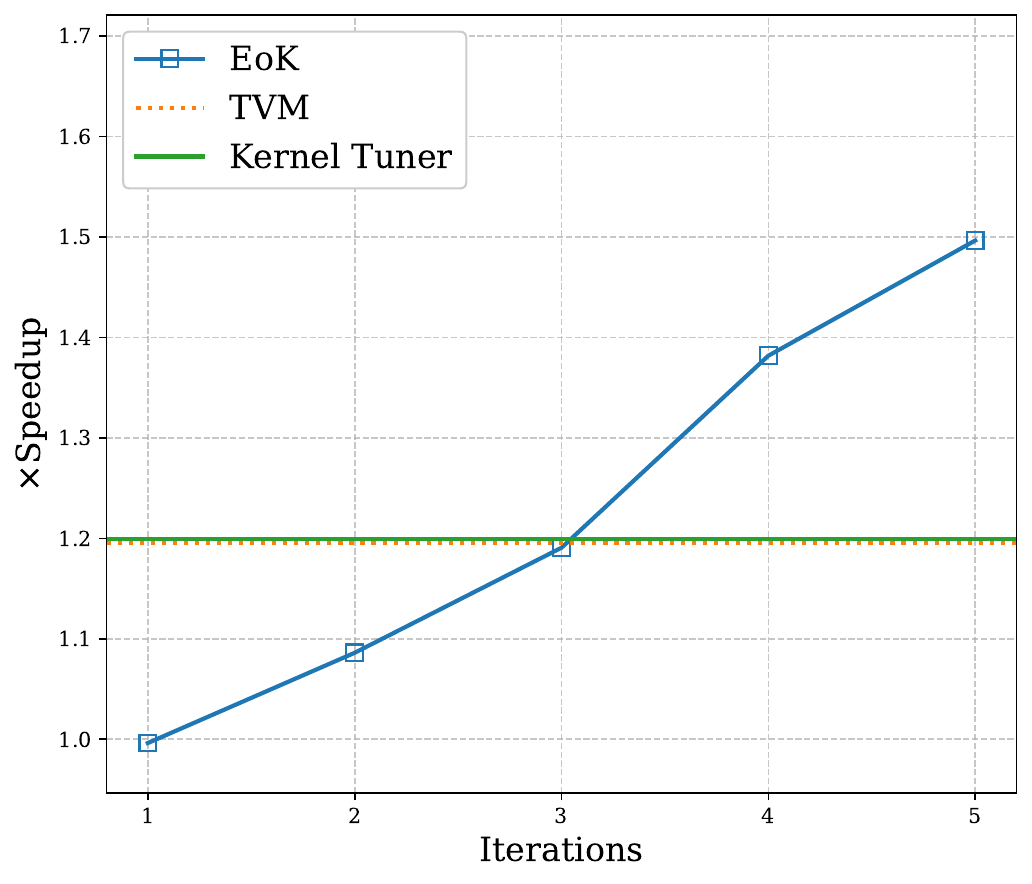}
         \caption*{GRU}        
     \end{subfigure}\\
     \begin{subfigure}[b]{0.14\textwidth}
         \centering
         \includegraphics[width=.95\textwidth]{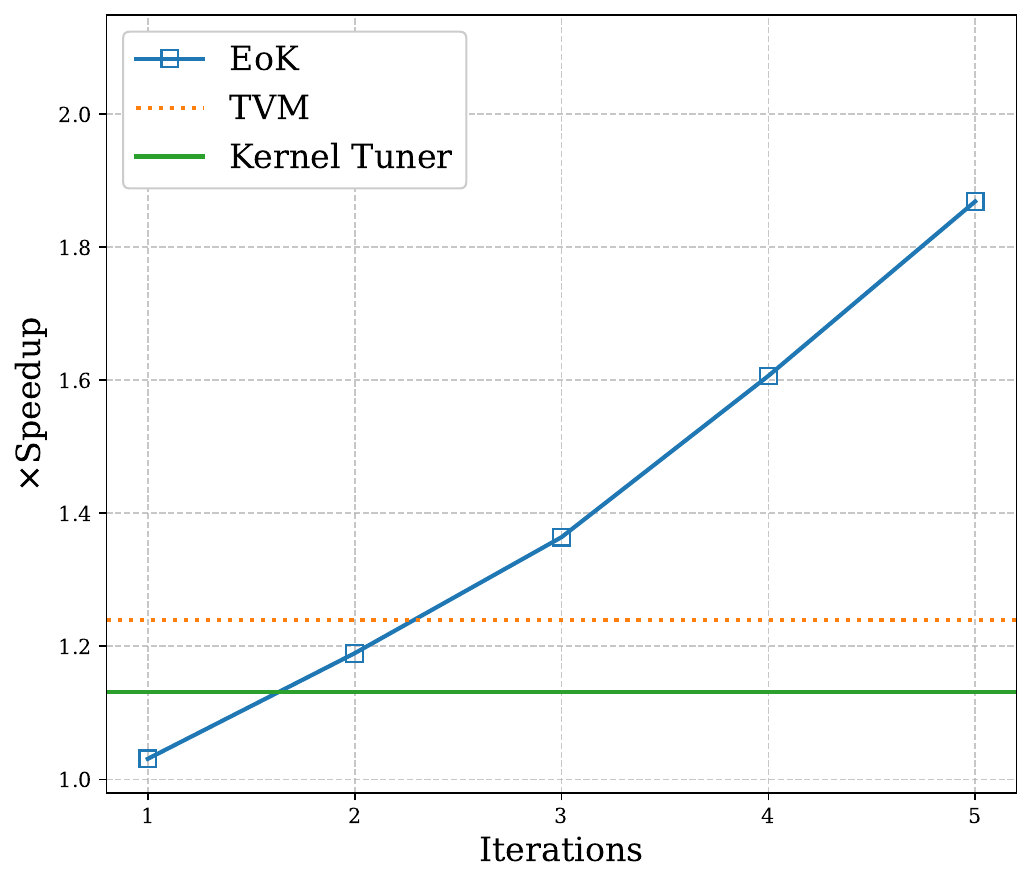}
         \caption*{Hard Sigmoid}         
     \end{subfigure}\hfill
     \begin{subfigure}[b]{0.14\textwidth}
         \centering
         \includegraphics[width=.95\textwidth]{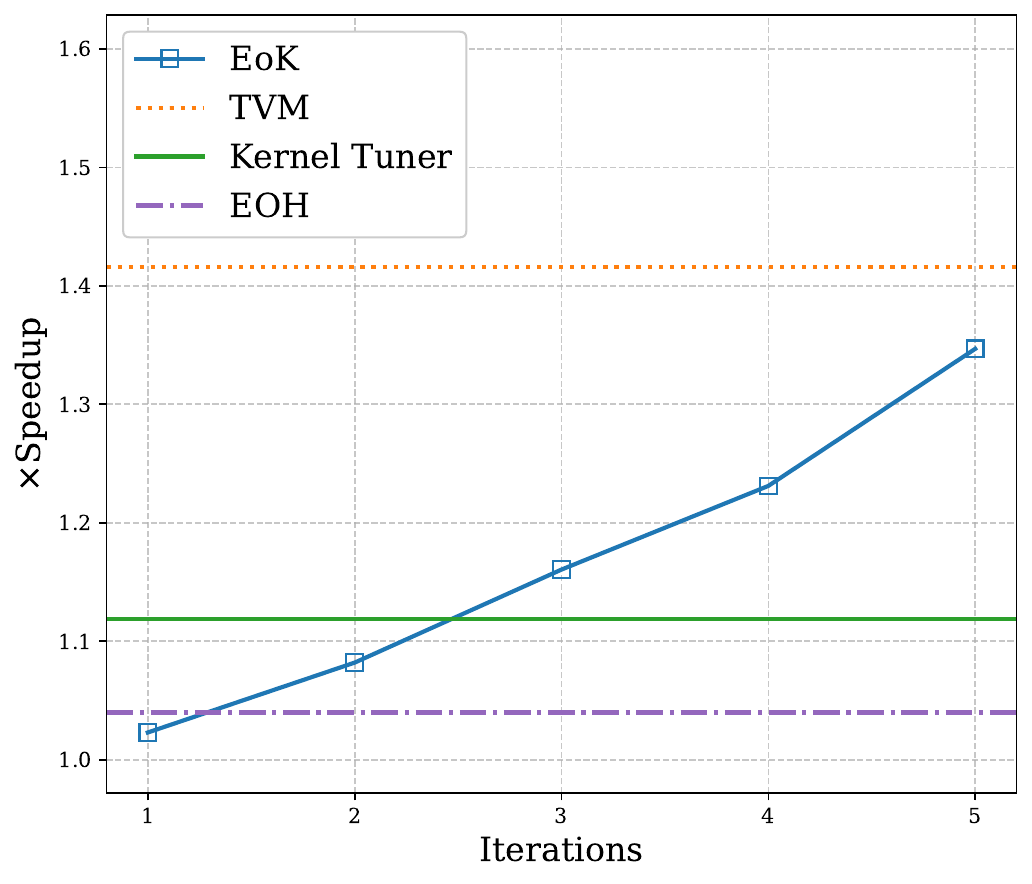}
         \caption*{Hard Swish}        
     \end{subfigure}\hfill
     \begin{subfigure}[b]{0.14\textwidth}
         \centering
         \includegraphics[width=.95\textwidth]{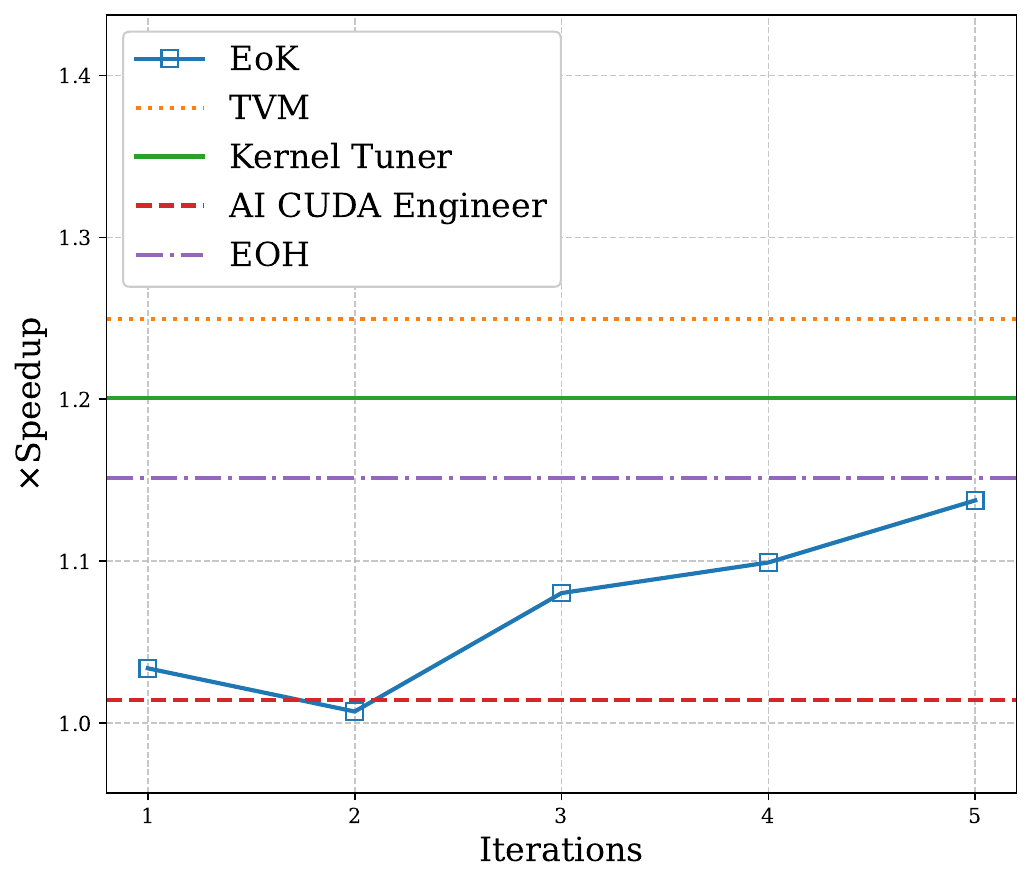}
         \caption*{Inner Product}         
     \end{subfigure}\hfill
     \begin{subfigure}[b]{0.14\textwidth}
         \centering
         \includegraphics[width=.95\textwidth]{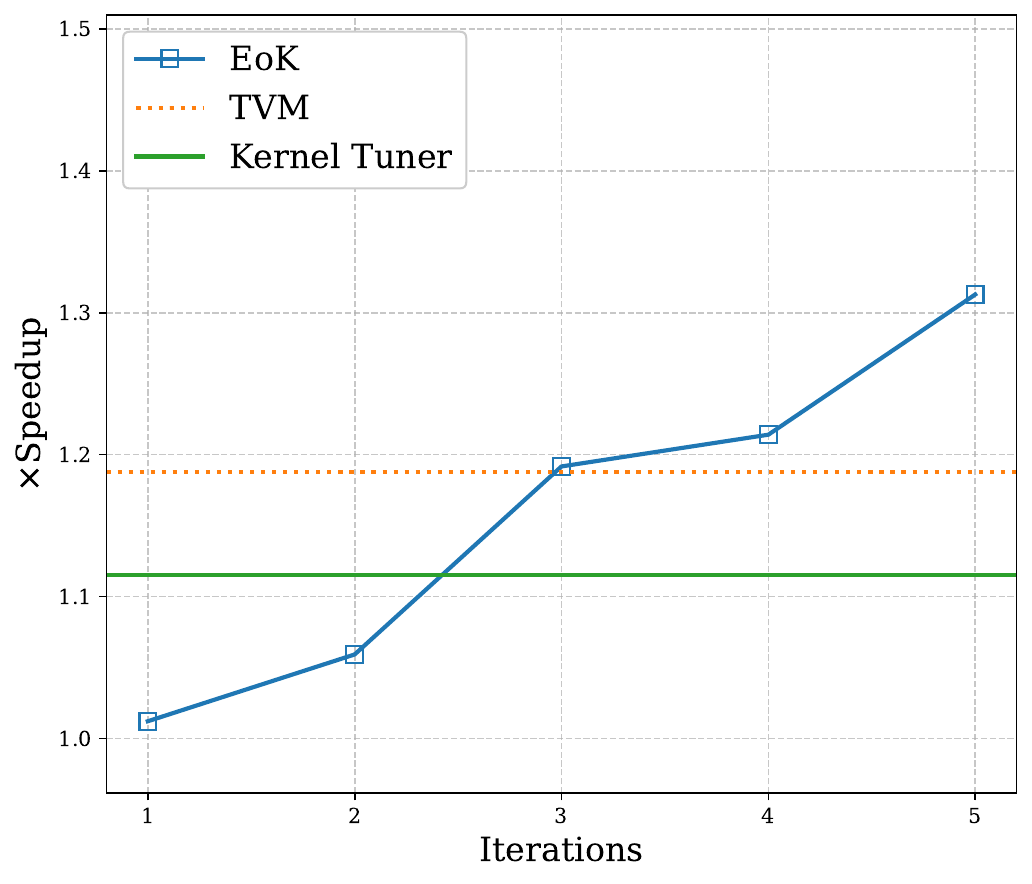}
         \caption*{InstanceNorm}         
     \end{subfigure}\hfill
     \begin{subfigure}[b]{0.14\textwidth}
         \centering
         \includegraphics[width=.95\textwidth]{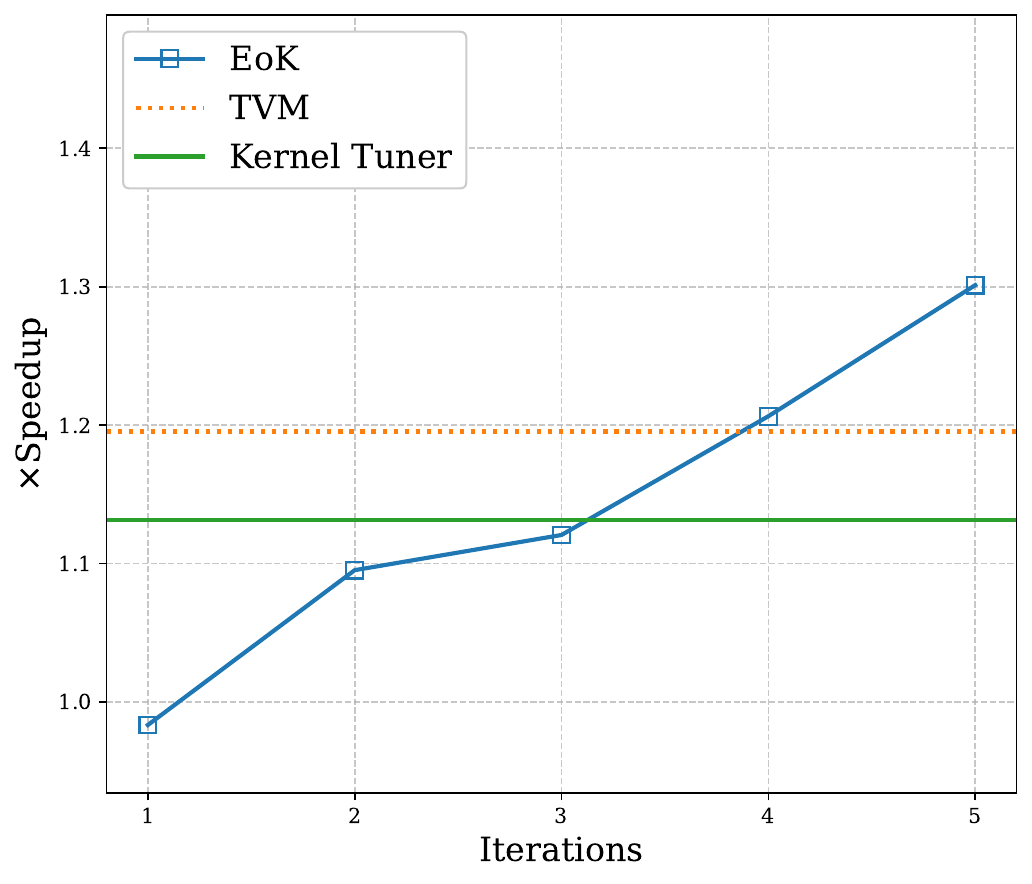}
         \caption*{Interp}        
     \end{subfigure}\hfill
     \begin{subfigure}[b]{0.14\textwidth}
         \centering
         \includegraphics[width=.95\textwidth]{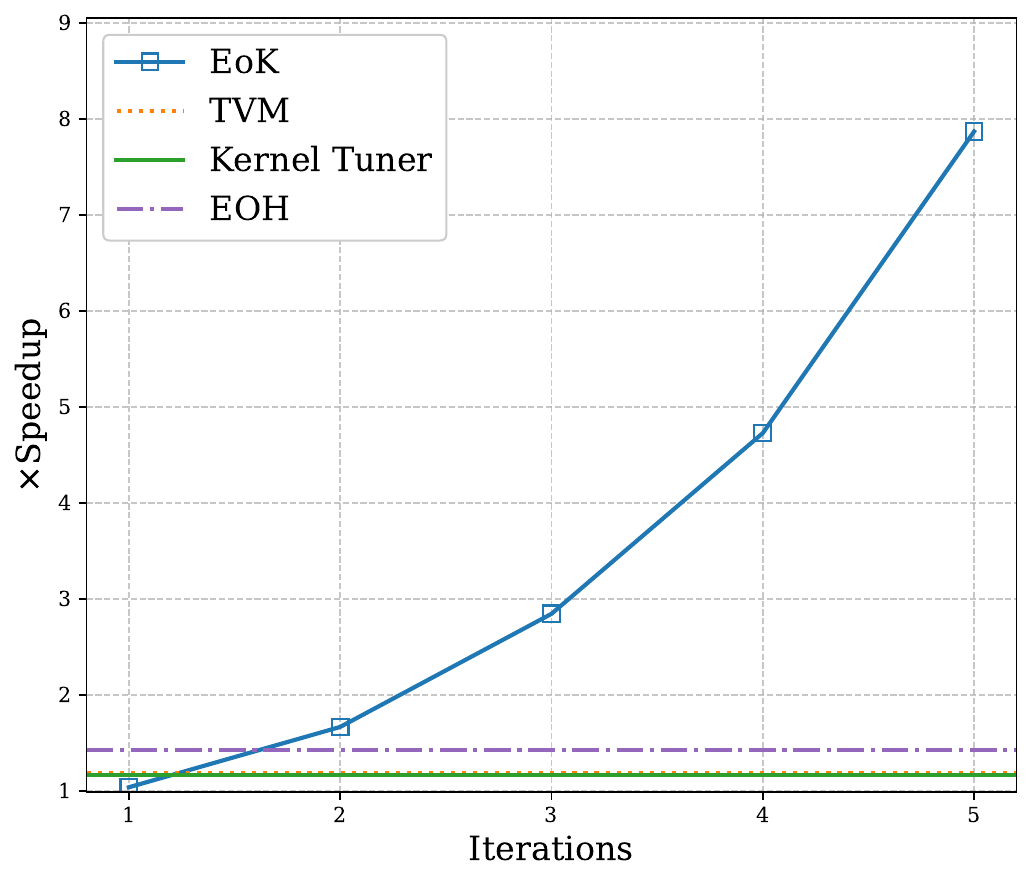}
         \caption*{Mish}         
     \end{subfigure}\hfill
     \begin{subfigure}[b]{0.14\textwidth}
         \centering
         \includegraphics[width=.95\textwidth]{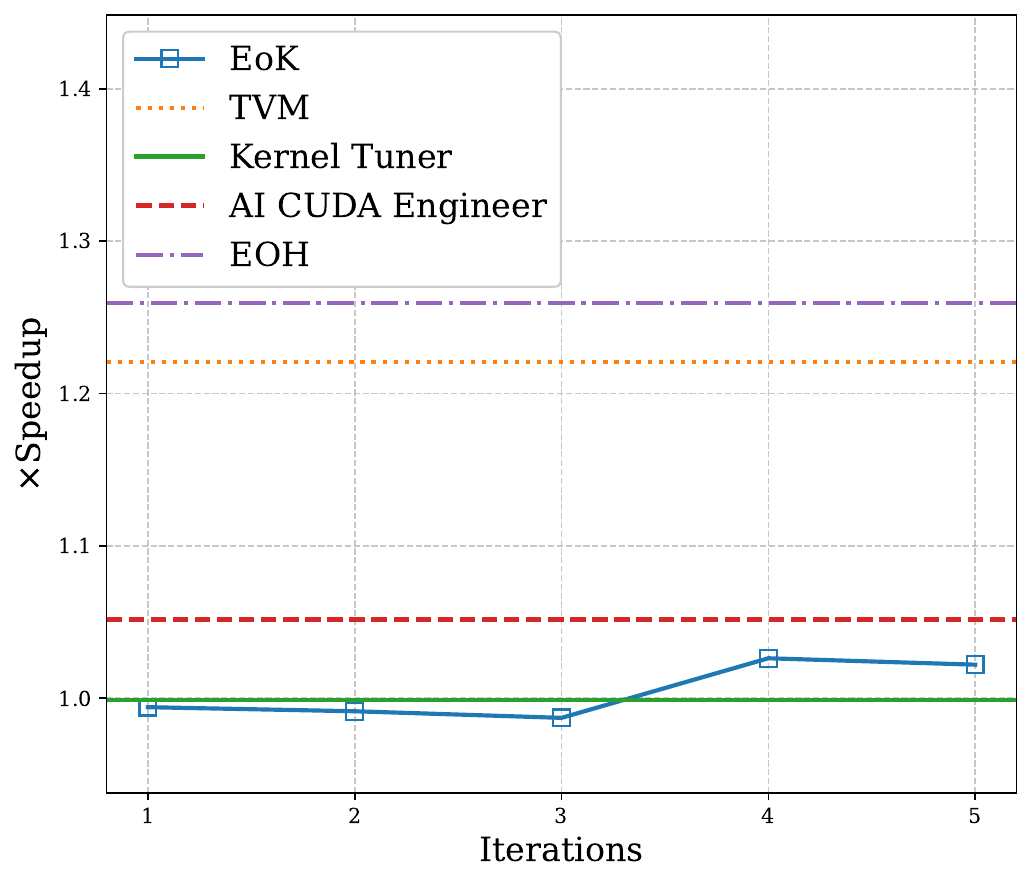}
         \caption*{Packing}         
     \end{subfigure}\\
     \begin{subfigure}[b]{0.14\textwidth}
         \centering
         \includegraphics[width=.95\textwidth]{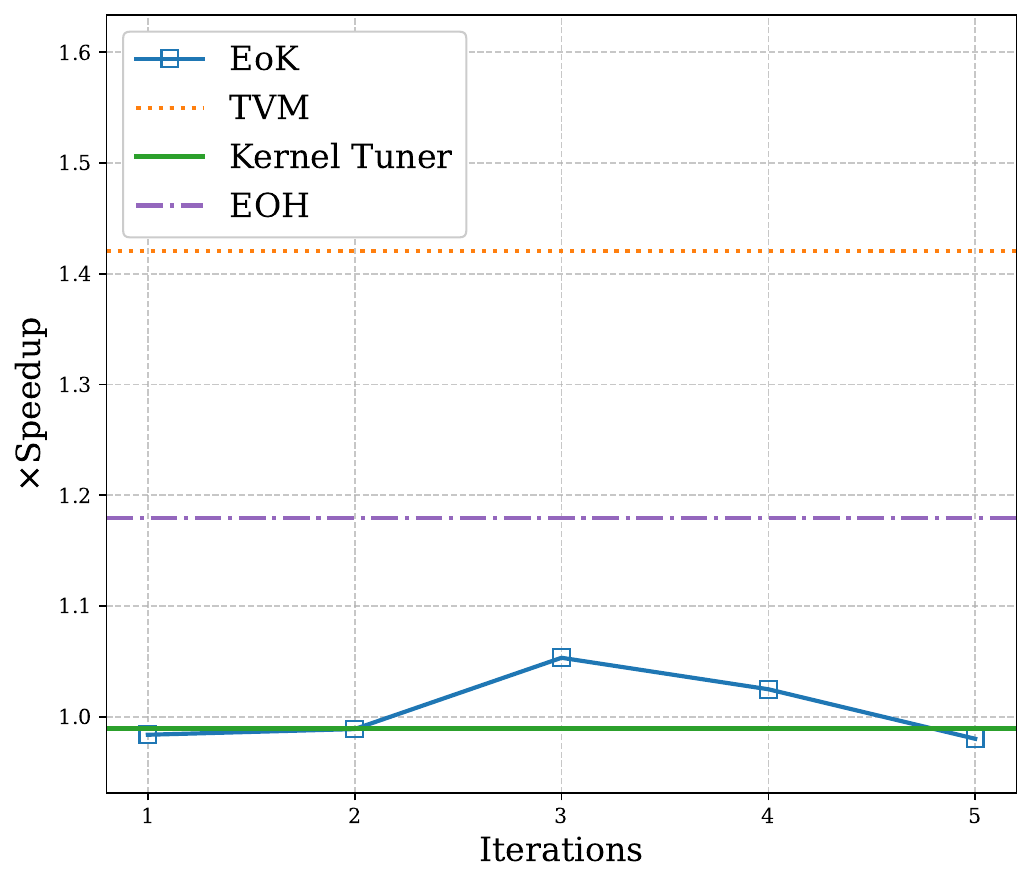}
         \caption*{Padding}         
     \end{subfigure}\hfill
     \begin{subfigure}[b]{0.14\textwidth}
         \centering
         \includegraphics[width=.95\textwidth]{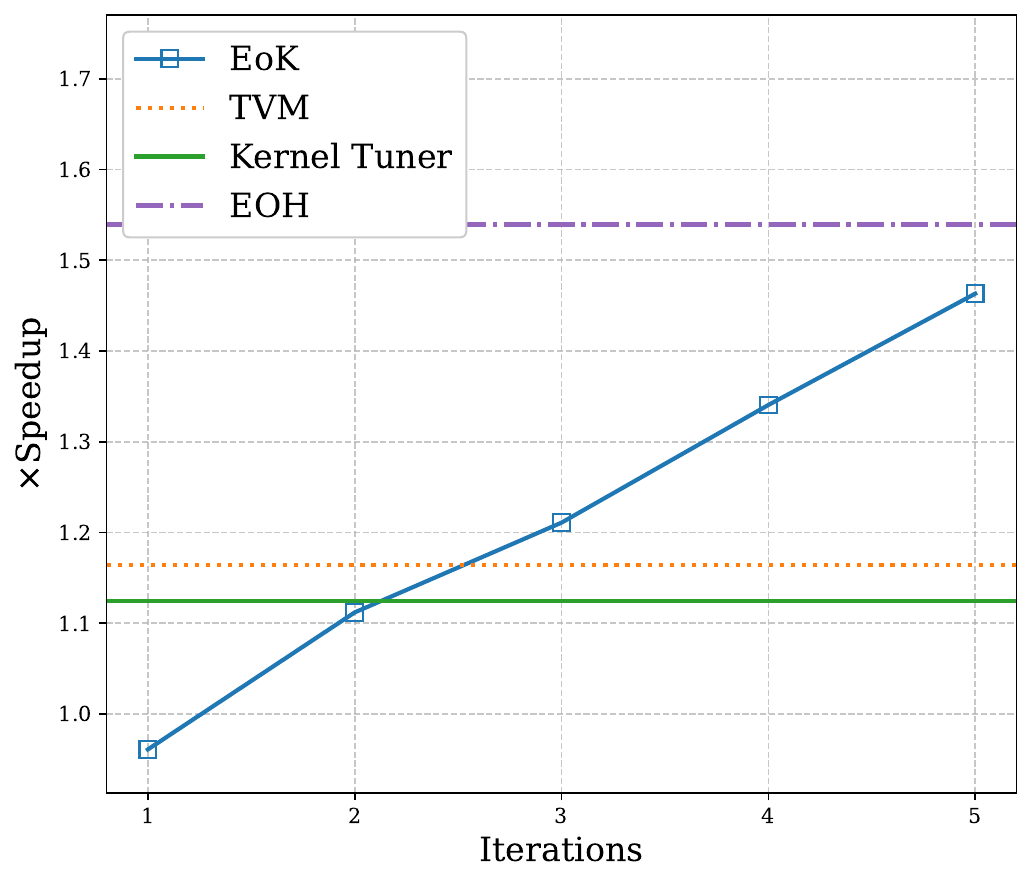}
         \caption*{Pooling}         
     \end{subfigure}\hfill
     \begin{subfigure}[b]{0.14\textwidth}
         \centering
         \includegraphics[width=.95\textwidth]{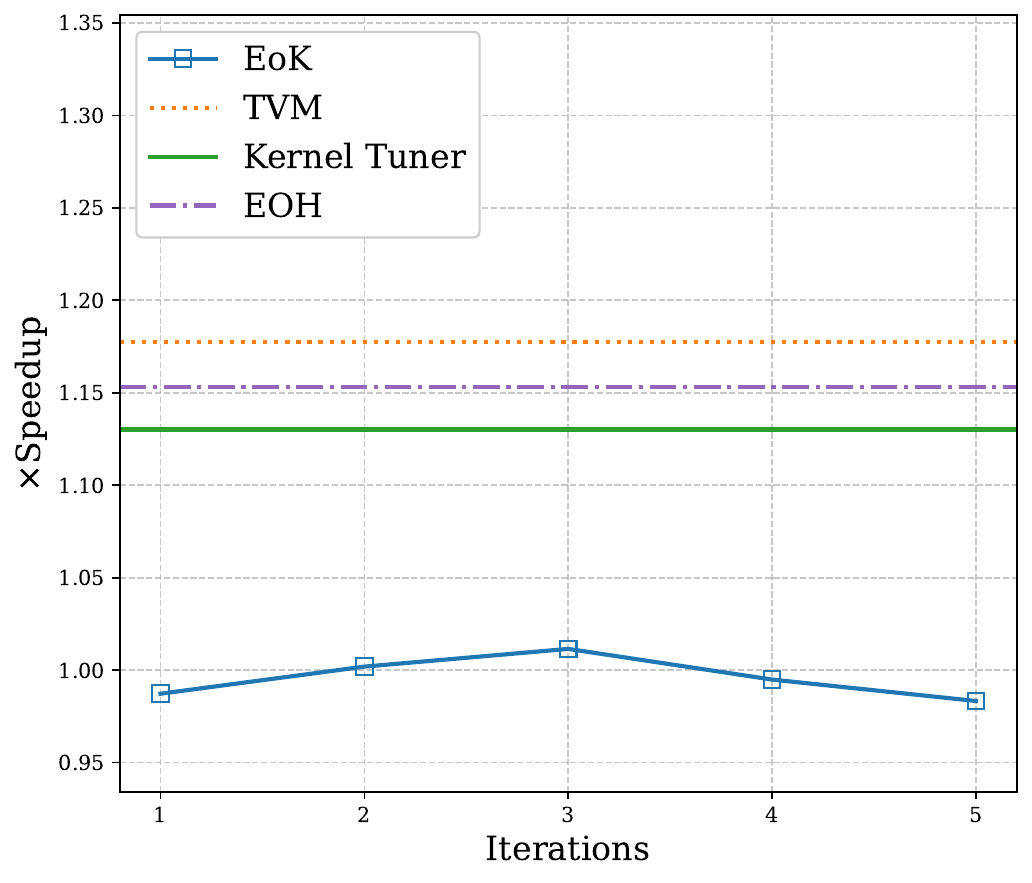}
         \caption*{PReLU}         
     \end{subfigure}\hfill
     \begin{subfigure}[b]{0.14\textwidth}
         \centering
         \includegraphics[width=.95\textwidth]{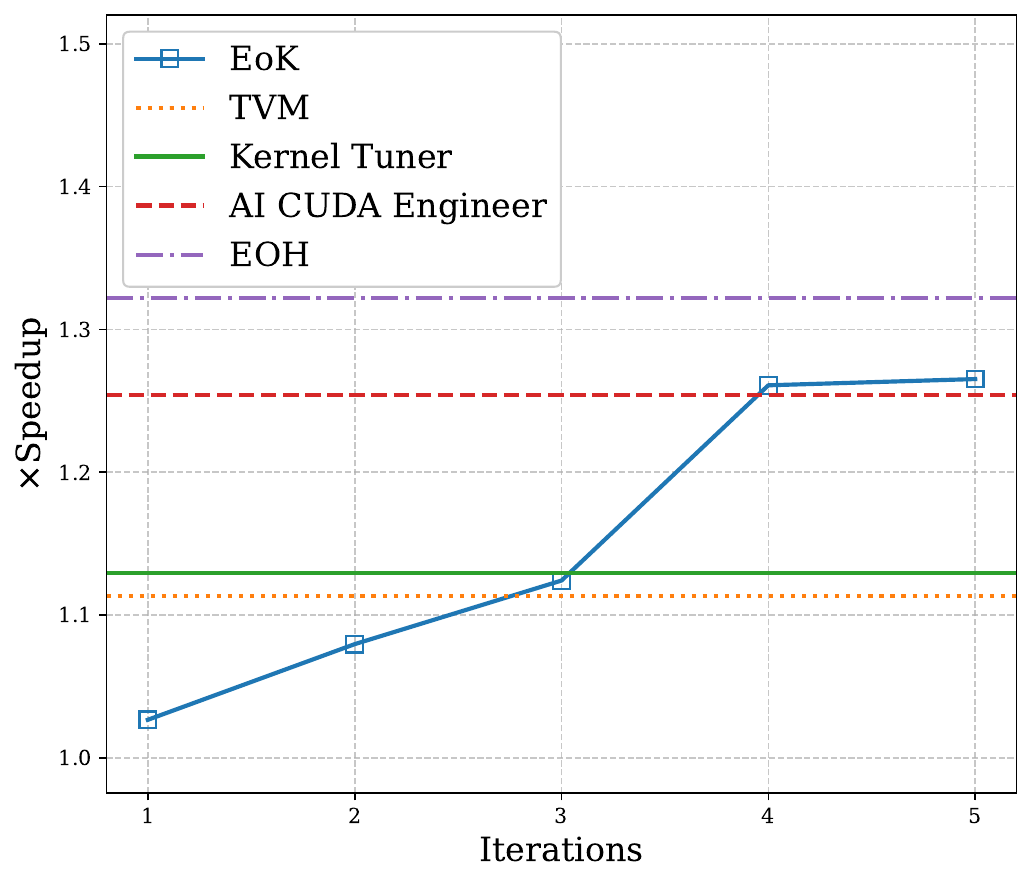}
         \caption*{ReLU}         
     \end{subfigure}\hfill
     \begin{subfigure}[b]{0.14\textwidth}
         \centering
         \includegraphics[width=.95\textwidth]{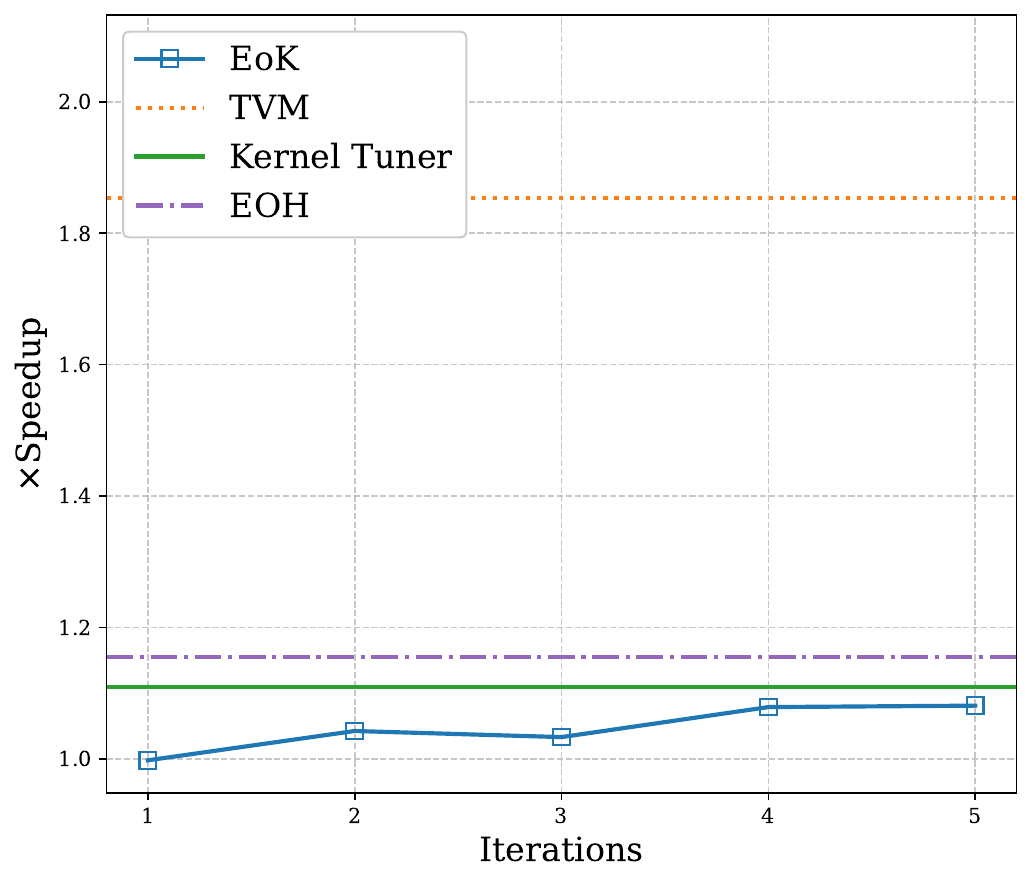}
         \caption*{SeLU}         
     \end{subfigure}\hfill
     \begin{subfigure}[b]{0.14\textwidth}
         \centering
         \includegraphics[width=.95\textwidth]{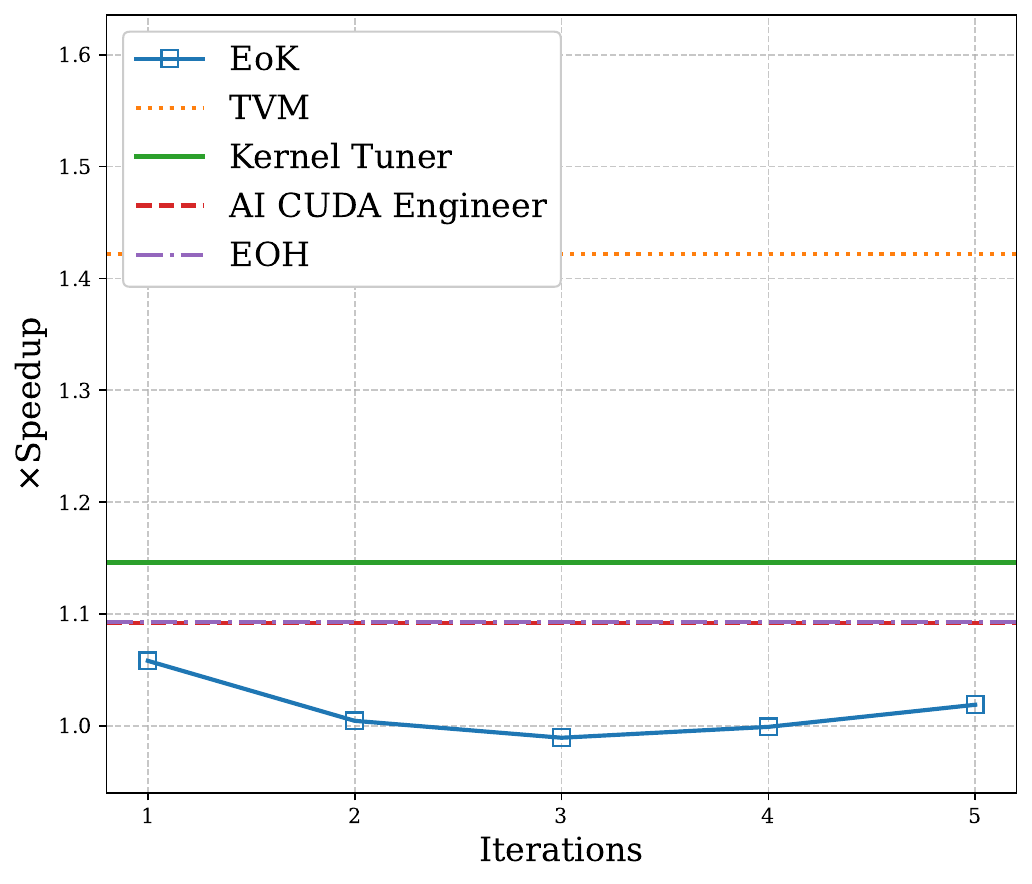}
         \caption*{Sigmoid}         
     \end{subfigure}\hfill
     \begin{subfigure}[b]{0.14\textwidth}
         \centering
         \includegraphics[width=.95\textwidth]{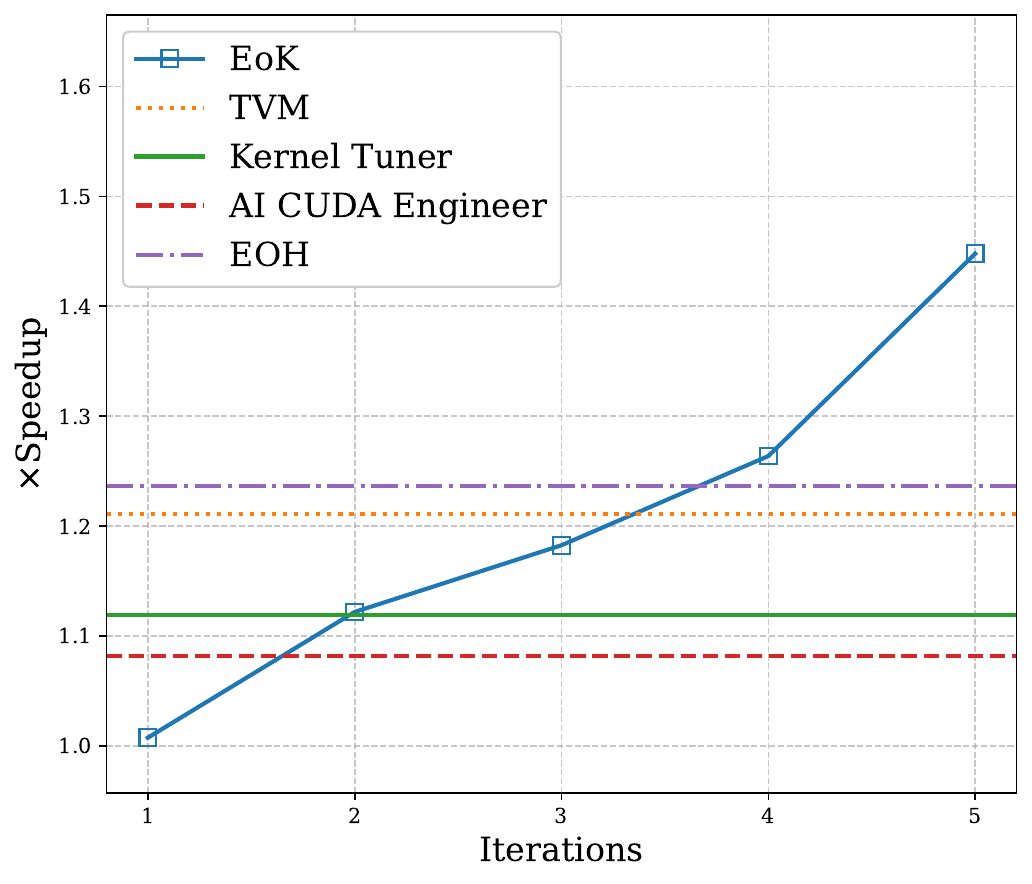}
         \caption*{Softmax}         
     \end{subfigure}\\
     \begin{subfigure}[b]{0.14\textwidth}
         \centering
         \includegraphics[width=.95\textwidth]{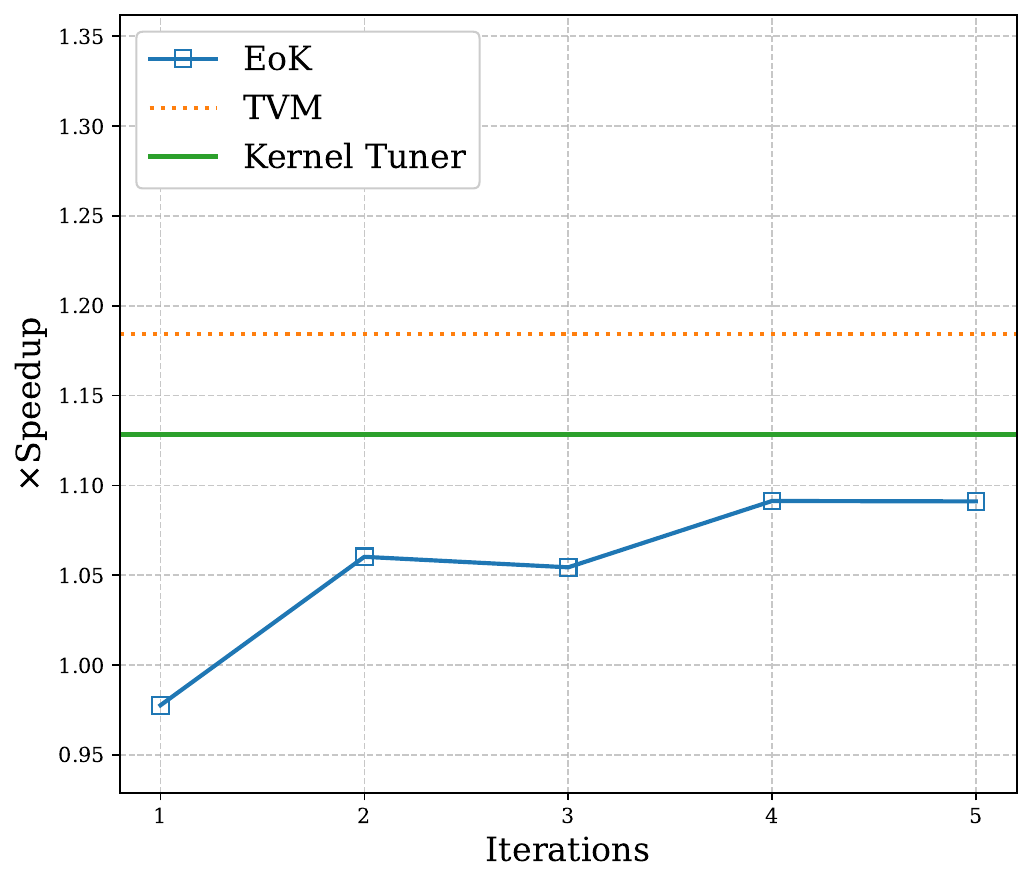}
         \caption*{Swish}         
     \end{subfigure}
     \begin{subfigure}[b]{0.14\textwidth}
         \centering
         \includegraphics[width=.95\textwidth]{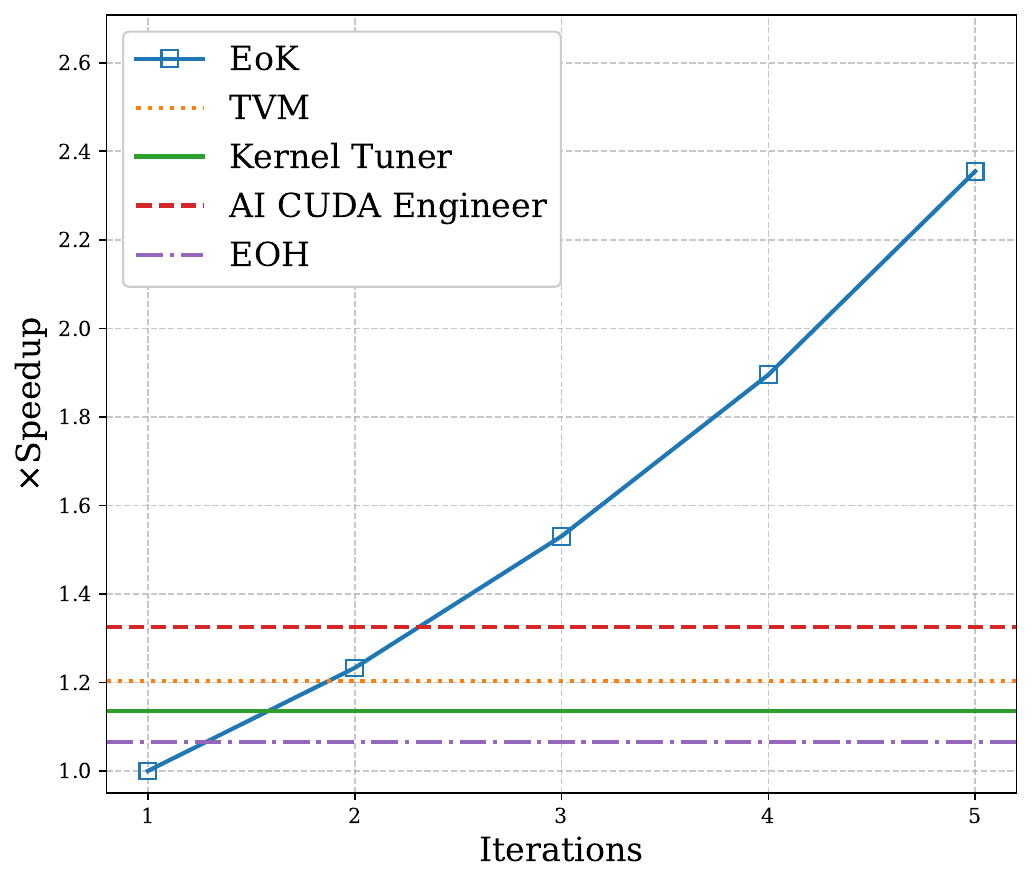}
         \caption*{Tanh}         
     \end{subfigure}
     \begin{subfigure}[b]{0.14\textwidth}
         \centering
         \includegraphics[width=.95\textwidth]{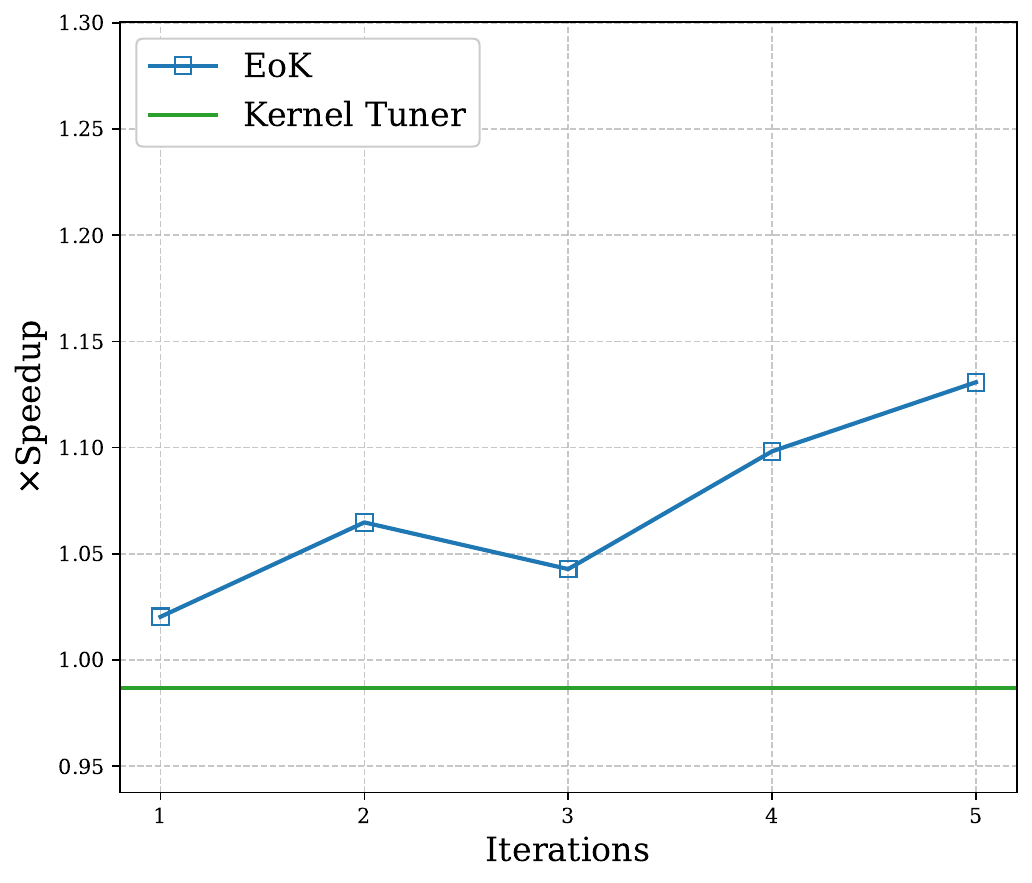}
         \caption*{Unary Op}         
     \end{subfigure}\hfill
     \caption{Plots of iteration versus speedup for \ourmethod{} across 38 full-precision neural network kernels. Speedups from other baseline methods are shown for a subset of kernels where these methods achieved at least one successful implementation.}
     \label{appxfig:nn_kernel_convergence_individual}
\end{figure}

Detailed comparison between \ourmethod{} and other baselines on 14 general-purpose and 66 neural network kernels provided in Figure~\ref{appxfig:gen _kernel_convergence_individual}, \ref{appxfig:nn_zfh_kernel_convergence_individual}, and ~\ref{appxfig:nn_kernel_convergence_individual}. 

\end{document}